\documentclass[letterpaper]{IEEEtran}%
\usepackage{amsfonts}
\usepackage{amsmath}
\usepackage{amssymb}
\usepackage{mathtools}
\usepackage{graphicx}
\usepackage[colorlinks=true,linkcolor=blue,citecolor=red,plainpages=false,pdfpagelabels]%
{hyperref}%
\providecommand{\U}[1]{\protect\rule{.1in}{.1in}}
\newtheorem{theorem}{Theorem}

\newtheorem{corollary}[theorem]{Corollary}

\newtheorem{definition}[theorem]{Definition}

\newtheorem{lemma}[theorem]{Lemma}

\newtheorem{proposition}[theorem]{Proposition}
\newtheorem{remark}[theorem]{Remark}

\allowdisplaybreaks
\begin{document}

\title{Bounding the forward classical capacity of bipartite quantum channels}
\author{Dawei Ding, Sumeet Khatri, Yihui Quek, Peter W.~Shor, Xin Wang, Mark
M.~Wilde\thanks{Dawei Ding is with Alibaba Quantum Laboratory, Alibaba Group,
Bellevue, Washington, USA. Mark M.~Wilde is with the Hearne Institute for
Theoretical Physics, Department of Physics and Astronomy, and Center for
Computation and Technology, Louisiana State University, Baton Rouge, Louisiana
70803, USA, and Sumeet Khatri had the same affiliation when this research was conducted. Yihui Quek
was with the Information Systems Laboratory, Stanford University, Stanford,
California 94305, USA, when this research was conducted. Sumeet Khatri and
Yihui Quek are now with the Dahlem Center for Complex Quantum Systems, Freie
Universit\"at Berlin, Berlin, Germany. Peter Shor is with the Center for
Theoretical Physics and Department of Mathematics, Massachusetts Institute of
Technology, Cambridge, Massachusetts 02139, USA. Xin Wang is with the
Institute for Quantum Computing, Baidu Research, Beijing 100193, China. Mark
M.~Wilde was also affiliated with the Stanford Institute for Theoretical
Physics, Stanford University, Stanford, California 94305, USA, when this
research was conducted. This work was presented in part at the 2021 IEEE
International Symposium on Information Theory, Melbourne, Victoria, Australia.
}}
\maketitle

\begin{abstract}
We introduce various measures of forward classical communication for bipartite
quantum channels. Since a point-to-point channel is a special case of a
bipartite channel, the measures reduce to measures of classical communication
for point-to-point channels. As it turns out, these reduced measures have been
reported in prior work of Wang \textit{et al}.~on bounding the classical
capacity of a quantum channel. As applications, we show that the measures are
upper bounds on the forward classical capacity of a bipartite channel. The
reduced measures are upper bounds on the classical capacity of a
point-to-point quantum channel assisted by a classical feedback channel. Some
of the various measures can be computed by semi-definite programming.

\end{abstract}
\tableofcontents


\section{Introduction}

The goal of quantum Shannon theory \cite{H17,H13book,Wat18,Wbook17} is to
characterize the information-processing capabilities of quantum states and
channels, for various tasks such as distillation of randomness, secret key,
entanglement or communication of classical, private, and quantum information.
With the goal of simplifying the theory or relating to practical communication
scenarios, often assisting resources are allowed, such as shared entanglement
\cite{PhysRevLett.69.2881,bennett2002entanglement,Hol01a}\ or public classical
communication \cite{BDSW96,FN98,BN05}.

One of the earliest information-theoretic tasks studied in quantum Shannon
theory is the capacity of a point-to-point quantum channel for transmitting
classical information or generating shared randomness
\cite{Holevo73,holevo1998capacity,schumacher1997sending}. It is well known
that the capacity for these two tasks is equal, and so we just refer to them
both as the classical capacity of a quantum channel. A formal expression for
the classical capacity of a quantum channel is known, given by what is called
the regularized Holevo information of a quantum channel (see, e.g.,
\cite{H17,H13book,Wat18,Wbook17} for reviews). On the one hand, it is unclear
whether this formal expression is computable for all quantum channels
\cite{WCP11}, and it is also known that the Holevo information formula is
generally non-additive \cite{Hastings09}. On the other hand, for some special
classes of channels \cite{S02,King03,KMNR07}, the regularized Holevo
information simplifies to what is known as the single-letter Holevo
information. Even when this simplification happens, it is not necessarily
guaranteed that the resulting capacity formula is efficiently computable
\cite{BS07}.

This difficulty in calculating the classical capacity of a quantum channel has
spurred the investigation of efficiently computable upper bounds on it
\cite{WXD18,WFT18,Fang2019a}. The main idea driving these bounds
\cite{WXD18}\ is to imagine a scenario in which the sender and receiver of a
quantum channel could be supplemented by the help of a coding scheme
that is simultaneously non-signaling and \textquotedblleft
positive-partial-transpose\textquotedblright\ (PPT)\ assisted
\cite{LM15,WXD18}. These latter coding scenarios could be considered
fictitious from a physical perspective, but the perspective is actually
extremely powerful when trying to provide an upper bound on the classical
capacity, while being faced with the aforementioned difficulties. The reason
is that the simultaneous constraints of being non-signaling and PPT-assisted
are semi-definite constraints and ultimately lead to upper bounds that are
efficiently computable by semi-definite programming.

Another thread that has been pursued on the topic of classical capacity, after
the initial investigations of
\cite{Holevo73,holevo1998capacity,schumacher1997sending}, is the classical
capacity of a quantum channel assisted by a classical feedback channel. This
direction started with \cite{FN98,BN05} and was ultimately inspired by Shannon's
work on the feedback-assisted capacity of a classical channel \cite{S56}, in
which it was shown that feedback does not increase capacity.\footnote{Note that the model of classical feedback considered in \cite{FN98} is more restrictive than the general model considered in \cite{BN05}---it is such that the decoding measurement of the receiver is restricted to one-way local operations and classical communication. See Section~4.2 of \cite{H17} for a review of the feedback scheme of \cite{FN98}.} For the quantum
case, it is known that a classical feedback channel does not enhance the
classical capacity of
\begin{enumerate}
\item an entanglement-breaking channel \cite{BN05},
\item a
pure-loss bosonic channel \cite{DQSW19}, and
\item a quantum erasure channel
\cite{DQSW19}.
\end{enumerate}
The first aforementioned result has been strengthened to a
strong-converse statement \cite{DingW18}. However, due to the quantum effect
of entanglement, it is also known that feedback can significantly increase the
classical capacity of certain quantum channels \cite{SS09}. More generally,
\cite{BDSS06}\ discussed several inequalities relating the classical capacity
assisted by classical feedback to other capacities, and \cite{GMW18}%
\ established inequalities relating classical capacities assisted by classical
communication to other notions of feedback-assisted capacities.

\begin{figure}[ptb]
\begin{center}
\includegraphics[
width=\linewidth
]{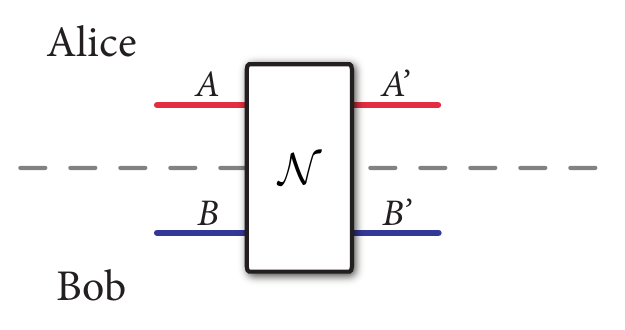}
\end{center}
\caption{Depiction of a bipartite channel. A bipartite channel is a
four-terminal channel in which Alice has access to the input system $A$ and
the output system $A^{\prime}$, and Bob has access to the input system $B$ and
the output system $B^{\prime}$.}%
\label{fig:bipartite-channel}%
\end{figure}

In this paper, we generalize these tasks further by considering the forward
classical capacity of a bipartite quantum channel, and we develop various
upper bounds on this operational communication measure for a bipartite
channel. This communication task has previously been studied for the special
case of bipartite unitary channels \cite{BHLS03} (see also
\cite{HL05,HS10,HL11}\ for studies of classical communication over bipartite
unitary channels). To be clear, a bipartite channel is a four-terminal channel
involving two parties, who each have access to one input port and one output
port of the channel \cite{CLL06} (see Figure~\ref{fig:bipartite-channel}). The
forward classical capacity of a bipartite channel is the optimal rate at which
Alice can communicate classical bits to Bob, with error probability tending to
zero as the number of channel uses becomes large.

A bipartite channel is an interesting communication scenario on its own, but
it is also a generalization of a point-to-point channel, with the latter being
a special case with one input port trivial and the output port for the other
party trivial. In the same way, it is a generalization of a classical feedback
channel. This relationship allows us to conclude upper bounds on the classical
capacity of a point-to-point channel assisted by classical feedback.
Interestingly, we prove that the most recent upper bound from \cite{Fang2019a}%
, on the unassisted classical capacity, is actually an upper bound on the
classical capacity assisted by classical feedback. As such, we now have an
efficiently computable upper bound on this feedback-assisted capacity. By combining our results with  \cite[Section~VI]{WFT18}, we establish the strong converse for the classical capacity of the quantum erasure channel assisted by a classical feedback channel, thus improving the weak converse result of \cite{DQSW19}.

The rest of the paper is structured as follows. We begin in
Section~\ref{sec:notation} by establishing some notation. In
Section~\ref{sec:bipartite-channel-measure}, we introduce a measure of forward
classical communication for a bipartite channel. Therein, we establish several
of its properties, including the fact that it is equal to zero for a product
of local channels, equal to zero for a classical feedback channel, and that it
is subadditive under serial compositions of bipartite channels. We then
introduce several variants of the basic measure and show how they reduce to
previous measures from \cite{WXD18,WFT18,Fang2019a}\ for point-to-point
channels. In Section~\ref{sec:apps}, we detail several of the applications
mentioned above:\ we establish that our measures of forward classical
communication (in particular the ones based on geometric R\'{e}nyi relative
entropy) serve as upper bounds on the forward classical capacity of a
bipartite channel and on the classical capacity of a point-to-point channel
assisted by a classical feedback channel. In
Section~\ref{sec:employ-sandwichedR}, we explore the same applications but
using the sandwiched R\'{e}nyi relative entropy instead of the geometric
R\'{e}nyi relative entropy, and in Section~\ref{sec:symmetries}, we show how
these bounds simplify if the channels possess symmetry. In
Section~\ref{sec:examples}, we evaluate our bounds for several examples of
bipartite and point-to-point channels. We finally conclude in
Section~\ref{sec:conclusion}\ with a summary and some open questions for
future work.

\section{Notation}

\label{sec:notation}

Here we list various notations and concepts that we use throughout the paper.
A quantum channel is a completely positive and trace-preserving map. We denote
the unnormalized maximally entangled operator by%
\begin{align}
\Gamma_{RA}  &  \coloneqq|\Gamma\rangle\!\langle\Gamma|_{RA},\\
|\Gamma\rangle_{RA}  &  \coloneqq\sum_{i=0}^{d-1}|i\rangle_{R}|i\rangle_{A},
\label{eq:gamma-def-unnorm}
\end{align}
where $R\simeq A$ with dimension $d$ and $\{|i\rangle_{R}\}_{i=0}^{d-1}$ and
$\{|i\rangle_{A}\}_{i=0}^{d-1}$ are orthonormal bases. The notation $R\simeq
A$\ means that the systems $R$ and $A$ are isomorphic. The maximally entangled
state is denoted by%
\begin{equation}
\Phi_{RA}\coloneqq\frac{1}{d}\Gamma_{RA},
\end{equation}
and the maximally mixed state by%
\begin{equation}
\pi_{A}\coloneqq\frac{1}{d}I_{A}.
\end{equation}
The Choi operator of a quantum channel $\mathcal{N}_{A\rightarrow B}$ (and
more generally a linear map) is denoted by%
\begin{equation}
\Gamma_{RB}^{\mathcal{N}}\coloneqq\mathcal{N}_{A\rightarrow B}(\Gamma_{RA}).
\end{equation}
A linear map $\mathcal{M}_{A\rightarrow B}$ is completely positive if and only
if its Choi operator $\Gamma_{RB}^{\mathcal{M}}$ is positive semi-definite,
and $\mathcal{M}_{A\rightarrow B}$ is trace preserving if and only if its Choi
operator satisfies $\operatorname{Tr}_{B}[\Gamma_{RB}^{\mathcal{M}}]=I_{R}$.

We denote the transpose map acting on the quantum system~$A$ by%
\begin{equation}
T_{A}(\cdot)\coloneqq\sum_{i,j=0}^{d-1}|i\rangle\!\langle j|_{A}%
(\cdot)|i\rangle\!\langle j|_{A}.
\end{equation}
A state $\rho_{AB}$ is a positive partial transpose (PPT) state if $T_{B}%
(\rho_{AB})$ is positive semi-definite. The partial transpose is its own
adjoint, in the sense that%
\begin{equation}
\operatorname{Tr}[Y_{AB}T_{A}(X_{AB})]=\operatorname{Tr}[T_{A}(Y_{AB})X_{AB}]
\end{equation}
for all linear operators $X_{AB}$ and $Y_{AB}$.

The following post-selected teleportation identity \cite{B05}\ plays a role in
our analysis:%
\begin{equation}
\mathcal{N}_{A\rightarrow B}(\rho_{SA})=\langle\Gamma|_{AR}\rho_{SA}%
\otimes\Gamma_{RB}^{\mathcal{N}}|\Gamma\rangle_{AR},
\label{eq:post-selected-TP-id}%
\end{equation}
as it has in previous works on feedback-assisted capacities
\cite{BW17,BDW18,Das2018thesis,KDWW19,DBW20,Fang2019a}. We also make frequent
use of the identities%
\begin{align}
\operatorname{Tr}_{A}[X_{AB}]  &  =\langle\Gamma|_{RA}(I_{R}\otimes
X_{AB})|\Gamma\rangle_{RA},\\
X_{AB}|\Gamma\rangle_{AR}  &  =T_{R}(X_{RB})|\Gamma\rangle_{AR}.
\end{align}
Given channels $\mathcal{N}_{A\rightarrow B}$ and $\mathcal{M}_{B\rightarrow
C}$, the Choi operator $\Gamma_{RC}^{\mathcal{M}\circ\mathcal{N}}$ of the
serial composition $\mathcal{M}_{B\rightarrow C}\circ\mathcal{N}_{A\rightarrow
B}$ is given by%
\begin{align}
\Gamma_{RC}^{\mathcal{M}\circ\mathcal{N}}  &  =\langle\Gamma|_{BS}\Gamma
_{RB}^{\mathcal{N}}\otimes\Gamma_{SC}^{\mathcal{M}}|\Gamma\rangle
_{BS}\label{eq:serial-compose-Choi}\\
&  =\operatorname{Tr}_{B}[\Gamma_{RB}^{\mathcal{N}}T_{B}(\Gamma_{BC}%
^{\mathcal{M}})],
\end{align}
where $B\simeq S$, the operator $\Gamma_{RB}^{\mathcal{N}}$ is the Choi
operator of $\mathcal{N}_{A\rightarrow B}$, and $\Gamma_{SC}^{\mathcal{M}}$ is
the Choi operator of $\mathcal{M}_{B\rightarrow C}$.

\section{Measures of forward classical communication for a bipartite
channel\label{sec:bipartite-channel-measure}}

\subsection{Basic measure}

Before defining the basic measure of forward classical communication for a
bipartite channel, let us recall some established concepts from quantum
information theory.

A bipartite channel $\mathcal{M}_{AB\rightarrow A^{\prime}B^{\prime}}$ is a
completely positive-partial-transpose preserving (C-PPT-P) channel if the
output state $\omega_{R_{A}A^{\prime}B^{\prime}R_{B}}\coloneqq \mathcal{M}%
_{AB\rightarrow A^{\prime}B^{\prime}}(\rho_{R_{A}ABR_{B}})$ is a PPT\ state
for every PPT input state $\rho_{R_{A}ABR_{B}}$ \cite{R99,R01,LM15,CVGG17}. To
be clear, the channel $\mathcal{M}_{AB\rightarrow A^{\prime}B^{\prime}}$ is
defined to be C-PPT-P if $T_{B^{\prime}R_{B}}(\omega_{R_{A}A^{\prime}%
B^{\prime}R_{B}})\geq0$ for every input state $\rho_{R_{A}ABR_{B}}$ that
satisfies $T_{BR_{B}}(\rho_{R_{A}ABR_{B}})\geq0$. Let $\Gamma_{AA^{\prime
}BB^{\prime}}^{\mathcal{M}}$ denote the Choi operator of $\mathcal{M}%
_{AB\rightarrow A^{\prime}B^{\prime}}$:%
\begin{equation}
\Gamma_{AA^{\prime}BB^{\prime}}^{\mathcal{M}}\coloneqq\mathcal{M}_{\hat{A}%
\hat{B}\rightarrow A^{\prime}B^{\prime}}(\Gamma_{A\hat{A}}\otimes\Gamma
_{B\hat{B}}),
\end{equation}
where $\hat{A}\simeq A$ and $\hat{B}\simeq B$. It is known that $\mathcal{M}%
_{AB\rightarrow A^{\prime}B^{\prime}}$ is C-PPT-P if and only if its Choi
operator $\Gamma_{AA^{\prime}BB^{\prime}}^{\mathcal{M}}$ is PPT\ (i.e.,
$T_{BB^{\prime}}(\Gamma_{AA^{\prime}BB^{\prime}}^{\mathcal{M}})\geq0$)
\cite{LM15,CVGG17}. Equivalently, it is known that $\mathcal{M}_{AB\rightarrow
A^{\prime}B^{\prime}}$ is C-PPT-P if and only if the map $T_{B^{\prime}}%
\circ\mathcal{M}_{AB\rightarrow A^{\prime}B^{\prime}}\circ T_{B}$ is
completely positive.

A C-PPT-P channel is not capable of generating entanglement shared between
Alice and Bob at a non-trivial rate when used many times \cite{BDW18,DBW20}.
As such, Alice cannot reliably communicate quantum information to Bob at a
non-zero rate when using a C-PPT-P channel. This feature is helpful for us in
devising a measure that serves as an upper bound on the forward classical
capacity of a bipartite channel or on the classical capacity of a
point-to-point channel assisted by classical feedback.

A bipartite channel $\mathcal{M}_{AB\rightarrow A^{\prime}B^{\prime}}$ is
non-signaling from Alice to Bob \cite{BGNP01,ESW02} if the following condition
holds \cite{PHHH06}%
\begin{equation}
\operatorname{Tr}_{A^{\prime}}\circ\mathcal{M}_{AB\rightarrow A^{\prime
}B^{\prime}}=\operatorname{Tr}_{A^{\prime}}\circ\mathcal{M}_{AB\rightarrow
A^{\prime}B^{\prime}}\circ\mathcal{R}_{A}^{\pi}, \label{eq:NS-A-to-B-cond}%
\end{equation}
where $\mathcal{R}_{A}^{\pi}$ is a replacer channel, defined as $\mathcal{R}%
_{A}^{\pi}(\cdot)\coloneqq \operatorname{Tr}_{A}[\cdot]\pi_{A}$, with $\pi
_{A}\coloneqq I_{A}/d_{A}$ the maximally mixed state on system $A$. To
interpret this condition, consider the following. For Bob, the reduced state
of his output system $B^{\prime}$ is obtained by tracing out Alice's output
system $A^{\prime}$. Note that the reduced state on $B^{\prime}$ is all that
Bob can access at the output in this scenario. If the condition in
\eqref{eq:NS-A-to-B-cond} holds, then the reduced state on Bob's output system
$B^{\prime}$ has no dependence on Alice's input system. Thus, if
\eqref{eq:NS-A-to-B-cond} holds, then Alice cannot use $\mathcal{M}%
_{AB\rightarrow A^{\prime}B^{\prime}}$ to send a signal to Bob.

One of our main interests in this paper is to bound the classical capacity of
a quantum channel assisted by a classical feedback channel from Bob to Alice.
In such a protocol, local channels are allowed for free, as well as the use of
a classical feedback channel. Both of these actions can be considered as
particular kinds of bipartite channels and both of them fall into the class of
bipartite channels that are non-signaling from Alice to Bob and C-PPT-P (call
this class NS$_{A\not \rightarrow B}~\cap$~PPT). As such, if we employ a
measure of bipartite channels that involves a comparison between a bipartite
channel of interest to all bipartite channels in NS$_{A\not \rightarrow
B}~\cap$~PPT, then the two kinds of free channels would have zero value and
the measure would indicate how different the channel of interest is from this
set (i.e., how different it is from a channel that has no ability to send
quantum information and no ability to signal from Alice to Bob). This is the
main idea behind the measure that we propose below in
Definition~\ref{def:beta-measure-bipartite-ch}, but one should keep in mind
that the measure below does not follow this reasoning precisely.

In Definition~\ref{def:beta-measure-bipartite-ch}, although we motivated the
measure for bipartite channels, we define it more generally for completely
positive bipartite maps, as it turns out to be useful to do so when we later
define other measures.

\begin{definition}
\label{def:beta-measure-bipartite-ch}
Let $\mathcal{M}_{AB\rightarrow
A^{\prime}B^{\prime}}$ be a completely positive bipartite map. Then we define%
\begin{equation}
C_{\beta}(\mathcal{M}_{AB\rightarrow A^{\prime}B^{\prime}})\coloneqq\log
_{2}\beta(\mathcal{M}_{AB\rightarrow A^{\prime}B^{\prime}}),
\end{equation}
\begin{multline}
\beta(\mathcal{M}_{AB\rightarrow A^{\prime}B^{\prime}})\coloneqq\\
\inf_{\substack{S_{AA^{\prime}BB^{\prime}},\\V_{AA^{\prime}BB^{\prime}}%
\in\operatorname{Herm}}}\left\{
\begin{array}
[c]{c}%
\left\Vert \operatorname{Tr}_{A^{\prime}B^{\prime}}[S_{AA^{\prime}BB^{\prime}%
}]\right\Vert _{\infty}:\\
T_{BB^{\prime}}(V_{AA^{\prime}BB^{\prime}}\pm\Gamma_{AA^{\prime}BB^{\prime}%
}^{\mathcal{M}})\geq0,\\
S_{AA^{\prime}BB^{\prime}}\pm V_{AA^{\prime}BB^{\prime}}\geq0,\\
\operatorname{Tr}_{A^{\prime}}[S_{AA^{\prime}BB^{\prime}}]=\\
\pi_{A}\otimes\operatorname{Tr}_{AA^{\prime}}[S_{AA^{\prime}BB^{\prime}}]
\end{array}
\right\}  , \label{eq:basic-measure-NS-PPT-bi}%
\end{multline}
where $\operatorname{Herm}$ denotes the set of Hermitian operators and
$\Gamma_{AA^{\prime}BB^{\prime}}^{\mathcal{M}}$ is the Choi operator of
$\mathcal{M}_{AB\rightarrow A^{\prime}B^{\prime}}$:%
\begin{equation}
\Gamma_{AA^{\prime}BB^{\prime}}^{\mathcal{M}}\coloneqq\mathcal{M}_{\hat{A}%
\hat{B}\rightarrow A^{\prime}B^{\prime}}(\Gamma_{A\hat{A}}\otimes\Gamma
_{B\hat{B}}).
\end{equation}
In the above, $\hat{A}\simeq A$, $\hat{B}\simeq B$,%
\begin{equation}
\Gamma_{A\hat{A}}\coloneqq\sum_{i,j=0}^{d_{A}-1}|i\rangle\!\langle
j|_{A}\otimes|i\rangle\!\langle j|_{\hat{A}},\quad\Gamma_{B\hat{B}%
}\coloneqq\sum_{i,j=0}^{d_{B}-1}|i\rangle\!\langle j|_{B}\otimes
|i\rangle\!\langle j|_{\hat{B}},
\end{equation}
and $\pi_{A}\coloneqq I_{A}/d_{A}$.
\end{definition}

Just before Definition~\ref{def:beta-measure-bipartite-ch}, we discussed how the $\beta$ measure incorporates PPT constraints, as well as non-signaling constraints. The constraint $T_{BB^{\prime}}(V_{AA^{\prime}BB^{\prime}}\pm\Gamma_{AA^{\prime}BB^{\prime}%
}^{\mathcal{M}})\geq0$ involves a PPT condition, and the constraint $\operatorname{Tr}_{A^{\prime}}[S_{AA^{\prime}BB^{\prime}}]=
\pi_{A}\otimes\operatorname{Tr}_{AA^{\prime}}[S_{AA^{\prime}BB^{\prime}}]
$ involves a non-signaling condition. Since $S_{AA^{\prime}BB^{\prime}}\pm V_{AA^{\prime}BB^{\prime}}\geq0$, it follows
that $S_{AA^{\prime}BB^{\prime}}\geq0$, implying that the operator $S_{AA^{\prime}BB^{\prime}}$ corresponds to a completely positive map. The definition above becomes more transparent, but however does not decrease, if we simply set $V_{AA^{\prime}BB^{\prime}} = S_{AA^{\prime}BB^{\prime}}$. Then it is clear that there is just a PPT constraint and non-signaling constraint corresponding to a single completely positive map. Furthermore, as we discuss below, the objective function 
$\left\Vert \operatorname{Tr}_{A^{\prime}B^{\prime}}[S_{AA^{\prime}BB^{\prime}%
}]\right\Vert _{\infty}$ measures how close $S_{AA^{\prime}BB^{\prime}}$ is to being a trace preserving map, and Proposition~\ref{prop:lower-bound-ch-non-neg} states that the minimum value of the objective function is one, in which case $S_{AA^{\prime}BB^{\prime}}$ corresponds to a quantum channel.

In Appendix~\ref{app:cp-formulation-beta}, we prove that $\beta(\mathcal{M}%
_{AB\rightarrow A^{\prime}B^{\prime}})$ can alternatively be expressed as
follows:%
\begin{equation}
\beta(\mathcal{M}_{AB\rightarrow A^{\prime}B^{\prime}})= \inf
_{\substack{\mathcal{S}_{AB\rightarrow A^{\prime}B^{\prime}},\\\mathcal{V}%
_{AB\rightarrow A^{\prime}B^{\prime}}\in\operatorname{HermP}}} \left\Vert
\mathcal{S}_{AB\rightarrow A^{\prime}B^{\prime}}\right\Vert _{1}%
\end{equation}
subject to
\begin{equation}%
\begin{array}
[c]{c}%
T_{B^{\prime}}\circ(\mathcal{V}_{AB\rightarrow A^{\prime}B^{\prime}}%
\pm\mathcal{M}_{AB\rightarrow A^{\prime}B^{\prime}})\circ T_{B}\geq0,\\
\mathcal{S}_{AB\rightarrow A^{\prime}B^{\prime}}\pm\mathcal{V}_{AB\rightarrow
A^{\prime}B^{\prime}}\geq0,\\
\operatorname{Tr}_{A^{\prime}}\circ\mathcal{S}_{AB\rightarrow A^{\prime
}B^{\prime}}=\operatorname{Tr}_{A^{\prime}}\circ\mathcal{S}_{AB\rightarrow
A^{\prime}B^{\prime}}\circ\mathcal{R}_{A}^{\pi}%
\end{array}
, \label{eq:CP-exp-beta}%
\end{equation}
where $\operatorname{HermP}$ is the set of Hermiticity preserving maps,
$\left\Vert \mathcal{S}_{AB\rightarrow A^{\prime}B^{\prime}}\right\Vert _{1}$
is the trace norm of the bipartite map $\mathcal{S}_{AB\rightarrow A^{\prime
}B^{\prime}}$, and the notation $\mathcal{L}_{AB\rightarrow A^{\prime
}B^{\prime}}\geq0$ means that the Hermiticity preserving map $\mathcal{L}%
_{AB\rightarrow A^{\prime}B^{\prime}}$ is completely positive. Related to how $S_{AA^{\prime}BB^{\prime}} \geq 0$ as discussed above, the constraint $\mathcal{S}_{AB\rightarrow A^{\prime}B^{\prime}}\pm\mathcal{V}_{AB\rightarrow
A^{\prime}B^{\prime}}\geq0$ implies that $\mathcal{S}_{AB\rightarrow A^{\prime}B^{\prime}}\geq0$, which is the same as $\mathcal{S}_{AB\rightarrow A^{\prime}B^{\prime}}$ being a completely positive map. Thus, the trace norm objective function $\left\Vert
\mathcal{S}_{AB\rightarrow A^{\prime}B^{\prime}}\right\Vert _{1}$ measures how close $\mathcal{S}_{AB\rightarrow A^{\prime}B^{\prime}}$ can be to a trace preserving map, i.e., a quantum channel, while satisfying the constraints given. With the
expression in \eqref{eq:CP-exp-beta}, it might become more clear that
$\beta(\mathcal{M}_{AB\rightarrow A^{\prime}B^{\prime}})$ involves a
comparison of $\mathcal{M}_{AB\rightarrow A^{\prime}B^{\prime}}$ to other
Hermiticity-preserving bipartite maps, which involves the C-PPT-P condition
and the non-signaling constraint. In Appendix~\ref{app:cp-formulation-beta},
not only do we prove the equality above, but we also explain these concepts in
more detail.

We can also express $\beta
(\mathcal{M}_{AB\rightarrow A^{\prime}B^{\prime}})$ as follows:%
\begin{equation}
\inf_{\substack{\lambda\geq 0, S_{AA^{\prime}BB^{\prime}},\\V_{AA^{\prime}BB^{\prime}}%
\in\operatorname{Herm}}}\left\{
\begin{array}
[c]{c}%
\lambda:\\
\operatorname{Tr}_{A^{\prime}B^{\prime}}[S_{AA^{\prime}BB^{\prime}}%
]\leq\lambda I_{AB}\\
T_{BB^{\prime}}(V_{AA^{\prime}BB^{\prime}}\pm\Gamma_{AA^{\prime}BB^{\prime}%
}^{\mathcal{M}})\geq0,\\
S_{AA^{\prime}BB^{\prime}}\pm V_{AA^{\prime}BB^{\prime}}\geq0,\\
\operatorname{Tr}_{A^{\prime}}[S_{AA^{\prime}BB^{\prime}}]=\\
\pi_{A}\otimes\operatorname{Tr}_{AA^{\prime}}[S_{AA^{\prime}BB^{\prime}}]
\end{array}
\right\} .
\end{equation}

By exploiting the equality constraint $\operatorname{Tr}_{A^{\prime}%
}[S_{AA^{\prime}BB^{\prime}}]=\pi_{A}\otimes\operatorname{Tr}_{AA^{\prime}%
}[S_{AA^{\prime}BB^{\prime}}]$, we find that%
\begin{align}
&  \left\Vert \operatorname{Tr}_{A^{\prime}B^{\prime}}[S_{AA^{\prime
}BB^{\prime}}]\right\Vert _{\infty}\nonumber\\
&  =\left\Vert \operatorname{Tr}_{B^{\prime}}[\operatorname{Tr}_{A^{\prime}%
}[S_{AA^{\prime}BB^{\prime}}]]\right\Vert _{\infty}\\
&  =\left\Vert \operatorname{Tr}_{B^{\prime}}[\pi_{A}\otimes\operatorname{Tr}%
_{AA^{\prime}}[S_{AA^{\prime}BB^{\prime}}]]\right\Vert _{\infty}\\
&  =\left\Vert \pi_{A}\otimes\operatorname{Tr}_{AA^{\prime}B^{\prime}%
}[S_{AA^{\prime}BB^{\prime}}]\right\Vert _{\infty}\\
&  =\frac{1}{d_{A}}\left\Vert \operatorname{Tr}_{AA^{\prime}B^{\prime}%
}[S_{AA^{\prime}BB^{\prime}}]\right\Vert _{\infty}.
\end{align}
Then we find that%
\begin{multline}
\beta(\mathcal{M}_{AB\rightarrow A^{\prime}B^{\prime}}) =\\
\inf_{\substack{S_{AA^{\prime}BB^{\prime}},\\V_{AA^{\prime}BB^{\prime}}%
\in\operatorname{Herm}}}\left\{
\begin{array}
[c]{c}%
\frac{1}{d_{A}}\left\Vert \operatorname{Tr}_{AA^{\prime}B^{\prime}%
}[S_{AA^{\prime}BB^{\prime}}]\right\Vert _{\infty}:\\
T_{BB^{\prime}}(V_{AA^{\prime}BB^{\prime}}\pm\Gamma_{AA^{\prime}BB^{\prime}%
}^{\mathcal{M}})\geq0,\\
S_{AA^{\prime}BB^{\prime}}\pm V_{AA^{\prime}BB^{\prime}}\geq0,\\
\operatorname{Tr}_{A^{\prime}}[S_{AA^{\prime}BB^{\prime}}]=\\
\pi_{A}\otimes\operatorname{Tr}_{AA^{\prime}}[S_{AA^{\prime}BB^{\prime}}]
\end{array}
\right\}  .
\end{multline}
Since $S_{AA^{\prime}BB^{\prime}}\pm V_{AA^{\prime}BB^{\prime}}\geq0$ implies
that $S_{AA^{\prime}BB^{\prime}}\geq0$, we can also rewrite $\beta
(\mathcal{M}_{AB\rightarrow A^{\prime}B^{\prime}})$ as%
\begin{multline}
\beta(\mathcal{M}_{AB\rightarrow A^{\prime}B^{\prime}})=\\
\inf_{\substack{\lambda,S_{AA^{\prime}BB^{\prime}}\geq0,\\V_{AA^{\prime
}BB^{\prime}}\in\operatorname{Herm}}}\left\{
\begin{array}
[c]{c}%
\lambda:\\
\frac{1}{d_{A}}\operatorname{Tr}_{AA^{\prime}B^{\prime}}[S_{AA^{\prime
}BB^{\prime}}]\leq\lambda I_{B},\\
T_{BB^{\prime}}(V_{AA^{\prime}BB^{\prime}}\pm\Gamma_{AA^{\prime}BB^{\prime}%
}^{\mathcal{M}})\geq0,\\
S_{AA^{\prime}BB^{\prime}}\pm V_{AA^{\prime}BB^{\prime}}\geq0,\\
\operatorname{Tr}_{A^{\prime}}[S_{AA^{\prime}BB^{\prime}}]=\\
\pi_{A}\otimes\operatorname{Tr}_{AA^{\prime}}[S_{AA^{\prime}BB^{\prime}}]
\end{array}
\right\}  . \label{eq:basic-measure-NS-PPT-bi-rewrite}%
\end{multline}

\subsection{Properties of the basic measure\label{sec:props-basic-measure}}

We now establish several properties of $C_{\beta}(\mathcal{N}_{AB\rightarrow
A^{\prime}B^{\prime}})$, which are basic properties that we might expect of a
measure of forward classical communication for a bipartite channel. These
include the following:

\begin{enumerate}
\item non-negativity (Proposition~\ref{prop:lower-bound-ch-non-neg}),

\item stability under tensoring with identity channels
(Proposition~\ref{prop:stability}),

\item zero value for classical feedback channels
(Proposition~\ref{prop:zero-feedback}),

\item zero value for a tensor product of local channels
(Proposition~\ref{prop:zero-local-chs}),

\item subadditivity under serial composition
(Proposition~\ref{prop:subadditivity-bipartite-CP-maps}),

\item data processing under pre- and post-processing by local channels
(Corollary~\ref{cor:DP-LC-beta-p-bipartite-ch}),

\item invariance under local unitary channels
(Corollary~\ref{cor:I-LUC-beta-p-bipartite-ch}),

\item convexity of $\beta$ (Proposition~\ref{prop:convexity-beta}).
\end{enumerate}

All of the properties above hold for bipartite channels, while the second and
fifth through eighth hold more generally for completely positive bipartite maps.

\begin{proposition}
[Non-negativity]\label{prop:lower-bound-ch-non-neg}Let $\mathcal{N}%
_{AB\rightarrow A^{\prime}B^{\prime}}$ be a bipartite channel. Then%
\begin{equation}
C_{\beta}(\mathcal{N}_{AB\rightarrow A^{\prime}B^{\prime}})\geq0.
\end{equation}

\end{proposition}

\begin{IEEEproof}
We prove the equivalent statement $\beta(\mathcal{N}_{AB\rightarrow A^{\prime
}B^{\prime}})\geq1$. Let $\lambda$, $S_{AA^{\prime}BB^{\prime}}$, and
$V_{AA^{\prime}BB^{\prime}}$ be arbitrary Hermitian operators satisfying the
constraints in \eqref{eq:basic-measure-NS-PPT-bi-rewrite}. Then consider that%
\begin{align}
\lambda d_{B}  &  =\lambda\operatorname{Tr}_{B}[I_{B}]\\
&  \geq\frac{1}{d_{A}}\operatorname{Tr}_{AA^{\prime}BB^{\prime}}%
[S_{AA^{\prime}BB^{\prime}}]\\
&  \geq\frac{1}{d_{A}}\operatorname{Tr}_{AA^{\prime}BB^{\prime}}%
[V_{AA^{\prime}BB^{\prime}}]\\
&  =\frac{1}{d_{A}}\operatorname{Tr}_{AA^{\prime}BB^{\prime}}[T_{BB^{\prime}%
}(V_{AA^{\prime}BB^{\prime}})]\\
&  \geq \frac{1}{d_{A}} \operatorname{Tr}_{AA^{\prime}BB^{\prime}}[T_{BB^{\prime}}(\Gamma_{AA^{\prime
}BB^{\prime}}^{\mathcal{N}})]\\
&  = \frac{1}{d_{A}} \operatorname{Tr}_{AA^{\prime}BB^{\prime}}[\Gamma_{AA^{\prime
}BB^{\prime}}^{\mathcal{N}}]\\
&  =\frac{1}{d_{A}}\operatorname{Tr}_{AB}[I_{AB}]\\
&  =d_{B}.
\end{align}
This implies that $\lambda\geq1$. Since the inequality holds for all $\lambda
$, $S_{AA^{\prime}BB^{\prime}}$, and $V_{AA^{\prime}BB^{\prime}}$ satisfying
the constraints in \eqref{eq:basic-measure-NS-PPT-bi-rewrite}, we conclude the
statement above.
\end{IEEEproof}

\bigskip

\begin{proposition}
[Stability]\label{prop:stability}Let $\mathcal{M}_{AB\rightarrow A^{\prime
}B^{\prime}}$ be a completely positive bipartite map.\ Then%
\begin{equation}
C_{\beta}(\operatorname{id}_{\bar{A}\rightarrow\tilde{A}}\otimes
\mathcal{M}_{AB\rightarrow A^{\prime}B^{\prime}}\otimes\operatorname{id}%
_{\bar{B}\rightarrow\tilde{B}})=C_{\beta}(\mathcal{M}_{AB\rightarrow
A^{\prime}B^{\prime}}).
\end{equation}

\end{proposition}

\begin{IEEEproof}
Let $S_{AA^{\prime}BB^{\prime}}$ and $V_{AA^{\prime}BB^{\prime}}$ be arbitrary
Hermitian operators satisfying the constraints in
\eqref{eq:basic-measure-NS-PPT-bi} for $\mathcal{M}_{AB\rightarrow A^{\prime
}B^{\prime}}$. The Choi operator of $\operatorname{id}_{\bar{A}\rightarrow
\tilde{A}}\otimes\mathcal{M}_{AB\rightarrow A^{\prime}B^{\prime}}%
\otimes\operatorname{id}_{\bar{B}\rightarrow\tilde{B}}$ is given by%
\begin{equation}
\Gamma_{\bar{A}\tilde{A}}\otimes\Gamma_{AA^{\prime}BB^{\prime}}^{\mathcal{M}%
}\otimes\Gamma_{\bar{B}\tilde{B}}.
\end{equation}
Let us show that $\Gamma_{\bar{A}\tilde{A}}\otimes S_{AA^{\prime}BB^{\prime}%
}\otimes\Gamma_{\bar{B}\tilde{B}}$ and $\Gamma_{\bar{A}\tilde{A}}\otimes
V_{AA^{\prime}BB^{\prime}}\otimes\Gamma_{\bar{B}\tilde{B}}$ satisfy the
constraints in \eqref{eq:basic-measure-NS-PPT-bi} for $\operatorname{id}%
_{\bar{A}\rightarrow\tilde{A}}\otimes\mathcal{M}_{AB\rightarrow A^{\prime
}B^{\prime}}\otimes\operatorname{id}_{\bar{B}\rightarrow\tilde{B}}$. Consider
that%
\begin{equation}
T_{BB^{\prime}}(V_{AA^{\prime}BB^{\prime}}\pm\Gamma_{AA^{\prime}BB^{\prime}%
}^{\mathcal{M}})\geq0
\end{equation}%
\begin{multline}
\Leftrightarrow\quad T_{BB^{\prime}}(\Gamma_{\bar{A}\tilde{A}}\otimes
V_{AA^{\prime}BB^{\prime}}\otimes\Gamma_{\bar{B}\tilde{B}})\geq\\
\pm T_{BB^{\prime}}(\Gamma_{\bar{A}\tilde{A}}\otimes\Gamma_{AA^{\prime
}BB^{\prime}}^{\mathcal{M}}\otimes\Gamma_{\bar{B}\tilde{B}})
\end{multline}%
\begin{multline}
\Leftrightarrow\quad T_{BB^{\prime}\bar{B}\tilde{B}}(\Gamma_{\bar{A}\tilde{A}%
}\otimes V_{AA^{\prime}BB^{\prime}}\otimes\Gamma_{\bar{B}\tilde{B}})\geq\\
\pm T_{BB^{\prime}\bar{B}\tilde{B}}(\Gamma_{\bar{A}\tilde{A}}\otimes
\Gamma_{AA^{\prime}BB^{\prime}}^{\mathcal{M}}\otimes\Gamma_{\bar{B}\tilde{B}})
\end{multline}
and%
\begin{equation}
S_{AA^{\prime}BB^{\prime}}\pm V_{AA^{\prime}BB^{\prime}}\geq0
\end{equation}%
\begin{multline}
\Leftrightarrow\quad\Gamma_{\bar{A}\tilde{A}}\otimes S_{AA^{\prime}BB^{\prime
}}\otimes\Gamma_{\bar{B}\tilde{B}}\geq\\
\pm\Gamma_{\bar{A}\tilde{A}}\otimes V_{AA^{\prime}BB^{\prime}}\otimes
\Gamma_{\bar{B}\tilde{B}}%
\end{multline}
and%
\begin{equation}
\operatorname{Tr}_{A^{\prime}}[S_{AA^{\prime}BB^{\prime}}]=\pi_{A}%
\otimes\operatorname{Tr}_{AA^{\prime}}[S_{AA^{\prime}BB^{\prime}}]
\end{equation}
\begin{align}
&  \Leftrightarrow\quad\operatorname{Tr}_{A^{\prime}\tilde{A}}[\Gamma_{\bar
{A}\tilde{A}}\otimes S_{AA^{\prime}BB^{\prime}}\otimes\Gamma_{\bar{B}\tilde
{B}}]\nonumber\\
&  =I_{\bar{A}}\otimes\pi_{A}\otimes\operatorname{Tr}_{AA^{\prime}%
}[S_{AA^{\prime}BB^{\prime}}\otimes\Gamma_{\bar{B}\tilde{B}}]\\
&  =\pi_{\bar{A}}\otimes\pi_{A}\otimes\operatorname{Tr}_{AA^{\prime}\bar
{A}\tilde{A}}[\Gamma_{\bar{A}\tilde{A}}\otimes S_{AA^{\prime}BB^{\prime}%
}\otimes\Gamma_{\bar{B}\tilde{B}}].
\end{align}
Also, consider that%
\begin{align}
&  \frac{1}{d_{A}d_{\bar{A}}}\left\Vert \operatorname{Tr}_{AA^{\prime}\bar
{A}\tilde{A}B^{\prime}\tilde{B}}[\Gamma_{\bar{A}\tilde{A}}\otimes
S_{AA^{\prime}BB^{\prime}}\otimes\Gamma_{\bar{B}\tilde{B}}]\right\Vert
_{\infty}\nonumber\\
&  =\frac{1}{d_{A}d_{\bar{A}}}\left\Vert d_{\bar{A}}\operatorname{Tr}%
_{AA^{\prime}B^{\prime}}[S_{AA^{\prime}BB^{\prime}}\otimes I_{\bar{B}%
}]\right\Vert _{\infty}\\
&  =\frac{1}{d_{A}}\left\Vert \operatorname{Tr}_{AA^{\prime}B^{\prime}%
}[S_{AA^{\prime}BB^{\prime}}]\otimes I_{\bar{B}}\right\Vert _{\infty}\\
&  =\frac{1}{d_{A}}\left\Vert \operatorname{Tr}_{AA^{\prime}B^{\prime}%
}[S_{AA^{\prime}BB^{\prime}}]\right\Vert _{\infty}.
\end{align}
Thus, it follows that%
\begin{equation}
\beta(\mathcal{M}_{AB\rightarrow A^{\prime}B^{\prime}})\geq\beta
(\operatorname{id}_{\bar{A}\rightarrow\tilde{A}}\otimes\mathcal{M}%
_{AB\rightarrow A^{\prime}B^{\prime}}\otimes\operatorname{id}_{\bar
{B}\rightarrow\tilde{B}}).
\end{equation}

Now let us show the opposite inequality. Let $S_{\bar{A}\tilde{A}AA^{\prime
}BB^{\prime}\bar{B}\tilde{B}}$ and $V_{\bar{A}\tilde{A}AA^{\prime}BB^{\prime
}\bar{B}\tilde{B}}$ be arbitrary Hermitian operators satisfying the
constraints in \eqref{eq:basic-measure-NS-PPT-bi} for $\operatorname{id}%
_{\bar{A}\rightarrow\tilde{A}}\otimes\mathcal{M}_{AB\rightarrow A^{\prime
}B^{\prime}}\otimes\operatorname{id}_{\bar{B}\rightarrow\tilde{B}}$. Set%
\begin{align}
S_{AA^{\prime}BB^{\prime}}^{\prime} &  \coloneqq\frac{1}{d_{\bar{A}}d_{\bar
{B}}}\operatorname{Tr}_{\bar{A}\tilde{A}\bar{B}\tilde{B}}[S_{\bar{A}\tilde
{A}AA^{\prime}BB^{\prime}\bar{B}\tilde{B}}],\\
V_{AA^{\prime}BB^{\prime}}^{\prime} &  \coloneqq\frac{1}{d_{\bar{A}}d_{\bar
{B}}}\operatorname{Tr}_{\bar{A}\tilde{A}\bar{B}\tilde{B}}[V_{\bar{A}\tilde
{A}AA^{\prime}BB^{\prime}\bar{B}\tilde{B}}].
\end{align}
We show that $S_{AA^{\prime}BB^{\prime}}^{\prime}$ and $V_{AA^{\prime}BB^{\prime}}^{\prime}$ satisfy the constraints in \eqref{eq:basic-measure-NS-PPT-bi} for $\mathcal{M}_{AB\rightarrow A^{\prime
}B^{\prime}}$. 
Consider that%
\begin{equation}
\Gamma_{\bar{A}\tilde{A}AA^{\prime}BB^{\prime}\bar{B}\tilde{B}}%
^{\operatorname{id}\otimes\mathcal{N}\otimes\operatorname{id}}=\Gamma_{\bar
{A}\tilde{A}}\otimes\Gamma_{AA^{\prime}BB^{\prime}}^{\mathcal{M}}\otimes
\Gamma_{\bar{B}\tilde{B}}.
\end{equation}
Then%
\begin{equation}
T_{BB^{\prime}\bar{B}\tilde{B}}(V_{\bar{A}\tilde{A}AA^{\prime}BB^{\prime}%
\bar{B}\tilde{B}}\pm\Gamma_{\bar{A}\tilde{A}}\otimes\Gamma_{AA^{\prime
}BB^{\prime}}^{\mathcal{M}}\otimes\Gamma_{\bar{B}\tilde{B}})\geq0
\end{equation}%
\begin{multline}
\Rightarrow\quad\operatorname{Tr}_{\bar{A}\tilde{A}\bar{B}\tilde{B}%
}[T_{BB^{\prime}\bar{B}\tilde{B}}(V_{\bar{A}\tilde{A}AA^{\prime}BB^{\prime
}\bar{B}\tilde{B}}\\
\pm\Gamma_{\bar{A}\tilde{A}}\otimes\Gamma_{AA^{\prime}BB^{\prime}%
}^{\mathcal{M}}\otimes\Gamma_{\bar{B}\tilde{B}})]\geq0
\end{multline}%
\begin{align}
\Leftrightarrow\quad T_{BB^{\prime}}(V_{AA^{\prime}BB^{\prime}}\pm d_{\bar{A}%
}d_{\bar{B}}\Gamma_{AA^{\prime}BB^{\prime}}^{\mathcal{M}}) &  \geq0\\
\Leftrightarrow\quad T_{BB^{\prime}}(V_{AA^{\prime}BB^{\prime}}^{\prime}%
\pm\Gamma_{AA^{\prime}BB^{\prime}}^{\mathcal{M}}) &  \geq0.
\end{align}
Also%
\begin{align}
S_{\bar{A}\tilde{A}AA^{\prime}BB^{\prime}\bar{B}\tilde{B}}\pm V_{\bar{A}%
\tilde{A}AA^{\prime}BB^{\prime}\bar{B}\tilde{B}} &  \geq0\\
\Rightarrow\quad\operatorname{Tr}_{\bar{A}\tilde{A}\bar{B}\tilde{B}}%
[S_{\bar{A}\tilde{A}AA^{\prime}BB^{\prime}\bar{B}\tilde{B}}\pm V_{\bar
{A}\tilde{A}AA^{\prime}BB^{\prime}\bar{B}\tilde{B}}] &  \geq0\\
\Leftrightarrow\quad S_{AA^{\prime}BB^{\prime}}^{\prime}\pm V_{AA^{\prime
}BB^{\prime}}^{\prime} &  \geq0,
\end{align}
and%
\begin{equation}
\operatorname{Tr}_{\tilde{A}A^{\prime}}[S_{\bar{A}\tilde{A}AA^{\prime
}BB^{\prime}\bar{B}\tilde{B}}]=\pi_{\bar{A}A}\otimes\operatorname{Tr}_{\bar
{A}\tilde{A}AA^{\prime}}[S_{\bar{A}\tilde{A}AA^{\prime}BB^{\prime}\bar
{B}\tilde{B}}]
\end{equation}
\begin{align}
\Rightarrow\quad &  \operatorname{Tr}_{\bar{A}\tilde{A}A^{\prime}\bar{B}%
\tilde{B}}[S_{\bar{A}\tilde{A}AA^{\prime}BB^{\prime}\bar{B}\tilde{B}%
}]\nonumber\\
&  =\operatorname{Tr}_{\bar{A}\bar{B}\tilde{B}}[\pi_{\bar{A}A}\otimes
\operatorname{Tr}_{\bar{A}\tilde{A}AA^{\prime}}[S_{\bar{A}\tilde{A}AA^{\prime
}BB^{\prime}\bar{B}\tilde{B}}]]\\
&  =\pi_{A}\otimes\operatorname{Tr}_{\bar{A}\tilde{A}AA^{\prime}\bar{B}%
\tilde{B}}[S_{\bar{A}\tilde{A}AA^{\prime}BB^{\prime}\bar{B}\tilde{B}}]
\end{align}%
\begin{equation}
\Leftrightarrow\quad\operatorname{Tr}_{A^{\prime}}[S_{AA^{\prime}BB^{\prime}%
}^{\prime}]=\pi_{A}\otimes\operatorname{Tr}_{AA^{\prime}}[S_{AA^{\prime
}BB^{\prime}}^{\prime}].
\end{equation}
Finally, let $\lambda$ be such that%
\begin{equation}
\frac{1}{d_{A}d_{\bar{A}}}\operatorname{Tr}_{\bar{A}\tilde{A}AA^{\prime
}B^{\prime}\tilde{B}}[S_{\bar{A}\tilde{A}AA^{\prime}BB^{\prime}\bar{B}%
\tilde{B}}]\leq\lambda I_{B\bar{B}}.
\end{equation}
Then it follows that%
\begin{align}
\operatorname{Tr}_{\bar{B}}\left[  \frac{1}{d_{A}d_{\bar{A}}}\operatorname{Tr}%
_{\bar{A}\tilde{A}AA^{\prime}B^{\prime}\tilde{B}}[S_{\bar{A}\tilde
{A}AA^{\prime}BB^{\prime}\bar{B}\tilde{B}}]\right]   &  \leq\operatorname{Tr}%
_{\bar{B}}[\lambda I_{B\bar{B}}]\\
\Leftrightarrow\quad\frac{1}{d_{A}d_{\bar{A}}}\operatorname{Tr}_{\bar{A}%
\tilde{A}AA^{\prime}B^{\prime}\bar{B}\tilde{B}}[S_{\bar{A}\tilde{A}AA^{\prime
}BB^{\prime}\bar{B}\tilde{B}}] &  \leq d_{\bar{B}}\lambda I_{B}\\
\Leftrightarrow\quad\frac{1}{d_{A}}\operatorname{Tr}_{AA^{\prime}B^{\prime}%
}[S_{AA^{\prime}BB^{\prime}}^{\prime}] &  \leq\lambda I_{B}.
\end{align}
Thus, we conclude that%
\begin{equation}
\beta(\mathcal{M}_{AB\rightarrow A^{\prime}B^{\prime}})\leq\beta
(\operatorname{id}_{\bar{A}\rightarrow\tilde{A}}\otimes\mathcal{M}%
_{AB\rightarrow A^{\prime}B^{\prime}}\otimes\operatorname{id}_{\bar
{B}\rightarrow\tilde{B}}).
\end{equation}
This concludes the proof.
\end{IEEEproof}

\bigskip

\begin{proposition}
[Zero on classical feedback channels]\label{prop:zero-feedback}Let
$\overline{\Delta}_{B\rightarrow A^{\prime}}$ be a classical feedback channel:%
\begin{equation}
\overline{\Delta}_{B\rightarrow A^{\prime}}(\cdot)\coloneqq\sum_{i=0}%
^{d-1}|i\rangle_{A^{\prime}}\langle i|_{B}(\cdot)|i\rangle_{B}\langle
i|_{A^{\prime}},
\end{equation}
where $A^{\prime}\simeq B$ and $d=d_{A^{\prime}}=d_{B}$. Then%
\begin{equation}
C_{\beta}(\overline{\Delta}_{B\rightarrow A^{\prime}})=0.
\label{eq:value-beta-p-feedback-ch}%
\end{equation}

\end{proposition}

\begin{IEEEproof}
We prove the equivalent statement that $\beta(\overline{\Delta}_{B\rightarrow
A^{\prime}})=1$. In this case, the $A$ and $B^{\prime}$ systems are trivial,
so that $d_{A}=1$, and the Choi operator of $\overline{\Delta}_{B\rightarrow
A^{\prime}}$ is given by%
\begin{equation}
\Gamma_{BA^{\prime}}^{\overline{\Delta}}=\overline{\Gamma}_{BA^{\prime}},
\end{equation}
where%
\begin{equation}
\overline{\Gamma}_{BA^{\prime}}\coloneqq\sum_{i=0}^{d_{B}-1}|i\rangle\!\langle
i|_{B}\otimes|i\rangle\!\langle i|_{A^{\prime}}.
\end{equation}
Pick $S_{BA^{\prime}}=V_{BA^{\prime}}=\overline{\Gamma}_{BA^{\prime}}$. Then
we need to check that the constraints in \eqref{eq:basic-measure-NS-PPT-bi}
are satisfied for these choices. Consider that%
\begin{align}
T_{B}(V_{BA^{\prime}}\pm\Gamma_{BA^{\prime}}^{\overline{\Delta}})  &  \geq0\\
\Leftrightarrow\quad T_{B}(\overline{\Gamma}_{BA^{\prime}}\pm\overline{\Gamma
}_{BA^{\prime}})  &  \geq0\\
\Leftrightarrow\quad\overline{\Gamma}_{BA^{\prime}}\pm\overline{\Gamma
}_{BA^{\prime}}  &  \geq0,
\end{align}
and the last inequality is trivially satisfied. Also,%
\begin{align}
S_{BA^{\prime}}\pm V_{BA^{\prime}}  &  \geq0\\
\Leftrightarrow\quad\overline{\Gamma}_{BA^{\prime}}\pm\overline{\Gamma
}_{BA^{\prime}}  &  \geq0,
\end{align}
and the no-signaling condition $\operatorname{Tr}_{A^{\prime}}[S_{AA^{\prime
}BB^{\prime}}]=\pi_{A}\otimes\operatorname{Tr}_{AA^{\prime}}[S_{AA^{\prime
}BB^{\prime}}]$ is trivially satisfied because the $A$ system is trivial,
having dimension equal to one. Finally, let us evaluate the objective function
for these choices:%
\begin{align}
\frac{1}{d_{A}}\left\Vert \operatorname{Tr}_{AA^{\prime}B^{\prime}%
}[S_{AA^{\prime}BB^{\prime}}]\right\Vert _{\infty}  &  =\left\Vert
\operatorname{Tr}_{A^{\prime}}[S_{A^{\prime}B}]\right\Vert _{\infty}\\
&  =\left\Vert \operatorname{Tr}_{A^{\prime}}[\overline{\Gamma}_{BA^{\prime}%
}]\right\Vert _{\infty}\\
&  =\left\Vert I_{B}\right\Vert _{\infty}\\
&  =1.
\end{align}
Combined with the general lower bound from
Proposition~\ref{prop:lower-bound-ch-non-neg}, we conclude \eqref{eq:value-beta-p-feedback-ch}.
\end{IEEEproof}

\bigskip

\begin{proposition}
[Zero on tensor product of local channels]\label{prop:zero-local-chs}Let
$\mathcal{E}_{A\rightarrow A^{\prime}}$ and $\mathcal{F}_{B\rightarrow
B^{\prime}}$ be quantum channels. Then%
\begin{equation}
C_{\beta}(\mathcal{E}_{A\rightarrow A^{\prime}}\otimes\mathcal{F}%
_{B\rightarrow B^{\prime}})=0. \label{eq:local-ch-beta-1}%
\end{equation}

\end{proposition}

\begin{IEEEproof}
We prove the equivalent statement that $\beta(\mathcal{E}_{A\rightarrow
A^{\prime}}\otimes\mathcal{F}_{B\rightarrow B^{\prime}})=1$. Set
$S_{AA^{\prime}BB^{\prime}}=V_{AA^{\prime}BB^{\prime}}=\Gamma_{AA^{\prime}%
}^{\mathcal{E}}\otimes\Gamma_{BB^{\prime}}^{\mathcal{F}}$, where
$\Gamma_{AA^{\prime}}^{\mathcal{E}}$ and $\Gamma_{BB^{\prime}}^{\mathcal{F}}$
are the Choi operators of $\mathcal{E}_{A\rightarrow A^{\prime}}$ and
$\mathcal{F}_{B\rightarrow B^{\prime}}$, respectively. We need to check that
the constraints in \eqref{eq:basic-measure-NS-PPT-bi} are satisfied for these
choices. Consider that%
\begin{align}
T_{BB^{\prime}}(V_{AA^{\prime}BB^{\prime}}\pm\Gamma_{AA^{\prime}}%
^{\mathcal{E}}\otimes\Gamma_{BB^{\prime}}^{\mathcal{F}})  &  \geq0\\
\Leftrightarrow\, T_{BB^{\prime}}(\Gamma_{AA^{\prime}}^{\mathcal{E}}%
\otimes\Gamma_{BB^{\prime}}^{\mathcal{F}}\pm\Gamma_{AA^{\prime}}^{\mathcal{E}%
}\otimes\Gamma_{BB^{\prime}}^{\mathcal{F}})  &  \geq0\\
\Leftrightarrow\, \Gamma_{AA^{\prime}}^{\mathcal{E}}\otimes T_{BB^{\prime}%
}(\Gamma_{BB^{\prime}}^{\mathcal{F}})\pm\Gamma_{AA^{\prime}}^{\mathcal{E}%
}\otimes T_{BB^{\prime}}(\Gamma_{BB^{\prime}}^{\mathcal{F}})  &  \geq0,
\end{align}
and the last inequality trivially holds because $T_{BB^{\prime}}$ acts as a
positive map on $\Gamma_{BB^{\prime}}^{\mathcal{F}}$. Also,%
\begin{align}
S_{AA^{\prime}BB^{\prime}}\pm V_{AA^{\prime}BB^{\prime}}  &  \geq0\\
\Leftrightarrow\quad\Gamma_{AA^{\prime}}^{\mathcal{E}}\otimes\Gamma
_{BB^{\prime}}^{\mathcal{F}}\pm\Gamma_{AA^{\prime}}^{\mathcal{E}}\otimes
\Gamma_{BB^{\prime}}^{\mathcal{F}}  &  \geq0,
\end{align}
and%
\begin{align}
\operatorname{Tr}_{A^{\prime}}[S_{AA^{\prime}BB^{\prime}}]  &
=\operatorname{Tr}_{A^{\prime}}[\Gamma_{AA^{\prime}}^{\mathcal{E}}%
\otimes\Gamma_{BB^{\prime}}^{\mathcal{F}}]\\
&  =I_{A}\otimes\Gamma_{BB^{\prime}}^{\mathcal{F}}\\
&  =\pi_{A}\otimes\operatorname{Tr}_{AA^{\prime}}[\Gamma_{AA^{\prime}%
}^{\mathcal{E}}\otimes\Gamma_{BB^{\prime}}^{\mathcal{F}}]\\
&  =\pi_{A}\otimes\operatorname{Tr}_{AA^{\prime}}[S_{AA^{\prime}BB^{\prime}}].
\end{align}
Finally, consider that the objective function evaluates to%
\begin{align}
\left\Vert \operatorname{Tr}_{A^{\prime}B^{\prime}}[S_{AA^{\prime}BB^{\prime}%
}]\right\Vert _{\infty}  &  =\left\Vert \operatorname{Tr}_{A^{\prime}%
B^{\prime}}[\Gamma_{AA^{\prime}}^{\mathcal{E}}\otimes\Gamma_{BB^{\prime}%
}^{\mathcal{F}}]\right\Vert _{\infty}\\
&  =\left\Vert I_{AB}\right\Vert _{\infty}\\
&  =1.
\end{align}
Combined with the general lower bound from
Proposition~\ref{prop:lower-bound-ch-non-neg}, we conclude \eqref{eq:local-ch-beta-1}.
\end{IEEEproof}

\bigskip

\begin{proposition}
[Subadditivity under composition]\label{prop:subadditivity-bipartite-CP-maps}%
Let $\mathcal{M}_{AB\rightarrow A^{\prime}B^{\prime}}^{1}$ and $\mathcal{M}%
_{A^{\prime}B^{\prime}\rightarrow A^{\prime\prime}B^{\prime\prime}}^{2}$ be
bipartite completely positive maps, and define%
\begin{equation}
\mathcal{M}_{AB\rightarrow A^{\prime\prime}B^{\prime\prime}}^{3}%
\coloneqq\mathcal{M}_{A^{\prime}B^{\prime}\rightarrow A^{\prime\prime
}B^{\prime\prime}}^{2}\circ\mathcal{M}_{AB\rightarrow A^{\prime}B^{\prime}%
}^{1}.
\end{equation}
Then%
\begin{equation}
C_{\beta}(\mathcal{M}_{AB\rightarrow A^{\prime\prime}B^{\prime\prime}}%
^{3})\leq C_{\beta}(\mathcal{M}_{A^{\prime}B^{\prime}\rightarrow
A^{\prime\prime}B^{\prime\prime}}^{2})+C_{\beta}(\mathcal{M}_{AB\rightarrow
A^{\prime}B^{\prime}}^{1}). \label{eq:subadd-serial-comp-maps}%
\end{equation}

\end{proposition}

\begin{IEEEproof}
We prove the equivalent statement that%
\begin{equation}
\beta(\mathcal{M}_{AB\rightarrow A^{\prime\prime}B^{\prime\prime}}^{3}%
)\leq\beta(\mathcal{M}_{A^{\prime}B^{\prime}\rightarrow A^{\prime\prime
}B^{\prime\prime}}^{2})\cdot\beta(\mathcal{M}_{AB\rightarrow A^{\prime
}B^{\prime}}^{1}).
\end{equation}
Let $S_{AA^{\prime}BB^{\prime}}^{1}$ and $V_{AA^{\prime}BB^{\prime}}^{1}$
satisfy%
\begin{align}
T_{BB^{\prime}}(V_{AA^{\prime}BB^{\prime}}^{1}\pm\Gamma_{AA^{\prime}%
BB^{\prime}}^{\mathcal{M}^{1}}) &  \geq0,\label{eq:map-1-constraint-1}\\
S_{AA^{\prime}BB^{\prime}}^{1}\pm V_{AA^{\prime}BB^{\prime}}^{1} &
\geq0,\label{eq:map-1-constraint-2}%
\end{align}%
\begin{equation}
\operatorname{Tr}_{A^{\prime}}[S_{AA^{\prime}BB^{\prime}}^{1}]=\pi_{A}%
\otimes\operatorname{Tr}_{AA^{\prime}}[S_{AA^{\prime}BB^{\prime}}%
^{1}],\label{eq:map-1-constraint-3}%
\end{equation}
and let $S_{A^{\prime}A^{\prime\prime}B^{\prime}B^{\prime\prime}}^{2}$ and
$V_{A^{\prime}A^{\prime\prime}B^{\prime}B^{\prime\prime}}^{2}$ satisfy%
\begin{align}
T_{B^{\prime}B^{\prime\prime}}(V_{A^{\prime}A^{\prime\prime}B^{\prime
}B^{\prime\prime}}^{2}\pm\Gamma_{A^{\prime}A^{\prime\prime}B^{\prime}%
B^{\prime\prime}}^{\mathcal{M}^{2}}) &  \geq0,\label{eq:map-2-constraint-1}\\
S_{A^{\prime}A^{\prime\prime}B^{\prime}B^{\prime\prime}}^{2}\pm V_{A^{\prime
}A^{\prime\prime}B^{\prime}B^{\prime\prime}}^{2} &  \geq
0,\label{eq:map-2-constraint-2}%
\end{align}%
\begin{equation}
\operatorname{Tr}_{A^{\prime\prime}}[S_{A^{\prime}A^{\prime\prime}B^{\prime
}B^{\prime\prime}}^{2}]=\pi_{A^{\prime}}\otimes\operatorname{Tr}_{A^{\prime
}A^{\prime\prime}}[S_{A^{\prime}A^{\prime\prime}B^{\prime}B^{\prime\prime}%
}^{2}].\label{eq:map-2-constraint-3}%
\end{equation}
Then it follows that%
\begin{multline}
T_{BB^{\prime}B^{\prime}B^{\prime\prime}}(V_{AA^{\prime}BB^{\prime}}%
^{1}\otimes V_{A^{\prime}A^{\prime\prime}B^{\prime}B^{\prime\prime}}%
^{2}\label{eq:subadd-proof-intermed-1}\\
\pm\Gamma_{AA^{\prime}BB^{\prime}}^{\mathcal{M}^{1}}\otimes\Gamma_{A^{\prime
}A^{\prime\prime}B^{\prime}B^{\prime\prime}}^{\mathcal{M}^{2}})\geq0,
\end{multline}%
\begin{equation}
S_{AA^{\prime}BB^{\prime}}^{1}\otimes S_{A^{\prime}A^{\prime\prime}B^{\prime
}B^{\prime\prime}}^{2}\pm V_{AA^{\prime}BB^{\prime}}^{1}\otimes V_{A^{\prime
}A^{\prime\prime}B^{\prime}B^{\prime\prime}}^{2}\geq
0.\label{eq:subadd-proof-intermed-2}%
\end{equation}
This latter statement is a consequence of the general fact that if $A$, $B$,
$C$, and $D$ are Hermitian operators satisfying $A\pm B\geq0$ and $C\pm
D\geq0$, then $A\otimes C\pm B\otimes D\geq0$. To see this, consider that the
original four operator inequalities imply the four operator inequalities
$\left(  A\pm B\right)  \otimes\left(  C\pm D\right)  \geq0$, and then summing
these four different operator inequalities in various ways leads to $A\otimes
C\pm B\otimes D\geq0$. See \eqref{eq:ABCD-1}--\eqref{eq:ABCD-last} for further clarification of this point. 

Now apply the following positive map to
\eqref{eq:subadd-proof-intermed-1}--\eqref{eq:subadd-proof-intermed-2}:%
\begin{equation}
(\cdot)\rightarrow(\langle\Gamma|_{A^{\prime}A^{\prime}}\otimes\langle
\Gamma|_{B^{\prime}B^{\prime}})(\cdot)(|\Gamma\rangle_{A^{\prime}A^{\prime}%
}\otimes|\Gamma\rangle_{B^{\prime}B^{\prime}}),
\end{equation}
where%
\begin{align}
|\Gamma\rangle_{A^{\prime}A^{\prime}} &  \coloneqq\sum_{i}|i\rangle
_{A^{\prime}}|i\rangle_{A^{\prime}},\\
|\Gamma\rangle_{B^{\prime}B^{\prime}} &  \coloneqq\sum_{i}|i\rangle
_{B^{\prime}}|i\rangle_{B^{\prime}}.
\end{align}
This gives%
\begin{align}
T_{BB^{\prime\prime}}(V_{AA^{\prime\prime}BB^{\prime\prime}}^{3}\pm
\Gamma_{AA^{\prime\prime}BB^{\prime\prime}}^{\mathcal{M}^{2}\circ
\mathcal{M}^{1}}) &  \geq0,\label{eq:concat-constraint-1}\\
S_{AA^{\prime\prime}BB^{\prime\prime}}^{3}\pm V_{AA^{\prime\prime}%
BB^{\prime\prime}}^{3} &  \geq0,\label{eq:concat-constraint-2}%
\end{align}
where%
\begin{align}
V_{AA^{\prime\prime}BB^{\prime\prime}}^{3} &  \coloneqq(\langle\Gamma
|_{A^{\prime}A^{\prime}}\otimes\langle\Gamma|_{B^{\prime}B^{\prime}%
})\nonumber\\
&  (V_{AA^{\prime}BB^{\prime}}^{1}\otimes V_{A^{\prime}A^{\prime\prime
}B^{\prime}B^{\prime\prime}}^{2})(|\Gamma\rangle_{A^{\prime}A^{\prime}}%
\otimes|\Gamma\rangle_{B^{\prime}B^{\prime}}),\\
\Gamma_{AA^{\prime\prime}BB^{\prime\prime}}^{\mathcal{M}^{2}\circ
\mathcal{M}^{1}} &  \coloneqq(\langle\Gamma|_{A^{\prime}A^{\prime}}%
\otimes\langle\Gamma|_{B^{\prime}B^{\prime}})\nonumber\\
&  (\Gamma_{AA^{\prime}BB^{\prime}}^{\mathcal{M}^{1}}\otimes\Gamma_{A^{\prime
}A^{\prime\prime}B^{\prime}B^{\prime\prime}}^{\mathcal{M}^{2}})(|\Gamma
\rangle_{A^{\prime}A^{\prime}}\otimes|\Gamma\rangle_{B^{\prime}B^{\prime}}),\\
S_{AA^{\prime\prime}BB^{\prime\prime}}^{3} &  \coloneqq(\langle\Gamma
|_{A^{\prime}A^{\prime}}\otimes\langle\Gamma|_{B^{\prime}B^{\prime}%
})\nonumber\\
&  (S_{AA^{\prime}BB^{\prime}}^{1}\otimes S_{A^{\prime}A^{\prime\prime
}B^{\prime}B^{\prime\prime}}^{2})(|\Gamma\rangle_{A^{\prime}A^{\prime}}%
\otimes|\Gamma\rangle_{B^{\prime}B^{\prime}}),
\end{align}
and we applied \eqref{eq:serial-compose-Choi}\ to conclude that%
\begin{multline}
(\langle\Gamma|_{A^{\prime}A^{\prime}}\otimes\langle\Gamma|_{B^{\prime
}B^{\prime}})(\Gamma_{AA^{\prime}BB^{\prime}}^{\mathcal{M}^{1}}\otimes
\Gamma_{A^{\prime}A^{\prime\prime}B^{\prime}B^{\prime\prime}}^{\mathcal{M}%
^{2}})\\
(|\Gamma\rangle_{A^{\prime}A^{\prime}}\otimes|\Gamma\rangle_{B^{\prime
}B^{\prime}})=\Gamma_{AA^{\prime\prime}BB^{\prime\prime}}^{\mathcal{M}%
^{2}\circ\mathcal{M}^{1}}.
\end{multline}
Also, consider that%
\begin{align}
&  \operatorname{Tr}_{A^{\prime\prime}}[S_{AA^{\prime\prime}BB^{\prime\prime}%
}^{3}]\nonumber\\
&  =\operatorname{Tr}_{A^{\prime\prime}}[(\langle\Gamma|_{A^{\prime}A^{\prime
}}\otimes\langle\Gamma|_{B^{\prime}B^{\prime}})(S_{AA^{\prime}BB^{\prime}}%
^{1}\otimes S_{A^{\prime}A^{\prime\prime}B^{\prime}B^{\prime\prime}}%
^{2})\nonumber\\
&  \qquad\qquad(|\Gamma\rangle_{A^{\prime}A^{\prime}}\otimes|\Gamma
\rangle_{B^{\prime}B^{\prime}})]\\
&  =(\langle\Gamma|_{A^{\prime}A^{\prime}}\otimes\langle\Gamma|_{B^{\prime
}B^{\prime}})(S_{AA^{\prime}BB^{\prime}}^{1}\otimes\operatorname{Tr}%
_{A^{\prime\prime}}[S_{A^{\prime}A^{\prime\prime}B^{\prime}B^{\prime\prime}%
}^{2}])\nonumber\\
&  \qquad\qquad(|\Gamma\rangle_{A^{\prime}A^{\prime}}\otimes|\Gamma
\rangle_{B^{\prime}B^{\prime}})\\
&  =(\langle\Gamma|_{A^{\prime}A^{\prime}}\otimes\langle\Gamma|_{B^{\prime
}B^{\prime}})\nonumber\\
&  \qquad\qquad(S_{AA^{\prime}BB^{\prime}}^{1}\otimes\pi_{A^{\prime}}%
\otimes\operatorname{Tr}_{A^{\prime}A^{\prime\prime}}[S_{A^{\prime}%
A^{\prime\prime}B^{\prime}B^{\prime\prime}}^{2}])\nonumber\\
&  \qquad\qquad(|\Gamma\rangle_{A^{\prime}A^{\prime}}\otimes|\Gamma
\rangle_{B^{\prime}B^{\prime}})\\
&  =\frac{1}{d_{A^{\prime}}}(\langle\Gamma|_{A^{\prime}A^{\prime}}%
\otimes\langle\Gamma|_{B^{\prime}B^{\prime}})\nonumber\\
&  \qquad\qquad(S_{AA^{\prime}BB^{\prime}}^{1}\otimes I_{A^{\prime}}%
\otimes\operatorname{Tr}_{A^{\prime}A^{\prime\prime}}[S_{A^{\prime}%
A^{\prime\prime}B^{\prime}B^{\prime\prime}}^{2}])\nonumber\\
&  \qquad\qquad(|\Gamma\rangle_{A^{\prime}A^{\prime}}\otimes|\Gamma
\rangle_{B^{\prime}B^{\prime}})
\end{align}%
\begin{align}
&  =\frac{1}{d_{A^{\prime}}}\langle\Gamma|_{B^{\prime}B^{\prime}%
}(\operatorname{Tr}_{A^{\prime}}[S_{AA^{\prime}BB^{\prime}}^{1}]\nonumber\\
&  \qquad\qquad\otimes\operatorname{Tr}_{A^{\prime}A^{\prime\prime}%
}[S_{A^{\prime}A^{\prime\prime}B^{\prime}B^{\prime\prime}}^{2}])|\Gamma
\rangle_{B^{\prime}B^{\prime}}\\
&  =\frac{1}{d_{A^{\prime}}}\langle\Gamma|_{B^{\prime}B^{\prime}}(\pi
_{A}\otimes\operatorname{Tr}_{AA^{\prime}}[S_{AA^{\prime}BB^{\prime}}%
^{1}]\nonumber\\
&  \qquad\qquad\otimes\operatorname{Tr}_{A^{\prime}A^{\prime\prime}%
}[S_{A^{\prime}A^{\prime\prime}B^{\prime}B^{\prime\prime}}^{2}])|\Gamma
\rangle_{B^{\prime}B^{\prime}}\\
&  =\pi_{A}\otimes\frac{1}{d_{A^{\prime}}}\langle\Gamma|_{B^{\prime}B^{\prime
}}(\operatorname{Tr}_{AA^{\prime}}[S_{AA^{\prime}BB^{\prime}}^{1}]\nonumber\\
&  \qquad\qquad\otimes\operatorname{Tr}_{A^{\prime}A^{\prime\prime}%
}[S_{A^{\prime}A^{\prime\prime}B^{\prime}B^{\prime\prime}}^{2}])|\Gamma
\rangle_{B^{\prime}B^{\prime}}.
\end{align}
Now consider that%
\begin{multline}
\operatorname{Tr}_{AA^{\prime\prime}}[S_{AA^{\prime\prime}BB^{\prime\prime}%
}^{3}]=\\
\frac{1}{d_{A^{\prime}}}\langle\Gamma|_{B^{\prime}B^{\prime}}%
(\operatorname{Tr}_{AA^{\prime}}[S_{AA^{\prime}BB^{\prime}}^{1}]\otimes
\operatorname{Tr}_{A^{\prime}A^{\prime\prime}}[S_{A^{\prime}A^{\prime\prime
}B^{\prime}B^{\prime\prime}}^{2}])|\Gamma\rangle_{B^{\prime}B^{\prime}}.
\end{multline}
So we conclude that%
\begin{equation}
\operatorname{Tr}_{A^{\prime\prime}}[S_{AA^{\prime\prime}BB^{\prime\prime}%
}^{3}]=\pi_{A}\otimes\operatorname{Tr}_{AA^{\prime\prime}}[S_{AA^{\prime
\prime}BB^{\prime\prime}}^{3}].\label{eq:concat-constraint-3}%
\end{equation}
Finally, consider that%
\begin{align}
&  \left\Vert \operatorname{Tr}_{A^{\prime\prime}B^{\prime\prime}%
}[S_{AA^{\prime\prime}BB^{\prime\prime}}^{3}]\right\Vert _{\infty}\nonumber\\
&  =\left\Vert
\begin{array}
[c]{c}%
\operatorname{Tr}_{A^{\prime\prime}B^{\prime\prime}}[\left(  \langle
\Gamma|_{A^{\prime}A^{\prime}}\otimes\langle\Gamma|_{B^{\prime}B^{\prime}%
}\right)  \\
\left(  S_{AA^{\prime}BB^{\prime}}^{1}\otimes S_{A^{\prime}A^{\prime\prime
}B^{\prime}B^{\prime\prime}}^{2}\right)  \\
(|\Gamma\rangle_{A^{\prime}A^{\prime}}\otimes|\Gamma\rangle_{B^{\prime
}B^{\prime}})]
\end{array}
\right\Vert _{\infty}\\
&  =\left\Vert
\begin{array}
[c]{c}%
\left(  \langle\Gamma|_{A^{\prime}A^{\prime}}\otimes\langle
\Gamma|_{B^{\prime}B^{\prime}}\right)  \\
\left(  S_{AA^{\prime}BB^{\prime}}^{1}\otimes\operatorname{Tr}_{A^{\prime
\prime}B^{\prime\prime}}[S_{A^{\prime}A^{\prime\prime}B^{\prime}%
B^{\prime\prime}}^{2}]\right)  \\
(|\Gamma\rangle_{A^{\prime}A^{\prime}}\otimes|\Gamma\rangle_{B^{\prime
}B^{\prime}})
\end{array}
\right\Vert _{\infty}\\
&  \leq\left\Vert \operatorname{Tr}_{A^{\prime\prime}B^{\prime\prime}%
}[S_{A^{\prime}A^{\prime\prime}B^{\prime}B^{\prime\prime}}^{2}]\right\Vert
_{\infty}\cdot\nonumber\\
&  \qquad \left\Vert
\begin{array}
[c]{c}%
\left(  \langle\Gamma|_{A^{\prime}A^{\prime}}\otimes\langle
\Gamma|_{B^{\prime}B^{\prime}}\right)  \left(  S_{AA^{\prime}BB^{\prime}}%
^{1}\otimes I_{A^{\prime}B^{\prime}}\right)  \\
(|\Gamma\rangle_{A^{\prime}A^{\prime}}\otimes|\Gamma\rangle_{B^{\prime
}B^{\prime}})
\end{array}
\right\Vert _{\infty}\\
&  =\left\Vert \operatorname{Tr}_{A^{\prime\prime}B^{\prime\prime}%
}[S_{A^{\prime}A^{\prime\prime}B^{\prime}B^{\prime\prime}}^{2}]\right\Vert
_{\infty}\left\Vert \operatorname{Tr}_{A^{\prime}B^{\prime}}[S_{AA^{\prime
}BB^{\prime}}^{1}]\right\Vert _{\infty}.
\end{align}
Since $S_{AA^{\prime\prime}BB^{\prime\prime}}^{3}$ and $V_{AA^{\prime\prime
}BB^{\prime\prime}}^{3}$ are particular choices that satisfy the constraints
in \eqref{eq:concat-constraint-1}, \eqref{eq:concat-constraint-2}, and \eqref{eq:concat-constraint-3}, we conclude
that%
\begin{multline}
\beta(\mathcal{M}_{AB\rightarrow A^{\prime\prime}B^{\prime\prime}}^{3})\leq\\
\left\Vert \operatorname{Tr}_{A^{\prime\prime}B^{\prime\prime}}[S_{A^{\prime
}A^{\prime\prime}B^{\prime}B^{\prime\prime}}^{2}]\right\Vert _{\infty
}\left\Vert \operatorname{Tr}_{A^{\prime}B^{\prime}}[S_{AA^{\prime}BB^{\prime
}}^{1}]\right\Vert _{\infty}.
\end{multline}
Since $S_{AA^{\prime}BB^{\prime}}^{1}$ and $V_{AA^{\prime}BB^{\prime}}^{1}$
are arbitrary Hermitian operators satisfying the constraints in
\eqref{eq:map-1-constraint-1}--\eqref{eq:map-1-constraint-3} and
$S_{A^{\prime}A^{\prime\prime}B^{\prime}B^{\prime\prime}}^{2}$ and
$V_{A^{\prime}A^{\prime\prime}B^{\prime}B^{\prime\prime}}^{2}$ are arbitrary
Hermitian operators satisfying the constraints in
\eqref{eq:map-2-constraint-1}--\eqref{eq:map-2-constraint-3}, we conclude \eqref{eq:subadd-serial-comp-maps}.
\end{IEEEproof}

\bigskip

\begin{corollary}
[Data processing under local channels]\label{cor:DP-LC-beta-p-bipartite-ch}Let
$\mathcal{M}_{AB\rightarrow A^{\prime}B^{\prime}}$ be a completely positive
bipartite map. Let $\mathcal{K}_{\hat{A}\rightarrow A}$, $\mathcal{L}_{\hat
{B}\rightarrow B}$, $\mathcal{N}_{A^{\prime}\rightarrow A^{\prime\prime}}$,
and $\mathcal{P}_{B^{\prime}\rightarrow B^{\prime\prime}}$ be local quantum
channels, and define the bipartite completely positive map $\mathcal{F}%
_{\hat{A}\hat{B}\rightarrow A^{\prime\prime}B^{\prime\prime}}$ as follows:%
\begin{multline}
\mathcal{F}_{\hat{A}\hat{B}\rightarrow A^{\prime\prime}B^{\prime\prime}%
}\coloneqq\\
(\mathcal{N}_{A^{\prime}\rightarrow A^{\prime\prime}}\otimes\mathcal{P}%
_{B^{\prime}\rightarrow B^{\prime\prime}})\mathcal{M}_{AB\rightarrow
A^{\prime}B^{\prime}}(\mathcal{K}_{\hat{A}\rightarrow A}\otimes\mathcal{L}%
_{\hat{B}\rightarrow B}).
\end{multline}
Then%
\begin{equation}
C_{\beta}(\mathcal{F}_{\hat{A}\hat{B}\rightarrow A^{\prime\prime}%
B^{\prime\prime}})\leq C_{\beta}(\mathcal{M}_{AB\rightarrow A^{\prime
}B^{\prime}}).
\end{equation}

\end{corollary}

\begin{IEEEproof}
Apply Propositions~\ref{prop:zero-local-chs}\ and
\ref{prop:subadditivity-bipartite-CP-maps}\ to find that%
\begin{align}
&  C_{\beta}(\mathcal{F}_{\hat{A}\hat{B}\rightarrow A^{\prime\prime}%
B^{\prime\prime}})\nonumber\\
&  \leq C_{\beta}(\mathcal{N}_{A^{\prime}\rightarrow A^{\prime\prime}}%
\otimes\mathcal{P}_{B^{\prime}\rightarrow B^{\prime\prime}})\nonumber\\
&  \qquad+C_{\beta}(\mathcal{M}_{AB\rightarrow A^{\prime}B^{\prime}}%
)+C_{\beta}(\mathcal{K}_{\hat{A}\rightarrow A}\otimes\mathcal{L}_{\hat
{B}\rightarrow B})\\
&  =C_{\beta}(\mathcal{M}_{AB\rightarrow A^{\prime}B^{\prime}}).
\end{align}
This concludes the proof.
\end{IEEEproof}

\bigskip

\begin{corollary}
[Invariance under local unitary channels]\label{cor:I-LUC-beta-p-bipartite-ch}%
Let $\mathcal{M}_{AB\rightarrow A^{\prime}B^{\prime}}$ be a completely
positive bipartite map. Let $\mathcal{U}_{A}$, $\mathcal{V}_{B}$,
$\mathcal{W}_{A^{\prime}}$, and $\mathcal{Y}_{B^{\prime}}$ be local unitary
channels, and define the bipartite completely positive map $\mathcal{F}%
_{\hat{A}\hat{B}\rightarrow A^{\prime\prime}B^{\prime\prime}}$ as follows:%
\begin{equation}
\mathcal{F}_{AB\rightarrow A^{\prime}B^{\prime}}\coloneqq(\mathcal{W}%
_{A^{\prime}}\otimes\mathcal{Y}_{B^{\prime}})\mathcal{M}_{AB\rightarrow
A^{\prime}B^{\prime}}(\mathcal{U}_{A}\otimes\mathcal{V}_{B}).
\end{equation}
Then%
\begin{equation}
C_{\beta}(\mathcal{F}_{AB\rightarrow A^{\prime}B^{\prime}})=C_{\beta
}(\mathcal{M}_{AB\rightarrow A^{\prime}B^{\prime}}).
\end{equation}

\end{corollary}

\begin{IEEEproof}
Apply Corollary~\ref{cor:DP-LC-beta-p-bipartite-ch} twice to conclude that
$C_{\beta}(\mathcal{M}_{AB\rightarrow A^{\prime}B^{\prime}})\geq C_{\beta
}(\mathcal{F}_{AB\rightarrow A^{\prime}B^{\prime}})$ and $C_{\beta
}(\mathcal{F}_{AB\rightarrow A^{\prime}B^{\prime}})\geq C_{\beta}%
(\mathcal{M}_{AB\rightarrow A^{\prime}B^{\prime}})$.
\end{IEEEproof}

%

\bigskip

\begin{proposition}
[Convexity]\label{prop:convexity-beta}The measure $\beta$ is convex, in the
following sense:%
\begin{multline}
\beta(\mathcal{M}_{AB\rightarrow A^{\prime}B^{\prime}}^{\lambda}%
)\leq\label{eq:convexity-beta}\\
\lambda\beta(\mathcal{M}_{AB\rightarrow A^{\prime}B^{\prime}}^{1})+\left(
1-\lambda\right)  \beta(\mathcal{M}_{AB\rightarrow A^{\prime}B^{\prime}}^{0}),
\end{multline}
where $\mathcal{M}_{AB\rightarrow A^{\prime}B^{\prime}}^{0}$ and
$\mathcal{M}_{AB\rightarrow A^{\prime}B^{\prime}}^{1}$ are completely positive
bipartite maps, $\lambda\in\left[  0,1\right]  $, and%
\begin{equation}
\mathcal{M}_{AB\rightarrow A^{\prime}B^{\prime}}^{\lambda}\coloneqq\lambda
\mathcal{M}_{AB\rightarrow A^{\prime}B^{\prime}}^{1}+\left(  1-\lambda\right)
\mathcal{M}_{AB\rightarrow A^{\prime}B^{\prime}}^{0}.
\end{equation}

\end{proposition}

\begin{IEEEproof}
Let $S_{AA^{\prime}BB^{\prime}}^{x}$ and $V_{AA^{\prime}BB^{\prime}}^{x}$
satisfy the constraints in \eqref{eq:basic-measure-NS-PPT-bi}\ for
$\mathcal{M}_{AB\rightarrow A^{\prime}B^{\prime}}^{x}$ for $x\in\left\{
0,1\right\}  $. Then%
\begin{align}
S_{AA^{\prime}BB^{\prime}}^{\lambda} &  \coloneqq\lambda S_{AA^{\prime
}BB^{\prime}}^{1}+\left(  1-\lambda\right)  S_{AA^{\prime}BB^{\prime}}^{0},\\
V_{AA^{\prime}BB^{\prime}}^{\lambda} &  \coloneqq\lambda V_{AA^{\prime
}BB^{\prime}}^{1}+\left(  1-\lambda\right)  V_{AA^{\prime}BB^{\prime}}^{0},
\end{align}
satisfy the constraints in \eqref{eq:basic-measure-NS-PPT-bi}\ for
$\mathcal{M}_{AB\rightarrow A^{\prime}B^{\prime}}^{\lambda}$.\ Then it follows
that%
\begin{align}
\beta(\mathcal{M}_{AB\rightarrow A^{\prime}B^{\prime}}^{\lambda}) &
\leq\left\Vert \operatorname{Tr}_{A^{\prime}B^{\prime}}[S_{AA^{\prime
}BB^{\prime}}^{\lambda}]\right\Vert _{\infty}\\
&  \leq\lambda\left\Vert \operatorname{Tr}_{A^{\prime}B^{\prime}%
}[S_{AA^{\prime}BB^{\prime}}^{1}]\right\Vert _{\infty}\nonumber\\
&  \qquad+\left(  1-\lambda\right)  \left\Vert \operatorname{Tr}_{A^{\prime
}B^{\prime}}[S_{AA^{\prime}BB^{\prime}}^{0}]\right\Vert _{\infty},
\end{align}
where the second inequality follows from convexity of the $\infty$-norm. Since
the inequality holds for all $S_{AA^{\prime}BB^{\prime}}^{x}$ and
$V_{AA^{\prime}BB^{\prime}}^{x}$ satisfying the constraints in
\eqref{eq:basic-measure-NS-PPT-bi}\ for $\mathcal{M}_{AB\rightarrow A^{\prime
}B^{\prime}}^{x}$ for $x\in\left\{  0,1\right\}  $, we conclude \eqref{eq:convexity-beta}.
\end{IEEEproof}

\subsection{Related measures}

We now define variations of the bipartite channel measure from
\eqref{eq:basic-measure-NS-PPT-bi}. We employ generalized divergences to do
so, and in doing so, we arrive at a large number of variations of the basic
bipartite channel measure.

Let $\boldsymbol{D}$ denote a generalized divergence \cite{PV10,SW12}, which
is a function that satisfies the following data-processing inequality, for
every state $\rho$, positive semi-definite operator $\sigma$, and quantum
channel $\mathcal{N}$:%
\begin{equation}
\boldsymbol{D}(\rho\Vert\sigma)\geq\boldsymbol{D}(\mathcal{N}(\rho
)\Vert\mathcal{N}(\sigma)). \label{eq:DP-gen-div}%
\end{equation}
In this paper, we make two additional minimal assumptions about a generalized divergence:

\begin{enumerate}
\item First, we assume that%
\begin{equation}
\boldsymbol{D}(1\Vert c)\geq0 \label{eq:min-prop-gen-div}%
\end{equation}
for $c\in(0,1]$. That is, if we plug in a trivial one-dimensional density
operator $\rho$ (i.e., the number $1$) and a trivial positive semi-definite
operator with trace less than or equal to one (i.e., $c\in(0,1]$), then the generalized divergence
evaluates to a non-negative real.

\item Next, we assume that%
\begin{equation}
\boldsymbol{D}(\rho\Vert\rho)=0 \label{eq:min-prop-gen-div-2}%
\end{equation}
for every state $\rho$. We should clarify that this assumption is quite
minimal. The reason is that it is essentially a direct consequence of
\eqref{eq:DP-gen-div} up to an inessential additive factor. That is,
\eqref{eq:DP-gen-div} implies that there exists a constant~$c$ such that%
\begin{equation}
\boldsymbol{D}(\rho\Vert\rho)=c \label{eq:gen-div-constant-on-same-state}%
\end{equation}
for every state $\rho$. To see this, consider that one can get from the state
$\rho$ to another state $\omega$ by means of a trace and replace channel
$\operatorname{Tr}[\cdot]\omega$, so that \eqref{eq:DP-gen-div} implies that%
\begin{equation}
\boldsymbol{D}(\rho\Vert\rho)\geq\boldsymbol{D}(\omega\Vert\omega).
\end{equation}
However, by the same argument, $\boldsymbol{D}(\omega\Vert\omega
)\geq\boldsymbol{D}(\rho\Vert\rho)$, so that the claim holds. So the
assumption in \eqref{eq:min-prop-gen-div-2} amounts to a redefinition of the
generalized divergence as%
\begin{equation}
\boldsymbol{D}^{\prime}(\rho\Vert\sigma)\coloneqq \boldsymbol{D}(\rho
\Vert\sigma)-c. \label{eq:alt-def-gen-div}%
\end{equation}

\end{enumerate}

Let us list particular choices of interest for a generalized divergence. The
quantum relative entropy \cite{U62}\ is defined as%
\begin{equation}
D(\rho\Vert\sigma)\coloneqq\operatorname{Tr}[\rho(\log_{2}\rho-\log_{2}%
\sigma)]
\label{eq:q-rel-entr}
\end{equation}
if $\operatorname{supp}(\rho)\subseteq\operatorname{supp}(\sigma)$ and it is
equal to $+\infty$ otherwise. The sandwiched R\'{e}nyi relative entropy is
defined for all $\alpha\in(0,1)\cup(1,\infty)$ as \cite{MDSFT13,WWY14}%
\begin{multline}
\widetilde{D}_{\alpha}(\rho\Vert\sigma)\coloneqq\\
\lim_{\varepsilon\rightarrow0^{+}}\frac{1}{\alpha-1}\log_{2}\operatorname{Tr}%
[(\sigma_{\varepsilon}^{-(1-\alpha)/2\alpha}\rho\sigma_{\varepsilon
}^{-(1-\alpha)/2\alpha})^{\alpha}],
\label{eq:sandwiched-Renyi-def}
\end{multline}
where $\sigma_{\varepsilon}\coloneqq\sigma+\varepsilon I$. In the case that
$\operatorname{supp}(\rho)\subseteq\operatorname{supp}(\sigma)$, we have the
following simplification:%
\begin{equation}
\widetilde{D}_{\alpha}(\rho\Vert\sigma)=\frac{1}{\alpha-1}\log_{2}%
\operatorname{Tr}[(\sigma^{-(1-\alpha)/2\alpha}\rho\sigma^{-(1-\alpha
)/2\alpha})^{\alpha}].
\end{equation}
Note that $\widetilde{D}_{\alpha}(\rho\Vert\sigma) = +\infty$ if $\alpha > 1$ and $\operatorname{supp}(\rho)\not\subseteq \operatorname{supp}(\sigma)$.
The sandwiched R\'{e}nyi relative entropy obeys the data-processing inequality
in \eqref{eq:DP-gen-div}\ for $\alpha\in\lbrack1/2,1)\cup(1,\infty)$
\cite{FL13,W18opt}. Some basic properties of the sandwiched R\'{e}nyi relative
entropy are as follows \cite{MDSFT13,WWY14}:\ for all $\alpha>\beta>0$%
\begin{equation}
\widetilde{D}_{\alpha}(\rho\Vert\sigma)\geq\widetilde{D}_{\beta}(\rho
\Vert\sigma),\label{eq:SW-Renyi-monotone-alpha}%
\end{equation}
and%
\begin{equation}
\lim_{\alpha\rightarrow1}\widetilde{D}_{\alpha}(\rho\Vert\sigma)=D(\rho
\Vert\sigma).\label{eq:limit-SW-to-1}%
\end{equation}
The Belavkin--Staszewski relative entropy \cite{Belavkin1982}\ is defined as%
\begin{equation}
\widehat{D}(\rho\Vert\sigma)\coloneqq\operatorname{Tr}[\rho\log_{2}(\rho
^{1/2}\sigma^{-1}\rho^{1/2})]
\end{equation}
if $\operatorname{supp}(\rho)\subseteq\operatorname{supp}(\sigma)$ and it is
equal to $+\infty$ otherwise. The geometric R\'{e}nyi relative entropy is
defined for all $\alpha\in(0,1)\cup(1,\infty)$ as
\cite{PR98,M13,Matsumoto2018,KW20}%
\begin{equation}
\widehat{D}_{\alpha}(\rho\Vert\sigma)\coloneqq\lim_{\varepsilon\rightarrow
0^{+}}\frac{1}{\alpha-1}\log_{2}\operatorname{Tr}[\sigma_{\varepsilon}%
(\sigma_{\varepsilon}^{-1/2}\rho\sigma_{\varepsilon}^{-1/2})^{\alpha}],
\end{equation}
where $\sigma_{\varepsilon}\coloneqq\sigma+\varepsilon I$. In the case that
$\operatorname{supp}(\rho)\subseteq\operatorname{supp}(\sigma)$, we have the
following simplification:%
\begin{equation}
\widehat{D}_{\alpha}(\rho\Vert\sigma)=\frac{1}{\alpha-1}\log_{2}%
\operatorname{Tr}[\sigma(\sigma^{-1/2}\rho\sigma^{-1/2})^{\alpha}].
\end{equation}
The geometric R\'{e}nyi relative entropy obeys the data-processing inequality
in \eqref{eq:DP-gen-div}\ for $\alpha\in\left(  0,1\right)  \cup(1,2]$. Some
basic properties of the geometric R\'{e}nyi relative entropy are as follows
\cite{KW20}:\ for all $\alpha>\beta>0$%
\begin{equation}
\widehat{D}_{\alpha}(\rho\Vert\sigma)\geq\widehat{D}_{\beta}(\rho\Vert
\sigma),\label{eq:geometric-Renyi-monotone-alpha}%
\end{equation}
and%
\begin{equation}
\lim_{\alpha\rightarrow1}\widehat{D}_{\alpha}(\rho\Vert\sigma)=\widehat
{D}(\rho\Vert\sigma).\label{eq:limit-geo-to-1}%
\end{equation}
We are also interested in the hypothesis testing relative entropy
\cite{BD10,BD11,WR12}, defined for $\varepsilon\in\left[  0,1\right]  $ as%
\begin{equation}
D_{H}^{\varepsilon}(\rho\Vert\sigma)\coloneqq-\log_{2}\inf_{\Lambda\geq
0}\left\{  \operatorname{Tr}[\Lambda\sigma]:\operatorname{Tr}[\Lambda\rho
]\geq1-\varepsilon,\ \Lambda\leq I\right\}  .\label{eq:HTRE-def}%
\end{equation}
The property in \eqref{eq:min-prop-gen-div} holds for all of the relative
entropies that we have listed above, while the property in
\eqref{eq:min-prop-gen-div-2} holds for all of them except for the hypothesis
testing relative entropy. For the hypothesis testing relative entropy, the
constant $c$ in \eqref{eq:gen-div-constant-on-same-state}\ is equal to
$-\log_{2}(1-\varepsilon)$, and the alternative definition in
\eqref{eq:alt-def-gen-div}\ is sometimes used \cite{DKFRR13}.

A generalized channel divergence between a quantum channel $\mathcal{N}%
_{A\rightarrow B}$\ and a completely positive map $\mathcal{M}_{A\rightarrow
B}$ is defined from a generalized divergence as follows \cite{LKDW18}:%
\begin{equation}
\boldsymbol{D}(\mathcal{N}\Vert\mathcal{M})\coloneqq\sup_{\rho_{RA}%
}\boldsymbol{D}(\mathcal{N}_{A\rightarrow B}(\rho_{RA})\Vert\mathcal{M}%
_{A\rightarrow B}(\rho_{RA})), \label{eq:gen-ch-div}%
\end{equation}
where the optimization is with respect to every bipartite state $\rho_{RA}$,
with the system $R$ arbitrarily large. By a standard argument (detailed in
\cite{LKDW18}), the following simplification occurs%
\begin{equation}
\boldsymbol{D}(\mathcal{N}\Vert\mathcal{M})\coloneqq\sup_{\psi_{RA}%
}\boldsymbol{D}(\mathcal{N}_{A\rightarrow B}(\psi_{RA})\Vert\mathcal{M}%
_{A\rightarrow B}(\psi_{RA})),
\end{equation}
where the optimization is with respect to all pure bipartite states with
$R\simeq A$. Using this, we define the following:

\begin{definition}
\label{def:ups-bipartite-gen}For a bipartite channel $\mathcal{N}%
_{AB\rightarrow A^{\prime}B^{\prime}}$, we define the following measure of
forward classical communication:%
\begin{multline}
\mathbf{\Upsilon}(\mathcal{N}_{AB\rightarrow A^{\prime}B^{\prime}})\coloneqq\\
\inf_{\mathcal{M}_{AB\rightarrow A^{\prime}B^{\prime}}:\beta(\mathcal{M}%
_{AB\rightarrow A^{\prime}B^{\prime}})\leq1}\boldsymbol{D}(\mathcal{N}%
_{AB\rightarrow A^{\prime}B^{\prime}}\Vert\mathcal{M}_{AB\rightarrow
A^{\prime}B^{\prime}}),\label{eq:gen-div-ch-ups-meas-bi-map}%
\end{multline}
where the optimization is with respect to every completely positive bipartite map
$\mathcal{M}_{AB\rightarrow A^{\prime}B^{\prime}}$.
\end{definition}

Using the quantum relative entropy, the sandwiched R\'{e}nyi relative entropy,
the Belavkin--Staszewski relative entropy, and the geometric R\'{e}nyi
relative entropy, we then obtain the following respective channel measures:
\begin{align}
\Upsilon(\mathcal{N}_{AB\rightarrow A^{\prime}B^{\prime}}), \\
\widetilde
{\Upsilon}_{\alpha}(\mathcal{N}_{AB\rightarrow A^{\prime}B^{\prime}}),\\
\widehat{\Upsilon}(\mathcal{N}_{AB\rightarrow A^{\prime}B^{\prime}}), \\
\widehat{\Upsilon}_{\alpha}(\mathcal{N}_{AB\rightarrow A^{\prime}B^{\prime}%
}),
\end{align}
defined by substituting $\boldsymbol{D}$ with $D$, $\widetilde{D}%
_{\alpha}$, $\widehat{D}$, and $\widehat{D}_{\alpha}$ in \eqref{eq:gen-div-ch-ups-meas-bi-map}.

We now establish some properties of $\mathbf{\Upsilon}(\mathcal{N}%
_{AB\rightarrow A^{\prime}B^{\prime}})$, analogous to those established
earlier for $C_{\beta}(\mathcal{N}_{AB\rightarrow A^{\prime}B^{\prime}})$ in
Section~\ref{sec:props-basic-measure}.

\begin{proposition}
[Non-negativity]\label{prop:non-neg-gen-div-ups}Let $\mathcal{N}%
_{AB\rightarrow A^{\prime}B^{\prime}}$ be a bipartite channel. Then
\begin{equation}
\mathbf{\Upsilon}(\mathcal{N}_{AB\rightarrow A^{\prime}B^{\prime}})\geq0.
\end{equation}

\end{proposition}

\begin{IEEEproof}
Let $\mathcal{M}_{AB\rightarrow A^{\prime}B^{\prime}}$ be an arbitrary
completely positive bipartite map satisfying $\beta(\mathcal{M}_{AB\rightarrow
A^{\prime}B^{\prime}})\leq1$. Then consider that%
\begin{align}
&  \boldsymbol{D}(\mathcal{N}_{AB\rightarrow A^{\prime}B^{\prime}}%
\Vert\mathcal{M}_{AB\rightarrow A^{\prime}B^{\prime}})\nonumber\\
&  \geq\boldsymbol{D}(\mathcal{N}_{AB\rightarrow A^{\prime}B^{\prime}}%
(\Phi_{RA}\otimes\Phi_{BS})\Vert\mathcal{M}_{AB\rightarrow A^{\prime}%
B^{\prime}}(\Phi_{RA}\otimes\Phi_{BS}))\nonumber\\
&  \geq\boldsymbol{D}(\operatorname{Tr}[\mathcal{N}_{AB\rightarrow A^{\prime
}B^{\prime}}(\Phi_{RA}\otimes\Phi_{BS})]\Vert\nonumber\\
&  \qquad\qquad\operatorname{Tr}[\mathcal{M}_{AB\rightarrow A^{\prime
}B^{\prime}}(\Phi_{RA}\otimes\Phi_{BS})])\\
&  =\boldsymbol{D}(1\Vert\operatorname{Tr}[\mathcal{M}_{AB\rightarrow
A^{\prime}B^{\prime}}(\Phi_{RA}\otimes\Phi_{BS})]).
\end{align}
The first inequality follows because the quantity $\boldsymbol{D}(\mathcal{N}%
_{AB\rightarrow A^{\prime}B^{\prime}}\Vert\mathcal{M}_{AB\rightarrow
A^{\prime}B^{\prime}})$ involves an optimization over all possible input
states, and we have chosen the product of maximally entangled states. The
second inequality follows from the data-processing inequality for the
generalized divergence. It then follows from
Definition~\ref{def:ups-bipartite-gen} that%
\begin{multline}
\mathbf{\Upsilon}(\mathcal{N}_{AB\rightarrow A^{\prime}B^{\prime}})\geq\\
\inf_{\substack{\mathcal{M}_{AB\rightarrow A^{\prime}B^{\prime}}%
:\\\beta(\mathcal{M}_{AB\rightarrow A^{\prime}B^{\prime}})\leq1}%
}\boldsymbol{D}(1\Vert\operatorname{Tr}[\mathcal{M}_{AB\rightarrow A^{\prime
}B^{\prime}}(\Phi_{RA}\otimes\Phi_{BS})]).
\end{multline}
Thus, the inequality follows if we can show that%
\begin{equation}
\operatorname{Tr}[\mathcal{M}_{AB\rightarrow A^{\prime}B^{\prime}}(\Phi
_{RA}\otimes\Phi_{BS})]\leq1.\label{eq:trace-M-less-one}%
\end{equation}
Let $\lambda$, $S_{AA^{\prime}BB^{\prime}}$, and $V_{AA^{\prime}BB^{\prime}}$
be arbitrary Hermitian operators satisfying the constraints in
\eqref{eq:basic-measure-NS-PPT-bi-rewrite} for $\mathcal{M}_{AB\rightarrow
A^{\prime}B^{\prime}}$. Then, we find that%
\begin{align}
\lambda d_{A}d_{B} &  =\lambda\operatorname{Tr}_{AB}[I_{AB}]\\
&  \geq\operatorname{Tr}_{AA^{\prime}BB^{\prime}}[S_{AA^{\prime}BB^{\prime}%
}]\\
&  \geq\operatorname{Tr}_{AA^{\prime}BB^{\prime}}[V_{AA^{\prime}BB^{\prime}%
}]\\
&  =\operatorname{Tr}_{AA^{\prime}BB^{\prime}}[T_{BB^{\prime}}(V_{AA^{\prime
}BB^{\prime}})]\\
&  \geq\operatorname{Tr}_{AA^{\prime}BB^{\prime}}[T_{BB^{\prime}}%
(\Gamma_{AA^{\prime}BB^{\prime}}^{\mathcal{M}})]\\
&  =\operatorname{Tr}_{AA^{\prime}BB^{\prime}}[\Gamma_{AA^{\prime}BB^{\prime}%
}^{\mathcal{M}}]\\
&  =\operatorname{Tr}[\Gamma_{AA^{\prime}BB^{\prime}}^{\mathcal{M}}],
\end{align}
which is equivalent to%
\begin{equation}
\lambda\geq\operatorname{Tr}[\mathcal{M}_{AB\rightarrow A^{\prime}B^{\prime}%
}(\Phi_{RA}\otimes\Phi_{BS})].
\end{equation}
Taking an infimum over $\lambda$, $S_{AA^{\prime}BB^{\prime}}$, and
$V_{AA^{\prime}BB^{\prime}}$ satisfying the constraints in
\eqref{eq:basic-measure-NS-PPT-bi-rewrite} for $\mathcal{M}_{AB\rightarrow
A^{\prime}B^{\prime}}$ and applying the assumption $\beta(\mathcal{M}%
_{AB\rightarrow A^{\prime}B^{\prime}})\leq1$, we conclude \eqref{eq:trace-M-less-one}.
\end{IEEEproof}

\bigskip

\begin{proposition}
[Stability]\label{prop:stability-gen-div-ups}Let $\mathcal{N}_{AB\rightarrow
A^{\prime}B^{\prime}}$ be a bipartite channel. Then
\begin{equation}
\mathbf{\Upsilon}(\mathcal{N}_{AB\rightarrow A^{\prime}B^{\prime}%
})=\mathbf{\Upsilon}(\operatorname{id}_{\bar{A}\rightarrow\tilde{A}}%
\otimes\mathcal{N}_{AB\rightarrow A^{\prime}B^{\prime}}\otimes
\operatorname{id}_{\bar{B}\rightarrow\tilde{B}}).
\label{eq:stability-gen-div-ups}%
\end{equation}

\end{proposition}

\begin{IEEEproof}
The definition of the generalized channel divergence in
\eqref{eq:gen-ch-div}\ implies that it is stable, in the sense that%
\begin{multline}
\boldsymbol{D}(\mathcal{N}_{AB\rightarrow A^{\prime}B^{\prime}}\Vert
\mathcal{M}_{AB\rightarrow A^{\prime}B^{\prime}})=\\
\boldsymbol{D}(\operatorname{id}_{\bar{A}\rightarrow\tilde{A}}\otimes
\mathcal{N}_{AB\rightarrow A^{\prime}B^{\prime}}\otimes\operatorname{id}%
_{\bar{B}\rightarrow\tilde{B}}\Vert\\
\operatorname{id}_{\bar{A}\rightarrow\tilde{A}}\otimes\mathcal{M}%
_{AB\rightarrow A^{\prime}B^{\prime}}\otimes\operatorname{id}_{\bar
{B}\rightarrow\tilde{B}}),
\end{multline}
for every channel $\mathcal{N}_{AB\rightarrow A^{\prime}B^{\prime}}$ and
completely positive map $\mathcal{M}_{AB\rightarrow A^{\prime}B^{\prime}}$.
Combining with Proposition~\ref{prop:stability}\ and the definition in
\eqref{eq:gen-div-ch-ups-meas-bi-map}, we conclude \eqref{eq:stability-gen-div-ups}.
\end{IEEEproof}

\bigskip

\begin{proposition}
[Zero on classical feedback channels]\label{prop:zero-feedback-renyi-ups}Let
$\overline{\Delta}_{B\rightarrow A^{\prime}}$ be a classical feedback channel:%
\begin{equation}
\overline{\Delta}_{B\rightarrow A^{\prime}}(\cdot)\coloneqq\sum_{i=0}%
^{d-1}|i\rangle_{A^{\prime}}\langle i|_{B}(\cdot)|i\rangle_{B}\langle
i|_{A^{\prime}},
\end{equation}
where $A^{\prime}\simeq B$ and $d=d_{A^{\prime}}=d_{B}$. Then%
\begin{equation}
\mathbf{\Upsilon}(\overline{\Delta}_{B\rightarrow A^{\prime}})=0.
\end{equation}

\end{proposition}

\begin{IEEEproof}
This follows from Proposition~\ref{prop:zero-feedback}. Since $\beta
(\overline{\Delta}_{B\rightarrow A^{\prime}})=1$, we can pick $\mathcal{M}%
_{B\rightarrow A^{\prime}}=\overline{\Delta}_{B\rightarrow A^{\prime}}$, and
then%
\begin{equation}
\boldsymbol{D}(\overline{\Delta}_{B\rightarrow A^{\prime}}\Vert\mathcal{M}%
_{B\rightarrow A^{\prime}})=\boldsymbol{D}(\overline{\Delta}_{B\rightarrow
A^{\prime}}\Vert\overline{\Delta}_{B\rightarrow A^{\prime}})=0.
\end{equation}
So this establishes that $\mathbf{\Upsilon}(\overline{\Delta}_{B\rightarrow
A^{\prime}})\leq0$, and the other inequality $\mathbf{\Upsilon}(\overline
{\Delta}_{B\rightarrow A^{\prime}})\geq0$ follows from
Proposition~\ref{prop:non-neg-gen-div-ups}.
\end{IEEEproof}

\bigskip

\begin{proposition}
[Zero on tensor product of local channels]%
\label{prop:zero-local-chs-gen-div-ups}Let $\mathcal{E}_{A\rightarrow
A^{\prime}}$ and $\mathcal{F}_{B\rightarrow B^{\prime}}$ be quantum channels.
Then%
\begin{equation}
\mathbf{\Upsilon}(\mathcal{E}_{A\rightarrow A^{\prime}}\otimes\mathcal{F}%
_{B\rightarrow B^{\prime}})=0.
\end{equation}

\end{proposition}

\begin{IEEEproof}
Same argument as given for Proposition~\ref{prop:zero-feedback-renyi-ups}, but
use Proposition~\ref{prop:zero-local-chs}\ instead.
\end{IEEEproof}

\bigskip

We now establish some properties that are more specific to the
Belavkin--Staszewski and geometric R\'enyi relative entropies (however the
first actually holds for quantum relative entropy and other quantum R\'enyi
relative entropies).

\begin{proposition}
\label{prop:ups-to-cbeta-bipartite}Let $\mathcal{N}_{AB\rightarrow A^{\prime
}B^{\prime}}$ be a bipartite channel. Then for all $\alpha\in(1,2]$,%
\begin{equation}
\widehat{\Upsilon}(\mathcal{N}_{AB\rightarrow A^{\prime}B^{\prime}}%
)\leq\widehat{\Upsilon}_{\alpha}(\mathcal{N}_{AB\rightarrow A^{\prime
}B^{\prime}})\leq C_{\beta}(\mathcal{N}_{AB\rightarrow A^{\prime}B^{\prime}}).
\end{equation}

\end{proposition}

\begin{IEEEproof}
Pick $\mathcal{M}_{AB\rightarrow A^{\prime}B^{\prime}}=\frac{1}{\beta
(\mathcal{N}_{AB\rightarrow A^{\prime}B^{\prime}})}\mathcal{N}_{AB\rightarrow
A^{\prime}B^{\prime}}$ in the definition of $\widehat{\Upsilon}(\mathcal{N}%
_{AB\rightarrow A^{\prime}B^{\prime}})$ and $\widehat{\Upsilon}_{\alpha
}(\mathcal{N}_{AB\rightarrow A^{\prime}B^{\prime}})$ and use the fact that,
for $c>0$, $\widehat{D}(\rho\Vert c\sigma)=\widehat{D}(\rho\Vert\sigma
)-\log_{2}c$ and $\widehat{D}_{\alpha}(\rho\Vert c\sigma)=\widehat{D}_{\alpha
}(\rho\Vert\sigma)-\log_{2}c$ for all $\alpha\in(1,2]$. We also require \eqref{eq:geometric-Renyi-monotone-alpha}.
\end{IEEEproof}

\bigskip

\begin{proposition}
[Subadditivity]For bipartite channels $\mathcal{N}_{AB\rightarrow A^{\prime
}B^{\prime}}^{1}$ and $\mathcal{N}_{A^{\prime}B^{\prime}\rightarrow
A^{\prime\prime}B^{\prime\prime}}^{2}$, the following inequality holds for all
$\alpha\in(0,1)\cup(1,2]$:%
\begin{multline}
\widehat{\Upsilon}_{\alpha}(\mathcal{N}_{A^{\prime}B^{\prime}\rightarrow
A^{\prime\prime}B^{\prime\prime}}^{2}\circ\mathcal{N}_{AB\rightarrow
A^{\prime}B^{\prime}}^{1})\leq\\
\widehat{\Upsilon}_{\alpha}(\mathcal{N}_{A^{\prime}B^{\prime}\rightarrow
A^{\prime\prime}B^{\prime\prime}}^{2})+\widehat{\Upsilon}_{\alpha}%
(\mathcal{N}_{AB\rightarrow A^{\prime}B^{\prime}}^{1}).
\end{multline}

\end{proposition}

\begin{IEEEproof}
This inequality is a direct consequence of the subadditivity inequality in
Eq.~(18) of \cite{Fang2019a}, Proposition~47\ of \cite{KW20}, and the fact
that if $\mathcal{M}^{1}$ and $\mathcal{M}^{2}$ are completely positive
bipartite maps satisfying $\beta(\mathcal{M}^{1}),\beta(\mathcal{M}^{2})\leq
1$, then $\beta(\mathcal{M}^{2}\circ\mathcal{M}^{1})\leq1$ (see
Proposition~\ref{prop:subadditivity-bipartite-CP-maps}).
\end{IEEEproof}

\subsection{Measure of classical communication for a point-to-point channel}

Let $\mathcal{M}_{A\rightarrow B^{\prime}}$ be a point-to-point completely
positive map, which is a special case of a completely positive bipartite map
with the Bob input $B$ trivial and the Alice output $A^{\prime}$ trivial.\ We
first show that $\beta$ in \eqref{eq:basic-measure-NS-PPT-bi}\ reduces to the
measure from \cite{WXD18}.

\begin{proposition}
Let $\mathcal{M}_{A\rightarrow B^{\prime}}$ be a point-to-point completely
positive map. Then%
\begin{multline}
\beta(\mathcal{M}_{A\rightarrow B^{\prime}})=\\
\inf_{S_{B^{\prime}},V_{AB^{\prime}}\in\operatorname{Herm}}\left\{
\begin{array}
[c]{c}%
\operatorname{Tr}[S_{B^{\prime}}]:\\
T_{B^{\prime}}(V_{AB^{\prime}}\pm\Gamma_{AB^{\prime}}^{\mathcal{M}})\geq0,\\
I_{A}\otimes S_{B^{\prime}}\pm V_{AB^{\prime}}\geq0
\end{array}
\right\}  . \label{eq:beta-p2p-exp}%
\end{multline}

\end{proposition}

\begin{IEEEproof}
In this case, the systems $A^{\prime}$ and $B$ are trivial. So then the
definition in \eqref{eq:basic-measure-NS-PPT-bi} reduces to%
\begin{multline}
\beta(\mathcal{M}_{A\rightarrow B^{\prime}})=\\
\inf_{S_{AB^{\prime}},V_{AB^{\prime}}\in\operatorname{Herm}}\left\{
\begin{array}
[c]{c}%
\left\Vert \operatorname{Tr}_{B^{\prime}}[S_{AB^{\prime}}]\right\Vert
_{\infty}:\\
T_{B^{\prime}}(V_{AB^{\prime}}\pm\Gamma_{AB^{\prime}}^{\mathcal{M}})\geq0,\\
S_{AB^{\prime}}\pm V_{AB^{\prime}}\geq0,\\
S_{AB^{\prime}}=\pi_{A}\otimes\operatorname{Tr}_{A}[S_{AB^{\prime}}]
\end{array}
\right\}  .
\end{multline}
The last constraint implies that the optimization simplifies to%
\begin{align}
&  \beta(\mathcal{M}_{A\rightarrow B^{\prime}})\nonumber\\
&  = \inf_{S_{AB^{\prime}},V_{AB^{\prime}}\in\operatorname{Herm}}\left\{
\begin{array}
[c]{c}%
\left\Vert \operatorname{Tr}_{B^{\prime}}[\pi_{A}\otimes\operatorname{Tr}%
_{A}[S_{AB^{\prime}}]]\right\Vert _{\infty}:\\
T_{B^{\prime}}(V_{AB^{\prime}}\pm\Gamma_{AB^{\prime}}^{\mathcal{M}})\geq0,\\
\pi_{A}\otimes\operatorname{Tr}_{A}[S_{AB^{\prime}}]\pm V_{AB^{\prime}}\geq0
\end{array}
\right\} \\
&  =\inf_{S_{B^{\prime}}^{\prime},V_{AB^{\prime}}\in\operatorname{Herm}%
}\left\{
\begin{array}
[c]{c}%
\left\Vert \operatorname{Tr}_{B^{\prime}}[\pi_{A}\otimes S_{B^{\prime}%
}^{\prime}]\right\Vert _{\infty}:\\
T_{B^{\prime}}(V_{AB^{\prime}}\pm\Gamma_{AB^{\prime}}^{\mathcal{M}})\geq0,\\
\pi_{A}\otimes S_{B^{\prime}}^{\prime}\pm V_{AB^{\prime}}\geq0
\end{array}
\right\} \\
&  =\inf_{S_{B^{\prime}}^{\prime},V_{AB^{\prime}}\in\operatorname{Herm}%
}\left\{
\begin{array}
[c]{c}%
\operatorname{Tr}[S_{B^{\prime}}^{\prime}]\left\Vert \pi_{A}\right\Vert
_{\infty}:\\
T_{B^{\prime}}(V_{AB^{\prime}}\pm\Gamma_{AB^{\prime}}^{\mathcal{M}})\geq0,\\
\pi_{A}\otimes S_{B^{\prime}}^{\prime}\pm V_{AB^{\prime}}\geq0
\end{array}
\right\} \\
&  =\inf_{S_{B^{\prime}}^{\prime},V_{AB^{\prime}}\in\operatorname{Herm}%
}\left\{
\begin{array}
[c]{c}%
\frac{1}{d_{A}}\operatorname{Tr}[S_{B^{\prime}}^{\prime}]:\\
T_{B^{\prime}}(V_{AB^{\prime}}\pm\Gamma_{AB^{\prime}}^{\mathcal{M}})\geq0,\\
\pi_{A}\otimes S_{B^{\prime}}^{\prime}\pm V_{AB^{\prime}}\geq0
\end{array}
\right\} \\
&  =\inf_{S_{B^{\prime}},V_{AB^{\prime}}\in\operatorname{Herm}}\left\{
\begin{array}
[c]{c}%
\operatorname{Tr}[S_{B^{\prime}}]:\\
T_{B^{\prime}}(V_{AB^{\prime}}\pm\Gamma_{AB^{\prime}}^{\mathcal{M}})\geq0,\\
I_{A}\otimes S_{B^{\prime}}\pm V_{AB^{\prime}}\geq0
\end{array}
\right\}  .
\end{align}
This concludes the proof.
\end{IEEEproof}

\bigskip

More generally, consider that the definition in
\eqref{eq:gen-div-ch-ups-meas-bi-map}\ becomes as follows for a point-to-point
channel $\mathcal{N}_{A\rightarrow B^{\prime}}$:%
\begin{equation}
\mathbf{\Upsilon}(\mathcal{N}_{A\rightarrow B^{\prime}})\coloneqq\inf
_{\mathcal{M}_{A\rightarrow B^{\prime}}:\beta(\mathcal{M}_{A\rightarrow
B^{\prime}})\leq1}\boldsymbol{D}(\mathcal{N}_{A\rightarrow B^{\prime}}%
\Vert\mathcal{M}_{A\rightarrow B^{\prime}}), \label{eq:p2p-gen-div-ups}%
\end{equation}
which leads to the quantities $\widehat{\Upsilon}(\mathcal{N}_{A\rightarrow
B^{\prime}})$ and $\widehat{\Upsilon}_{\alpha}(\mathcal{N}_{A\rightarrow
B^{\prime}})$, for which we have the following bounds for $\alpha\in(1,2]$:%
\begin{equation}
\widehat{\Upsilon}(\mathcal{N}_{A\rightarrow B^{\prime}})\leq\widehat
{\Upsilon}_{\alpha}(\mathcal{N}_{A\rightarrow B^{\prime}})\leq C_{\beta
}(\mathcal{N}_{A\rightarrow B^{\prime}}).
\label{eq:ups-p2p-relations-bs-geo-c-beta}%
\end{equation}
Note that the quantities given just above were defined in
\cite{WFT18,Fang2019a}, and our observation here is that the definition in
\eqref{eq:gen-div-ch-ups-meas-bi-map} reduces to them.

The next proposition is critical for establishing our upper bound proofs in
Section~\ref{sec:apps}. It states that if one share of a maximally classically
correlated state passes through a completely positive\ map $\mathcal{M}%
_{A\rightarrow B^{\prime}}$\ for which $\beta(\mathcal{M}_{A\rightarrow
B^{\prime}})\leq1$, then the resulting operator has a very small chance of
passing the comparator test, as defined in \eqref{eq:comparator-test-def}.

\begin{proposition}
[Bound for comparator test success probability]\label{prop:comp-test}Let%
\begin{equation}
\overline{\Phi}_{\hat{A}A}\coloneqq\frac{1}{d}\sum_{i=0}^{d-1}|i\rangle
\!\langle i|_{\hat{A}}\otimes|i\rangle\!\langle i|_{A}%
\end{equation}
denote the maximally classically correlated state, and let $\mathcal{M}%
_{A\rightarrow B^{\prime}}$\ be a completely positive\ map  for which $\beta(\mathcal{M}_{A\rightarrow
B^{\prime}})\leq1$. Then%
\begin{equation}
\operatorname{Tr}[\Pi_{\hat{A}B^{\prime}}\mathcal{M}_{A\rightarrow B^{\prime}%
}(\overline{\Phi}_{\hat{A}A})]\leq\frac{1}{d},
\end{equation}
where $\Pi_{\hat{A}B^{\prime}}$ is the comparator test:%
\begin{equation}
\Pi_{\hat{A}B^{\prime}}\coloneqq\sum_{i=0}^{d-1}|i\rangle\!\langle i|_{\hat
{A}}\otimes|i\rangle\!\langle i|_{B^{\prime}}, \label{eq:comparator-test-def}%
\end{equation}
and $\hat{A}\simeq A\simeq B^{\prime}$.
\end{proposition}

\begin{IEEEproof}
Recall the expression for $\beta(\mathcal{M}_{A\rightarrow B^{\prime}})$ in
\eqref{eq:beta-p2p-exp}. Let $S_{B^{\prime}}$ and $V_{AB^{\prime}}$ be
arbitrary Hermitian operators satisfying the constraints for $\beta
(\mathcal{M}_{A\rightarrow B^{\prime}})$. An application of
\eqref{eq:post-selected-TP-id}\ implies that%
\begin{equation}
\mathcal{M}_{A\rightarrow B^{\prime}}(\overline{\Phi}_{\hat{A}A}%
)=\langle\Gamma|_{A\tilde{A}}\overline{\Phi}_{\hat{A}A}\otimes\Gamma
_{\tilde{A}B^{\prime}}^{\mathcal{M}}|\Gamma\rangle_{A\tilde{A}},
\end{equation}
where $\tilde{A}\simeq A$. This means that%
\begin{align}
&  \operatorname{Tr}[\Pi_{\hat{A}B^{\prime}}\mathcal{M}_{A\rightarrow
B^{\prime}}(\overline{\Phi}_{\hat{A}A})]\nonumber\\
&  =\operatorname{Tr}[\Pi_{\hat{A}B^{\prime}}\langle\Gamma|_{A\tilde{A}%
}\overline{\Phi}_{\hat{A}A}\otimes\Gamma_{\tilde{A}B^{\prime}}^{\mathcal{M}%
}|\Gamma\rangle_{A\tilde{A}}]\\
&  =\operatorname{Tr}[T_{B^{\prime}}(\Pi_{\hat{A}B^{\prime}})\langle
\Gamma|_{A\tilde{A}}\overline{\Phi}_{\hat{A}A}\otimes\Gamma_{\tilde
{A}B^{\prime}}^{\mathcal{M}}|\Gamma\rangle_{A\tilde{A}}]\\
&  =\operatorname{Tr}[\Pi_{\hat{A}B^{\prime}}\langle\Gamma|_{A\tilde{A}%
}\overline{\Phi}_{\hat{A}A}\otimes T_{B^{\prime}}(\Gamma_{\tilde{A}B^{\prime}%
}^{\mathcal{M}})|\Gamma\rangle_{A\tilde{A}}]\\
&  \leq\operatorname{Tr}[\Pi_{\hat{A}B^{\prime}}\langle\Gamma|_{A\tilde{A}%
}\overline{\Phi}_{\hat{A}A}\otimes T_{B^{\prime}}(V_{\tilde{A}B^{\prime}%
})|\Gamma\rangle_{A\tilde{A}}]\\
&  =\operatorname{Tr}[T_{B^{\prime}}(\Pi_{\hat{A}B^{\prime}})\langle
\Gamma|_{A\tilde{A}}\overline{\Phi}_{\hat{A}A}\otimes V_{\tilde{A}B^{\prime}%
}|\Gamma\rangle_{A\tilde{A}}]\\
&  =\operatorname{Tr}[\Pi_{\hat{A}B^{\prime}}\langle\Gamma|_{A\tilde{A}%
}\overline{\Phi}_{\hat{A}A}\otimes V_{\tilde{A}B^{\prime}}|\Gamma
\rangle_{A\tilde{A}}]\\
&  \leq\operatorname{Tr}[\Pi_{\hat{A}B^{\prime}}\langle\Gamma|_{A\tilde{A}%
}\overline{\Phi}_{\hat{A}A}\otimes I_{\tilde{A}}\otimes S_{B^{\prime}}%
|\Gamma\rangle_{A\tilde{A}}]\\
&  =\operatorname{Tr}[\Pi_{\hat{A}B^{\prime}}\langle\Gamma|_{A\tilde{A}%
}\overline{\Phi}_{\hat{A}A}\otimes I_{\tilde{A}}|\Gamma\rangle_{A\tilde{A}%
}\otimes S_{B^{\prime}}]\\
&  =\operatorname{Tr}[\Pi_{\hat{A}B^{\prime}}\operatorname{Tr}_{A}%
[\overline{\Phi}_{\hat{A}A}]\otimes S_{B^{\prime}}]\\
&  =\frac{1}{d}\operatorname{Tr}[\Pi_{\hat{A}B^{\prime}}I_{\hat{A}}\otimes
S_{B^{\prime}}]\\
&  =\frac{1}{d}\operatorname{Tr}[S_{B^{\prime}}].
\end{align}
Since this holds for all $S_{B^{\prime}}$ and $V_{AB^{\prime}}$ satisfying the
constraints for $\beta(\mathcal{M}_{A\rightarrow B^{\prime}})$, we conclude
that%
\begin{equation}
\operatorname{Tr}[\Pi_{\hat{A}B^{\prime}}\mathcal{M}_{A\rightarrow B^{\prime}%
}(\overline{\Phi}_{\hat{A}A})]\leq\frac{1}{d}.
\end{equation}
This concludes the proof.
\end{IEEEproof}

\bigskip

We finally state another proposition that plays an essential role in our upper
bound proofs in Section~\ref{sec:apps}.

\begin{proposition}
\label{prop:err-bound}Suppose that $\mathcal{N}_{A\rightarrow B}$ is a channel
with $A\simeq B$ that satisfies%
\begin{equation}
\frac{1}{2}\left\Vert \mathcal{N}_{A\rightarrow B}(\overline{\Phi}%
_{RA})-\overline{\Phi}_{RB}\right\Vert _{1}\leq\varepsilon,
\end{equation}
for $\varepsilon\in\lbrack0,1)$ and where $\overline{\Phi}_{RB}\coloneqq\frac
{1}{d}\sum_{i}|i\rangle\!\langle i|_{R}\otimes|i\rangle\!\langle i|_{B}$ and
$d=d_{R}=d_{A}=d_{B}$. Then%
\begin{multline}
\log_{2}d\leq\\
\inf_{\mathcal{M}_{A\rightarrow B}:\beta(\mathcal{M}_{A\rightarrow B})\leq
1}D_{H}^{\varepsilon}(\mathcal{N}_{A\rightarrow B}(\overline{\Phi}_{RA}%
)\Vert\mathcal{M}_{A\rightarrow B}(\overline{\Phi}_{RA})),
\label{eq:err-bound-HTRE}%
\end{multline}
and for all $\alpha\in(1,2]$,%
\begin{multline}
\log_{2}d\leq\\
\inf_{\mathcal{M}_{A\rightarrow B}:\beta(\mathcal{M}_{A\rightarrow B})\leq
1}\widehat{D}_{\alpha}(\mathcal{N}_{A\rightarrow B}(\overline{\Phi}_{RA}%
)\Vert\mathcal{M}_{A\rightarrow B}(\overline{\Phi}_{RA}))\\
+\frac{\alpha}{\alpha-1}\log_{2}\!\left(  \frac{1}{1-\varepsilon}\right)  .
\label{eq:err-bound}%
\end{multline}

\end{proposition}

\begin{IEEEproof}
We begin by proving \eqref{eq:err-bound-HTRE}. The condition%
\begin{equation}
\frac{1}{2}\left\Vert \mathcal{N}_{A\rightarrow B}(\overline{\Phi}%
_{RA})-\overline{\Phi}_{RB}\right\Vert _{1}\leq\varepsilon
\end{equation}
implies that%
\begin{equation}
\operatorname{Tr}[\Pi_{RB}\mathcal{N}_{A\rightarrow B}(\overline{\Phi}%
_{RA})]\geq1-\varepsilon,
\end{equation}
where $\Pi_{RB}\coloneqq\sum_{i}|i\rangle\!\langle i|_{R}\otimes
|i\rangle\!\langle i|_{B}$ is the comparator test. Indeed, applying a
completely dephasing channel $\overline{\Delta}_{B}(\cdot)\coloneqq\sum
_{i}|i\rangle\!\langle i|(\cdot)|i\rangle\!\langle i|$\ to the output of the
channel $\mathcal{N}_{A\rightarrow B}$ and applying the data-processing
inequality for trace distance, we conclude that%
\begin{align}
\varepsilon &  \geq\frac{1}{2}\left\Vert \mathcal{N}_{A\rightarrow
B}(\overline{\Phi}_{RA})-\overline{\Phi}_{RB}\right\Vert _{1}\\
&  \geq\frac{1}{2}\left\Vert (\overline{\Delta}_{B}\circ\mathcal{N}%
_{A\rightarrow B})(\overline{\Phi}_{RA})-\overline{\Delta}_{B}(\overline{\Phi
}_{RB})\right\Vert _{1}\\
&  =\frac{1}{2}\left\Vert (\overline{\Delta}_{B}\circ\mathcal{N}_{A\rightarrow
B})(\overline{\Phi}_{RA})-\overline{\Phi}_{RB}\right\Vert _{1}.
\end{align}
Let $\omega_{RB}\coloneqq(\overline{\Delta}_{B}\circ\mathcal{N}_{A\rightarrow
B})(\overline{\Phi}_{RA})$ and observe that it can be written as%
\begin{equation}
\omega_{RB}=\frac{1}{d}\sum_{i,j}p(j|i)|i\rangle\!\langle i|_{R}%
\otimes|j\rangle\!\langle j|_{B}%
\end{equation}
for some conditional probability distribution $p(j|i)$. Then%
\begin{align}
&  \frac{1}{2}\left\Vert (\overline{\Delta}_{B}\circ\mathcal{N}_{A\rightarrow
B})(\overline{\Phi}_{RA})-\overline{\Phi}_{RB}\right\Vert _{1}\nonumber\\
&  =\frac{1}{2}\left\Vert
\begin{array}
[c]{c}%
\frac{1}{d}\sum_{i,j}p(j|i)|i\rangle\!\langle i|_{R}\otimes|j\rangle\!\langle
j|_{B}\\
\quad-\frac{1}{d}\sum_{i,j}\delta_{i,j}|i\rangle\!\langle i|_{R}\otimes
|j\rangle\!\langle j|_{B}%
\end{array}
\right\Vert _{1}\\
&  =\frac{1}{2d}\sum_{i}\left\Vert \sum_{j}(p(j|i)-\delta
_{i,j})|j\rangle\!\langle j|_{B}\right\Vert _{1}\\
&  =\frac{1}{2d}\sum_{i}\left[  \left(  1-p(i|i)\right)
+\sum_{j\neq i}p(j|i)\right]  \\
&  =\frac{1}{d}\sum_{i}\left(  1-p(i|i)\right)  \\
&  =1-\sum_{i}\frac{1}{d}p(i|i).
\end{align}
This implies that%
\begin{equation}
\sum_{i}\frac{1}{d}p(i|i)\geq1-\varepsilon.
\end{equation}
Now consider that%
\begin{align}
&  \operatorname{Tr}[\Pi_{RB}\mathcal{N}_{A\rightarrow B}(\overline{\Phi}%
_{RA})]\nonumber\\
&  =\operatorname{Tr}[\overline{\Delta}_{B}(\Pi_{RB})\mathcal{N}_{A\rightarrow
B}(\overline{\Phi}_{RA})]\\
&  =\operatorname{Tr}[\Pi_{RB}(\overline{\Delta}_{B}\circ\mathcal{N}%
_{A\rightarrow B})(\overline{\Phi}_{RA})]\\
&  =\operatorname{Tr}[\Pi_{RB}\omega_{RB}]\\
&  =\sum_{i}\frac{1}{d}p(i|i).
\end{align}
So we conclude that%
\begin{equation}
\operatorname{Tr}[\Pi_{RB}\mathcal{N}_{A\rightarrow B}(\overline{\Phi}%
_{RA})]\geq1-\varepsilon.
\label{eq:achievable-HTRE-meta}
\end{equation}
Applying the definition of the hypothesis testing relative entropy from
\eqref{eq:HTRE-def}, we conclude that%
\begin{multline}
\inf_{\mathcal{M}_{A\rightarrow B}:\beta(\mathcal{M}_{A\rightarrow B})\leq
1}D_{H}^{\varepsilon}(\mathcal{N}_{A\rightarrow B}(\overline{\Phi}_{RA}%
)\Vert\mathcal{M}_{A\rightarrow B}(\overline{\Phi}_{RA}))=\nonumber\\
-\log_{2}\sup_{\substack{\mathcal{M}_{A\rightarrow B}:\\\beta(\mathcal{M}%
_{A\rightarrow B})\leq1}}\inf_{\Lambda_{RB}\geq0}\left\{
\begin{array}
[c]{c}%
\operatorname{Tr}[\Lambda_{RB}\mathcal{M}_{A\rightarrow B}(\overline{\Phi
}_{RA})]:\\
\operatorname{Tr}[\Lambda_{RB}\mathcal{N}_{A\rightarrow B}(\overline{\Phi
}_{RA})]\\
\geq1-\varepsilon,\\
\Lambda_{RB}\leq I_{RB}%
\end{array}
\right\}  .
\end{multline}
Now consider that%
\begin{align}
&  \sup_{\substack{\mathcal{M}_{A\rightarrow B}:\\\beta(\mathcal{M}%
_{A\rightarrow B})\leq1}}\inf_{\Lambda_{RB}\geq0}\left\{
\begin{array}
[c]{c}%
\operatorname{Tr}[\Lambda_{RB}\mathcal{M}_{A\rightarrow B}(\overline{\Phi
}_{RA})]:\\
\operatorname{Tr}[\Lambda_{RB}\mathcal{N}_{A\rightarrow B}(\overline{\Phi
}_{RA})]\geq1-\varepsilon,\\
\Lambda_{RB}\leq I_{RB}%
\end{array}
\right\}  \nonumber\\
&  \leq\sup_{\mathcal{M}_{A\rightarrow B}:\beta(\mathcal{M}_{A\rightarrow
B})\leq1}\operatorname{Tr}[\Pi_{RB}\mathcal{M}_{A\rightarrow B}(\overline
{\Phi}_{RA})]\\
&  \leq\frac{1}{d},
\end{align}
where the first inequality follows from \eqref{eq:achievable-HTRE-meta} and the definition in \eqref{eq:HTRE-def} and the last inequality follows from Proposition~\ref{prop:comp-test}. Then
applying a negative logarithm gives \eqref{eq:err-bound-HTRE}.

The inequality in \eqref{eq:err-bound}\ follows as a direct application of the
following relationship between the hypothesis testing relative entropy and the
geometric R\'{e}nyi relative entropy:%
\begin{equation}
D_{H}^{\varepsilon}(\rho\Vert\sigma)\leq\widehat{D}_{\alpha}(\rho\Vert
\sigma)+\frac{\alpha}{\alpha-1}\log_{2}\!\left(  \frac{1}{1-\varepsilon
}\right)  , \label{eq:HTRE-to-geo-Reny}%
\end{equation}
as well as the previous proposition. The proof of \eqref{eq:HTRE-to-geo-Reny}
follows the same proof given for \cite[Lemma~5]{CMW14}.
\end{IEEEproof}

\section{Applications\label{sec:apps}}

\subsection{Bounding the forward classical capacity of a bipartite channel}

We now apply the bipartite channel measure in
\eqref{eq:gen-div-ch-ups-meas-bi-map} to obtain an upper bound on the forward
classical capacity of a bipartite channel $\mathcal{N}_{AB\rightarrow
A^{\prime}B^{\prime}}$. We begin by describing a forward classical
communication protocol for a bipartite channel and then define the associated capacities.

\begin{figure*}[ptb]
\begin{center}
\includegraphics[
width=\linewidth
]{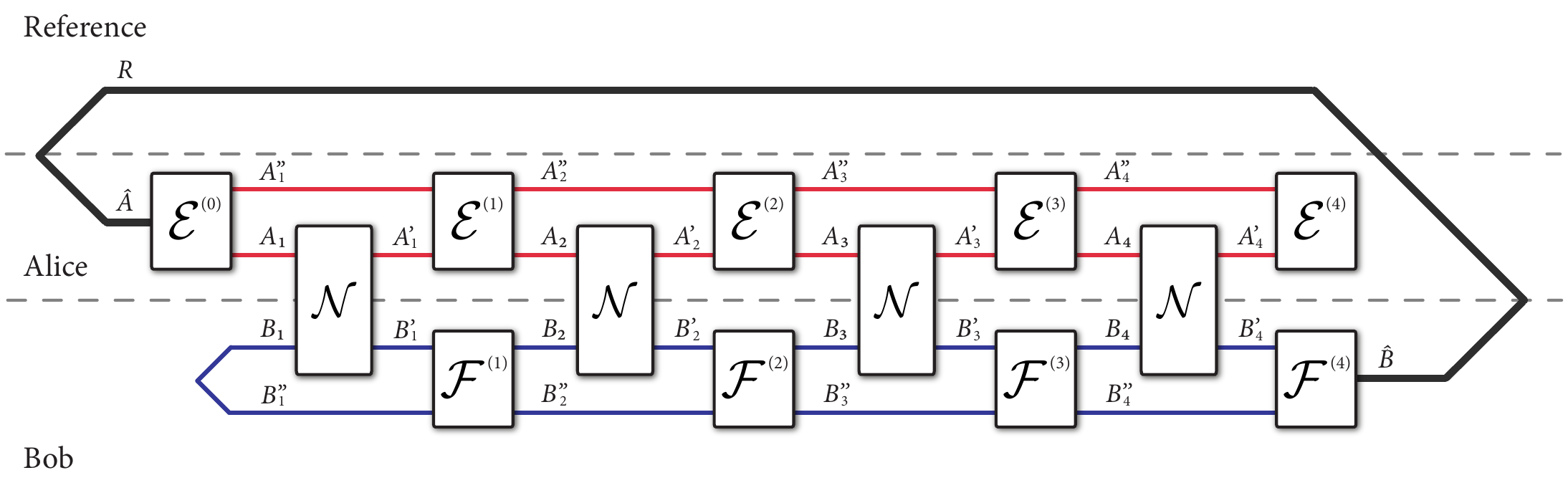}
\end{center}
\caption{Depiction of a protocol for forward classical communication over a
bipartite channel, by making $n=4$ uses of the bipartite channel
$\mathcal{N}_{AB \to A^{\prime}B^{\prime}}$.}%
\label{fig:forward-comm-bipartite-channel}%
\end{figure*}

Fix $n,M\in\mathbb{N}$ and $\varepsilon\in\left[  0,1\right]  $. An
$(n,M,\varepsilon)$ protocol for forward classical communication using a
bipartite channel $\mathcal{N}_{AB\rightarrow A^{\prime}B^{\prime}}$ begins
with a reference party preparing the state $\overline{\Phi}_{R\hat{A}}^{p}$
and sending the $\hat{A}$ system to Alice, where $\overline{\Phi}_{R\hat{A}%
}^{p}$ is the following classically correlated state:%
\begin{equation}
\overline{\Phi}_{R\hat{A}}^{p}\coloneqq\sum_{m=1}^{M}p(m)|m\rangle\!\langle
m|_{R}\otimes|m\rangle\!\langle m|_{\hat{A}}%
,\label{eq:initial-classically-corr-state-bipartite-comm-prot}%
\end{equation}
and $p(m)$ is a probability distribution over the messages. Alice acts on the
system $\hat{A}$ with a local encoding channel $\mathcal{E}_{\hat
{A}\rightarrow A_{1}^{\prime\prime}A_{1}}^{(0)}$, resulting in the following
state:%
\begin{equation}
\sigma_{RA_{1}^{\prime\prime}A_{1}}\coloneqq\mathcal{E}_{\hat{A}\rightarrow
A_{1}^{\prime\prime}A_{1}}^{(0)}(\overline{\Phi}_{R\hat{A}}^{p}).
\end{equation}
Bob prepares the local state $\tau_{B_{1}^{\prime\prime}B_{1}}$, so that the
initial global state of the reference, Alice, and Bob is%
\begin{equation}
\sigma_{RA_{1}^{\prime\prime}A_{1}}\otimes\tau_{B_{1}^{\prime\prime}B_{1}}.
\end{equation}
The systems $A_{1}B_{1}$ are then fed into the first use of the channel,
producing the output state%
\begin{align}
\rho_{RA_{1}^{\prime\prime}A_{1}^{\prime}B_{1}^{\prime\prime}B_{1}^{\prime}%
}^{(1)} &  \coloneqq\mathcal{N}_{A_{1}B_{1}\rightarrow A_{1}^{\prime}%
B_{1}^{\prime}}(\sigma_{RA_{1}^{\prime\prime}A_{1}B_{1}^{\prime\prime}B_{1}%
}^{(1)}),\\
\sigma_{RA_{1}^{\prime\prime}A_{1}B_{1}^{\prime\prime}B_{1}}^{(1)} &
\coloneqq\sigma_{A_{1}^{\prime\prime}A_{1}}\otimes\tau_{B_{1}^{\prime\prime
}B_{1}}.
\end{align}
Then Alice applies the local channel $\mathcal{E}_{A_{1}^{\prime\prime}%
A_{1}^{\prime}\rightarrow A_{2}^{\prime\prime}A_{2}}^{(1)}$ to her systems,
and Bob applies the local channel $\mathcal{F}_{B_{1}^{\prime\prime}%
B_{1}^{\prime}\rightarrow B_{2}^{\prime\prime}B_{2}}^{(1)}$ to his systems.
The systems $A_{2}B_{2}$ are fed into the next channel use, leading to the
state%
\begin{equation}
\rho_{RA_{2}^{\prime\prime}A_{2}^{\prime}B_{2}^{\prime\prime}B_{2}^{\prime}%
}^{(2)}\coloneqq\mathcal{N}_{A_{2}B_{2}\rightarrow A_{2}^{\prime}B_{2}%
^{\prime}}(\sigma_{RA_{2}^{\prime\prime}A_{2}B_{2}^{\prime\prime}B_{2}}%
^{(2)}),
\end{equation}%
\begin{multline}
\sigma_{RA_{2}^{\prime\prime}A_{2}B_{2}^{\prime\prime}B_{2}}^{(2)}\coloneqq\\
(\mathcal{E}_{A_{1}^{\prime\prime}A_{1}^{\prime}\rightarrow A_{2}%
^{\prime\prime}A_{2}}^{(1)}\otimes\mathcal{F}_{B_{1}^{\prime\prime}%
B_{1}^{\prime}\rightarrow B_{2}^{\prime\prime}B_{2}}^{(1)})(\rho
_{RA_{1}^{\prime\prime}A_{1}^{\prime}B_{1}^{\prime\prime}B_{1}^{\prime}}%
^{(1)}).
\end{multline}
This process iterates $n-2$ more times, and we define%
\begin{equation}
\rho_{RA_{i}^{\prime\prime}A_{i}^{\prime}B_{i}^{\prime\prime}B_{i}^{\prime}%
}^{(i)}\coloneqq\mathcal{N}_{A_{i}B_{i}\rightarrow A_{i}^{\prime}B_{i}%
^{\prime}}(\sigma_{RA_{i}^{\prime\prime}A_{i}B_{i}^{\prime\prime}B_{i}}%
^{(i)}),
\end{equation}%
\begin{multline}
\sigma_{RA_{i}^{\prime\prime}A_{i}B_{i}^{\prime\prime}B_{i}}^{(i)}\coloneqq\\
(\mathcal{E}_{A_{i-1}^{\prime\prime}A_{i-1}^{\prime}\rightarrow A_{i}%
^{\prime\prime}A_{i}}^{(i-1)}\otimes\mathcal{F}_{B_{i-1}^{\prime\prime}%
B_{i-1}^{\prime}\rightarrow B_{i}^{\prime\prime}B_{i}}^{(i-1)})(\rho
_{RA_{i-1}^{\prime\prime}A_{i-1}^{\prime}B_{i-1}^{\prime\prime}B_{i-1}%
^{\prime}}^{(i-1)}),
\end{multline}
for $i\in\left\{  3,\ldots,n\right\}  $. The final channel output state
$\rho_{RA_{n}^{\prime\prime}A_{n}^{\prime}B_{n}^{\prime\prime}B_{n}^{\prime}%
}^{(n)}$ is processed a final time with the local channels $\mathcal{E}%
_{A_{n}^{\prime\prime}A_{n}^{\prime}\rightarrow\emptyset}^{(n)}%
\coloneqq\operatorname{Tr}_{A_{n}^{\prime\prime}A_{n}^{\prime}}$ and
$\mathcal{F}_{B_{n}^{\prime\prime}B_{n}^{\prime}\rightarrow\hat{B}}^{(n)}$,
where $\emptyset$ indicates a trivial system, to produce the final protocol
state%
\begin{equation}
\omega_{R\hat{B}}^{p}\coloneqq(\mathcal{E}_{A_{n}^{\prime\prime}A_{n}^{\prime
}\rightarrow\emptyset}^{(n)}\otimes\mathcal{F}_{B_{n}^{\prime\prime}%
B_{n}^{\prime}\rightarrow\hat{B}}^{(n)})(\rho_{RA_{n}^{\prime\prime}%
A_{n}^{\prime}B_{n}^{\prime\prime}B_{n}^{\prime}}^{(n)}).
\end{equation}
For an $(n,M,\varepsilon)$ protocol, the final state $\omega_{R\hat{B}}^{p}$
satisfies the following condition%
\begin{equation}
\max_{p}\frac{1}{2}\left\Vert \omega_{R\hat{B}}^{p}-\overline{\Phi}_{R\hat{B}%
}^{p}\right\Vert _{1}\leq\varepsilon,\label{eq:RD-approx-crit}%
\end{equation}
where the maximization is over all probability distributions $p(m)$ and%
\begin{equation}
\overline{\Phi}_{R\hat{B}}^{p}\coloneqq\sum_{m=1}^{M}p(m)|m\rangle\!\langle
m|_{\hat{A}}\otimes|m\rangle\!\langle m|_{\hat{B}}%
.\label{eq:max-class-corr-state-proof}%
\end{equation}
Note that the condition in \eqref{eq:RD-approx-crit} is equivalent to the traditional condition on the decoding error probability (see \cite[Lemma~6.2]{KW20book}).
Figure~\ref{fig:forward-comm-bipartite-channel} depicts such a protocol with
$n=4$.

Let us denote the set consisting of the initially prepared state and the
sequence of local channels as the protocol $\mathcal{P}^{(n)}$:%
\begin{multline}
\mathcal{P}^{(n)}\coloneqq\Bigg\{\mathcal{E}_{\hat{A}\rightarrow A_{1}%
^{\prime\prime}A_{1}}^{(0)}\otimes\tau_{B_{1}^{\prime\prime}B_{1}},\\
\left\{  \mathcal{E}_{A_{i-1}^{\prime\prime}A_{i-1}^{\prime}\rightarrow
A_{i}^{\prime\prime}A_{i}}^{(i-1)}\otimes\mathcal{F}_{B_{i-1}^{\prime\prime
}B_{i-1}^{\prime}\rightarrow B_{i}^{\prime\prime}B_{i}}^{(i-1)}\right\}
_{i=2}^{n},\\
\mathcal{E}_{A_{n}^{\prime\prime}A_{n}^{\prime}\rightarrow\emptyset}%
^{(n)}\otimes\mathcal{F}_{B_{n}^{\prime\prime}B_{n}^{\prime}\rightarrow\hat
{B}}^{(n)}\Bigg\}.
\end{multline}
Then we can write the final state $\omega_{R\hat{B}}$ as%
\begin{equation}
\omega_{R\hat{B}}^{p}=\mathcal{C}_{\hat{A}\rightarrow\hat{B}}(\overline{\Phi
}_{R\hat{A}}^{p}),\label{eq:RD-final-state-simpl}%
\end{equation}
where%
\begin{equation}
\mathcal{C}_{\hat{A}\rightarrow\hat{B}}\coloneqq\mathcal{L}^{(n)}%
\circ\mathcal{N}\circ\mathcal{L}^{(n-1)}\cdots\circ\mathcal{L}^{(2)}%
\circ\mathcal{N}\circ\mathcal{L}^{(1)}\circ\mathcal{N}\circ\mathcal{L}^{(0)},
\end{equation}
$\mathcal{L}^{(0)}$ acts on system $\hat{A}$\ of $\overline{\Phi}_{R\hat{A}}$
to prepare the state $\sigma_{RA_{1}^{\prime\prime}A_{1}B_{1}^{\prime\prime
}B_{1}}^{(1)}$,%
\begin{equation}
\mathcal{L}^{(i-1)}\equiv\mathcal{E}_{A_{i-1}^{\prime\prime}A_{i-1}^{\prime
}\rightarrow A_{i}^{\prime\prime}A_{i}}^{(i-1)}\otimes\mathcal{F}%
_{B_{i-1}^{\prime\prime}B_{i-1}^{\prime}\rightarrow B_{i}^{\prime\prime}B_{i}%
}^{(i-1)},
\end{equation}
for $i\in\left\{  2,\ldots,n\right\}  $, and%
\begin{equation}
\mathcal{L}^{(n)}\equiv\mathcal{E}_{A_{n}^{\prime\prime}A_{n}^{\prime
}\rightarrow\emptyset}^{(n)}\otimes\mathcal{F}_{B_{n}^{\prime\prime}%
B_{n}^{\prime}\rightarrow\hat{B}}^{(n)}.
\end{equation}

The $n$-shot forward classical capacity of a bipartite channel $\mathcal{N}%
_{AB\rightarrow A^{\prime}B^{\prime}}$\ is then defined as follows:%
\begin{multline}
C^{n,\varepsilon}(\mathcal{N}_{AB\rightarrow A^{\prime}B^{\prime}})\coloneqq\\
\sup_{M\in\mathbb{N},\ \mathcal{P}^{(n)}}\left\{  \frac{1}{n}\log_{2}%
M:\exists(n,M,\varepsilon)~\text{protocol }\mathcal{P}^{(n)}\right\}  .
\end{multline}
The forward classical capacity and strong converse forward classical capacity
of the bipartite channel $\mathcal{N}_{AB\rightarrow A^{\prime}B^{\prime}}$
are defined as%
\begin{align}
C(\mathcal{N}_{AB\rightarrow A^{\prime}B^{\prime}}) &  \coloneqq\inf
_{\varepsilon\in(0,1)}\liminf_{n\rightarrow\infty}C^{n,\varepsilon
}(\mathcal{N}_{AB\rightarrow A^{\prime}B^{\prime}}),\\
\widetilde{C}(\mathcal{N}_{AB\rightarrow A^{\prime}B^{\prime}}) &
\coloneqq\sup_{\varepsilon\in(0,1)}\limsup_{n\rightarrow\infty}%
C^{n,\varepsilon}(\mathcal{N}_{AB\rightarrow A^{\prime}B^{\prime}}).
\end{align}
From the definitions, it is clear that%
\begin{equation}
C(\mathcal{N}_{AB\rightarrow A^{\prime}B^{\prime}})\leq\widetilde
{C}(\mathcal{N}_{AB\rightarrow A^{\prime}B^{\prime}}).
\end{equation}

An $(n,M,\varepsilon)$ randomness transmission protocol is exactly as
specified above, but with $p(m)=1/M$ (i.e., the uniform distribution) in
\eqref{eq:initial-classically-corr-state-bipartite-comm-prot}. Let us define%
\begin{equation}
\overline{\Phi}_{R\hat{A}}\coloneqq\frac{1}{M}\sum_{m=1}^{M}|m\rangle\!\langle
m|_{R}\otimes|m\rangle\!\langle m|_{\hat{A}}.
\end{equation}
Then the error criterion for such a protocol is%
\begin{equation}
\frac{1}{2}\left\Vert \omega_{R\hat{B}}-\overline{\Phi}_{R\hat{B}}\right\Vert
_{1}\leq\varepsilon,
\label{eq:avg-err-crit}
\end{equation}
where $\omega_{R\hat{B}}$ is defined as in \eqref{eq:RD-final-state-simpl} but
with $\overline{\Phi}_{R\hat{A}}^{p}$ replaced by $\overline{\Phi}_{R\hat{A}}%
$. Also,
\begin{equation}
\overline{\Phi}_{R\hat{B}}\coloneqq\frac{1}{M}\sum_{m=1}^{M}|m\rangle\!\langle
m|_{\hat{A}}\otimes|m\rangle\!\langle m|_{\hat{B}}.
\end{equation}
Note that the condition in \eqref{eq:avg-err-crit} is equivalent to the traditional condition on the average decoding error probability (see \cite[Lemma~6.2]{KW20book}).

We define the following quantities for the randomness transmission capacity of
$\mathcal{N}_{AB\rightarrow A^{\prime}B^{\prime}}$:%
\begin{multline}
R^{n,\varepsilon}(\mathcal{N}_{AB\rightarrow A^{\prime}B^{\prime}})=\\
\sup_{M\in\mathbb{N},\ \mathcal{P}^{(n)}}\left\{  \frac{1}{n}\log_{2}%
M:\exists(n,M,\varepsilon)~\text{RT protocol }\mathcal{P}^{(n)}\right\}  ,
\end{multline}
where RT is an abbreviation for \textquotedblleft randomness
transmission.\textquotedblright\ The randomness transmission capacity and
strong converse randomness transmission capacity of the bipartite channel
$\mathcal{N}_{AB\rightarrow A^{\prime}B^{\prime}}$ are defined as%
\begin{align}
R(\mathcal{N}_{AB\rightarrow A^{\prime}B^{\prime}})  &  \coloneqq\inf
_{\varepsilon\in(0,1)}\liminf_{n\rightarrow\infty}R^{n,\varepsilon
}(\mathcal{N}_{AB\rightarrow A^{\prime}B^{\prime}}),\\
\widetilde{R}(\mathcal{N}_{AB\rightarrow A^{\prime}B^{\prime}})  &
\coloneqq\sup_{\varepsilon\in(0,1)}\limsup_{n\rightarrow\infty}%
R^{n,\varepsilon}(\mathcal{N}_{AB\rightarrow A^{\prime}B^{\prime}}).
\end{align}
From the definitions, it is clear that%
\begin{equation}
R(\mathcal{N}_{AB\rightarrow A^{\prime}B^{\prime}})\leq\widetilde
{R}(\mathcal{N}_{AB\rightarrow A^{\prime}B^{\prime}}).
\end{equation}

Since every $(n,M,\varepsilon)$ forward classical communication protocol is an
$(n,M,\varepsilon)$ randomness transmission protocol, the following inequality
holds%
\begin{equation}
C^{n,\varepsilon}(\mathcal{N}_{AB\rightarrow A^{\prime}B^{\prime}})\leq
R^{n,\varepsilon}(\mathcal{N}_{AB\rightarrow A^{\prime}B^{\prime}}).
\end{equation}
By the standard expurgation argument (throwing away the worst half of the
codewords to give maximal error probability $\leq2\varepsilon$; see, e.g., \cite[Exercise~2.2.1]{Wbook17}), the following
inequality holds%
\begin{equation}
R^{n,\varepsilon}(\mathcal{N}_{AB\rightarrow A^{\prime}B^{\prime}})-\frac
{1}{n}\leq C^{n,2\varepsilon}(\mathcal{N}_{AB\rightarrow A^{\prime}B^{\prime}%
}).
\end{equation}
By employing definitions, we conclude that%
\begin{align}
C(\mathcal{N}_{AB\rightarrow A^{\prime}B^{\prime}})  &  =R(\mathcal{N}%
_{AB\rightarrow A^{\prime}B^{\prime}})\\
&  \leq\widetilde{C}(\mathcal{N}_{AB\rightarrow A^{\prime}B^{\prime}})\\
&  \leq\widetilde{R}(\mathcal{N}_{AB\rightarrow A^{\prime}B^{\prime}}).
\end{align}

In what follows, we establish an upper bound on the strong converse randomness
transmission capacity of $\mathcal{N}_{AB\rightarrow A^{\prime}B^{\prime}}$,
and by the inequalities above, this gives an upper bound on the forward
classical capacity and strong converse forward classical capacity of
$\mathcal{N}_{AB\rightarrow A^{\prime}B^{\prime}}$.

\begin{theorem}
\label{thm:n-shot-upper-bnd-bipartite}The following upper bound holds for the
$n$-shot randomness transmission capacity of a bipartite channel
$\mathcal{N}_{AB\rightarrow A^{\prime}B^{\prime}}$:%
\begin{multline}
R^{n,\varepsilon}(\mathcal{N}_{AB\rightarrow A^{\prime}B^{\prime}})\leq\\
\widehat{\Upsilon}_{\alpha}(\mathcal{N}_{AB\rightarrow A^{\prime}B^{\prime}%
})+\frac{\alpha}{n\left(  \alpha-1\right)  }\log_{2}\!\left(  \frac
{1}{1-\varepsilon}\right)  , \label{eq:n-shot-bound}%
\end{multline}
for all $\alpha\in(1,2]$ and $\varepsilon\in\lbrack0,1)$.
\end{theorem}

\begin{IEEEproof}
Consider an arbitrary $n$-shot randomness transmission protocol of the form
described above. Focusing in particular on \eqref{eq:RD-approx-crit} and
\eqref{eq:RD-final-state-simpl}, we apply \eqref{eq:err-bound}\ of
Proposition~\ref{prop:err-bound}\ to conclude that%
\begin{align}
&  \log_{2}M\nonumber\\
&  \leq\inf_{\mathcal{M}_{\hat{A}\rightarrow\hat{B}}:\beta(\mathcal{M}_{\hat{A}\rightarrow\hat{B}})\leq1}\widehat{D}_{\alpha
}(\mathcal{C}_{\hat{A}\rightarrow\hat{B}}(\overline{\Phi}_{R\hat{A}}%
)\Vert\mathcal{M}_{\hat{A}\rightarrow\hat{B}}(\overline{\Phi}_{R\hat{A}%
}))\nonumber\\
&  \qquad+\frac{\alpha}{\alpha-1}\log_{2}\!\left(  \frac{1}{1-\varepsilon
}\right) \\
&  \leq\widehat{\Upsilon}_{\alpha}(\mathcal{C}_{\hat{A}\rightarrow\hat{B}%
})+\frac{\alpha}{\alpha-1}\log_{2}\!\left(  \frac{1}{1-\varepsilon}\right)  ,
\end{align}
where the inequality follows from the definition in \eqref{eq:p2p-gen-div-ups}
with $\boldsymbol{D}$ set to $\widehat{D}_{\alpha}$.
Eq.~\eqref{eq:RD-final-state-simpl} indicates that the whole protocol is a
serial composition of bipartite channels. Then we find that%
\begin{align}
&  \widehat{\Upsilon}_{\alpha}(\mathcal{C}_{\hat{A}\rightarrow\hat{B}%
})\nonumber\\
&  =\widehat{\Upsilon}_{\alpha}(\mathcal{L}^{(n)}\circ\mathcal{N}%
\circ\mathcal{L}^{(n-1)}\cdots\circ\mathcal{L}^{(2)}\circ\mathcal{N}%
\circ\mathcal{L}^{(1)}\circ\mathcal{N}\circ\mathcal{L}^{(0)})\\
&  \leq n\widehat{\Upsilon}_{\alpha}(\mathcal{N})+\sum_{i=0}^{n}%
\widehat{\Upsilon}_{\alpha}(\mathcal{L}^{(i)})\\
&  =n\widehat{\Upsilon}_{\alpha}(\mathcal{N}).
\end{align}
The inequality follows from
Proposition~\ref{prop:subadditivity-bipartite-CP-maps}\ and the last equality
from Proposition~\ref{prop:zero-local-chs-gen-div-ups}. We also implicitly
used the stability property in Proposition~\ref{prop:stability-gen-div-ups}.
Then we find that%
\begin{equation}
\frac{1}{n}\log_{2}M\leq\widehat{\Upsilon}_{\alpha}(\mathcal{N})+\frac{\alpha
}{n(\alpha-1)}\log_{2}\!\left(  \frac{1}{1-\varepsilon}\right)  .
\end{equation}
Since the upper bound holds for an arbitrary protocol, this concludes the proof.
\end{IEEEproof}

\begin{theorem}
\label{thm:SC-RD-bipartite-ch}The following upper bound holds for the strong
converse randomness transmission capacity of a bipartite channel
$\mathcal{N}_{AB\rightarrow A^{\prime}B^{\prime}}$:%
\begin{equation}
\widetilde{R}(\mathcal{N}_{AB\rightarrow A^{\prime}B^{\prime}})\leq
\widehat{\Upsilon}(\mathcal{N}_{AB\rightarrow A^{\prime}B^{\prime}}),
\end{equation}
where $\widehat{\Upsilon}(\mathcal{N}_{AB\rightarrow A^{\prime}B^{\prime}})$
is defined from \eqref{eq:gen-div-ch-ups-meas-bi-map}\ using the
Belavkin--Staszewski relative entropy.
\end{theorem}

\begin{IEEEproof}
Applying the bound in \eqref{eq:n-shot-bound} and taking the $n\rightarrow
\infty$ limit, we find that the following holds for all $\alpha\in(1,2]$ and
$\varepsilon\in\lbrack0,1)$:%
\begin{align}
&  \limsup_{n\rightarrow\infty}R^{n,\varepsilon}(\mathcal{N}_{AB\rightarrow
A^{\prime}B^{\prime}})\nonumber\\
&  \leq\limsup_{n\rightarrow\infty}\left[  \widehat{\Upsilon}_{\alpha
}(\mathcal{N}_{AB\rightarrow A^{\prime}B^{\prime}})+\frac{\alpha}{n\left(
\alpha-1\right)  }\log_{2}\!\left(  \frac{1}{1-\varepsilon}\right)  \right] \\
&  =\widehat{\Upsilon}_{\alpha}(\mathcal{N}_{AB\rightarrow A^{\prime}%
B^{\prime}}).
\end{align}
Since the upper bound holds for all $\alpha\in(1,2]$, we can take the infimum
over all these values, and we conclude that the following holds for all
$\varepsilon\in\lbrack0,1)$:%
\begin{equation}
\limsup_{n\rightarrow\infty}R^{n,\varepsilon}(\mathcal{N}_{AB\rightarrow
A^{\prime}B^{\prime}})\leq\widehat{\Upsilon}(\mathcal{N}_{AB\rightarrow
A^{\prime}B^{\prime}}).
\end{equation}
Here we applied the definitions of $\widehat{\Upsilon}(\mathcal{N}%
_{AB\rightarrow A^{\prime}B^{\prime}})$ and $\widehat{\Upsilon}_{\alpha
}(\mathcal{N}_{AB\rightarrow A^{\prime}B^{\prime}})$
and Proposition~\ref{prop:lim-optimized-ch-div} in Appendix~\ref{app:renyi-ch-div-limits}.
The upper bound holds for all $\varepsilon\in(0,1)$ and so we
conclude the statement of the theorem.
\end{IEEEproof}

\subsection{Bounding the classical capacity of a point-to-point quantum
channel assisted by a classical feedback channel}

One of the main applications in our paper is an upper bound on the classical
capacity of a point-to-point quantum channel assisted by classical feedback.
For a point-to-point channel $\mathcal{N}_{A\rightarrow B^{\prime}}$, this
capacity is denoted by $C_{\leftarrow}(\mathcal{N}_{A\rightarrow B^{\prime}})$.

In what follows, we briefly define the classical capacity of a point-to-point
quantum channel $\mathcal{N}_{A\rightarrow B^{\prime}}$ assisted by classical
feedback. Before doing so, let us first expand the notion of an $n$-shot
protocol for forward classical communication from the previous section, such
that each use of the bipartite channel is no longer constrained to be
identical. The final state of such a protocol is then a generalization of that
in \eqref{eq:RD-final-state-simpl}:%
\begin{multline}
\omega_{R\hat{B}}^{p}= (\mathcal{L}^{(n)}\circ\mathcal{N}^{(n)}\circ
\mathcal{L}^{(n-1)}\circ\cdots\\
\circ\mathcal{L}^{(2)}\circ\mathcal{N}^{(2)}\circ\mathcal{L}^{(1)}%
\circ\mathcal{N}^{(1)}\circ\mathcal{L}^{(0)})(\overline{\Phi}_{R\hat{A}}^{p}),
\label{eq:most-gen-protocol}%
\end{multline}
and the protocol is an $(n,M,\varepsilon)$ protocol if the inequality%
\begin{equation}
\max_{p}\frac{1}{2}\left\Vert \omega_{R\hat{B}}^{p}-\overline{\Phi}_{R\hat{B}%
}^{p}\right\Vert _{1}\leq\varepsilon
\end{equation}
holds with $\overline{\Phi}_{R\hat{B}}^{p}$ the classically correlated state
as defined in \eqref{eq:max-class-corr-state-proof}. Note that the following
bound holds for all $(n,M,\varepsilon)$ forward classical communication
protocols, with $n,M\in\mathbb{N}$ and $\varepsilon\in(0,1]$, and $\alpha
\in(1,2]$:%
\begin{equation}
\log_{2}M\leq\sum_{i=1}^{n}\widehat{\Upsilon}_{\alpha}(\mathcal{N}%
_{AB\rightarrow A^{\prime}B^{\prime}}^{(i)})+\frac{\alpha}{\alpha-1}\log
_{2}\!\left(  \frac{1}{1-\varepsilon}\right)  ,
\label{eq:bound-generic-BIID-protocol}%
\end{equation}
by following the same steps given in the proof of
Theorem~\ref{thm:n-shot-upper-bnd-bipartite}.

\begin{figure*}[ptb]
\begin{center}
\includegraphics[
width=\linewidth
]{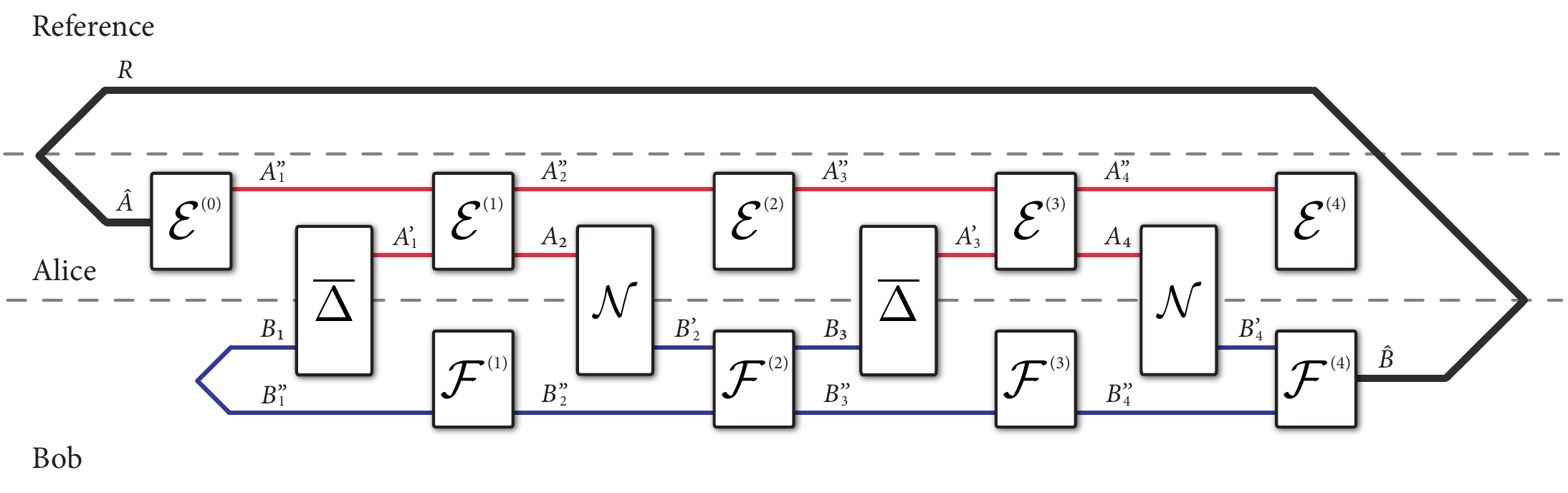}
\end{center}
\caption{Depiction of a protocol for classical communication over a
point-to-point channel with classical feedback, by making $n=2$ uses of the
point-to-point channel $\mathcal{N}_{A \to B^{\prime}}$. This protocol is
related to the one in Figure~\ref{fig:forward-comm-bipartite-channel}, by
replacing every odd use of $\mathcal{N}_{AB \to A^{\prime}B^{\prime}}$ with
the classical feedback channel $\overline{\Delta}_{B\to A^{\prime}}$ and every
even use of $\mathcal{N}_{AB \to A^{\prime}B^{\prime}}$ with the
point-to-point channel $\mathcal{N}_{A \to B^{\prime}}$.}%
\label{fig:feedback-comm-p2p-channel}%
\end{figure*}

With the more general definition in hand, we define an $(n,M,\varepsilon)$
protocol for classical communication over a point-to-point channel
$\mathcal{N}_{A\rightarrow B^{\prime}}$ assisted by a classical feedback
channel as a special case of a $(2n,M,\varepsilon)$ protocol of the form
above, in which every $\mathcal{N}^{(i)}$ with $i$ odd is replaced by a
classical feedback channel $\overline{\Delta}_{B_{i}\rightarrow A_{i}^{\prime
}}$ (with $d_{i}\coloneqq d_{B_{i}}=d_{A_{i}^{\prime}}$ and trivial input
system $A_{i}$ and trivial output system $B_{i}^{\prime}$) and every
$\mathcal{N}^{(i)}$ with $i$ even is replaced by the forward point-to-point
channel $\mathcal{N}_{A\rightarrow B^{\prime}}$ (such that the input system
$B_{i}$ and the output system $A_{i}^{\prime}$ are trivial). The final state
of the protocol is given by%
\begin{multline}
\omega_{R\hat{B}}\coloneqq\\
(\mathcal{L}^{(2n)}\circ\mathcal{N}_{A_{2n}\rightarrow B_{2n}^{\prime}}%
\circ\mathcal{L}^{(2n-1)}\circ\overline{\Delta}_{B_{2n-1}\rightarrow
A_{2n-1}^{\prime}}\circ\mathcal{L}^{(2n-2)}\circ
\label{eq:feedback-assisted-protocol}\\
\cdots\circ\mathcal{L}^{(2)}\circ\mathcal{N}_{A_{2}\rightarrow B_{2}^{\prime}%
}\circ\mathcal{L}^{(1)}\circ\overline{\Delta}_{B_{1}\rightarrow A_{1}^{\prime
}}\circ\mathcal{L}^{(1)})(\overline{\Phi}_{R\hat{A}}).
\end{multline}
Let $\mathcal{P}^{(2n)}$ denote the protocol, which consists of $\mathcal{L}%
^{(0)}$, $\mathcal{L}^{(1)}$, \ldots, $\mathcal{L}^{(2n)}$. This protocol is
depicted in Figure~\ref{fig:feedback-comm-p2p-channel}.

The $n$-shot classical capacity of the point-to-point channel $\mathcal{N}%
_{A\rightarrow B^{\prime}}$ assisted by classical feedback is defined as%
\begin{multline}
C_{\leftarrow}^{n,\varepsilon}(\mathcal{N}_{A\rightarrow B^{\prime}})=\\
\sup_{M\in\mathbb{N},\ \mathcal{P}^{(2n)}}\left\{  \frac{1}{n}\log
_{2}M:\exists(n,M,\varepsilon)~\text{protocol }\mathcal{P}^{(2n)}\right\}  .
\label{eq:def-n-shot-RD-prot-assisted-FB}%
\end{multline}
That is, it is the largest rate at which messages can be transmitted up to
an $\varepsilon$ error probability. The classical capacity of the
point-to-point channel $\mathcal{N}_{A\rightarrow B^{\prime}}$ assisted by
classical feedback is defined as the following limit:%
\begin{equation}
C_{\leftarrow}(\mathcal{N}_{A\rightarrow B^{\prime}})\coloneqq\inf
_{\varepsilon\in(0,1)}\liminf_{n\rightarrow\infty}C_{\leftarrow}%
^{n,\varepsilon}(\mathcal{N}_{A\rightarrow B^{\prime}}),
\end{equation}
and the strong converse classical capacity as%
\begin{equation}
\widetilde{C}_{\leftarrow}(\mathcal{N}_{A\rightarrow B^{\prime}}%
)\coloneqq\sup_{\varepsilon\in(0,1)}\limsup_{n\rightarrow\infty}C_{\leftarrow
}^{n,\varepsilon}(\mathcal{N}_{A\rightarrow B^{\prime}}).
\end{equation}
The following inequality is an immediate consequence of definitions:%
\begin{equation}
C_{\leftarrow}(\mathcal{N}_{A\rightarrow B^{\prime}})\leq\widetilde
{C}_{\leftarrow}(\mathcal{N}_{A\rightarrow B^{\prime}}).
\label{eq:weak-less-then-strong-cap}%
\end{equation}

\begin{theorem}
\label{thm:geo-renyi-bound-feedback-assisted-n-shot-cap}Fix $n\in\mathbb{N}$
and $\varepsilon\in\lbrack0,1)$. The $n$-shot classical capacity
$C_{\leftarrow}^{n,\varepsilon}(\mathcal{N}_{A\rightarrow B^{\prime}})$ of the
point-to-point channel $\mathcal{N}_{A\rightarrow B^{\prime}}$ assisted by
classical feedback is bounded from above as follows:%
\begin{equation}
C_{\leftarrow}^{n,\varepsilon}(\mathcal{N}_{A\rightarrow B^{\prime}}%
)\leq\widehat{\Upsilon}_{\alpha}(\mathcal{N}_{A\rightarrow B^{\prime}}%
)+\frac{\alpha}{n\left(  \alpha-1\right)  }\log_{2}\!\left(  \frac
{1}{1-\varepsilon}\right)  ,
\end{equation}
for all $\alpha\in(1,2]$.
\end{theorem}

\begin{IEEEproof}
Applying the bound in \eqref{eq:bound-generic-BIID-protocol} with the choices
in \eqref{eq:feedback-assisted-protocol}, we conclude that the following bound
holds for all $\alpha\in(1,2]$ and for an arbitrary $(n,M,\varepsilon)$
classical communication protocol assisted by a classical feedback channel:%
\begin{align}
&  \log_{2}M\nonumber\\
&  \leq n\widehat{\Upsilon}_{\alpha}(\mathcal{N}_{A\rightarrow B^{\prime}%
})+\sum_{i=1}^{n}\widehat{\Upsilon}_{\alpha}(\overline{\Delta}_{B_{2i-1}%
\rightarrow A_{2i-1}^{\prime}})\nonumber\\
&  \qquad+\frac{\alpha}{\alpha-1}\log_{2}\!\left(  \frac{1}{1-\varepsilon
}\right) \\
&  =n\widehat{\Upsilon}_{\alpha}(\mathcal{N}_{A\rightarrow B^{\prime}}%
)+\frac{\alpha}{\alpha-1}\log_{2}\!\left(  \frac{1}{1-\varepsilon}\right)  .
\end{align}
The equality follows from Proposition~\ref{prop:zero-feedback-renyi-ups}. This
then implies the following bound%
\begin{equation}
\frac{1}{n}\log_{2}M\leq\widehat{\Upsilon}_{\alpha}(\mathcal{N}_{A\rightarrow
B^{\prime}})+\frac{\alpha}{n(\alpha-1)}\log_{2}\!\left(  \frac{1}%
{1-\varepsilon}\right)  . \label{eq:SC-RD-FB-up-bnd-proof}%
\end{equation}
Since the bound in \eqref{eq:SC-RD-FB-up-bnd-proof}\ holds for an arbitrary
protocol, we conclude the statement of the theorem after applying the
definition in \eqref{eq:def-n-shot-RD-prot-assisted-FB}.
\end{IEEEproof}

\begin{theorem}
\label{thm:SC-DR-p2p-up-bound}The strong converse classical capacity of a
point-to-point quantum channel $\mathcal{N}_{A\rightarrow B^{\prime}}$
assisted by a classical feedback channel is bounded from above as follows:%
\begin{equation}
\widetilde{C}_{\leftarrow}(\mathcal{N}_{A\rightarrow B^{\prime}})\leq
\widehat{\Upsilon}(\mathcal{N}_{A\rightarrow B^{\prime}}).
\end{equation}

\end{theorem}

\begin{IEEEproof}
The reasoning here is the same as that given in the proof of
Theorem~\ref{thm:SC-RD-bipartite-ch}.
\end{IEEEproof}

\medskip

Before proceeding to the next section, let us finally note that our bounds above apply in a more general setting in which the classical feedback channel is replaced by an entanglement-breaking channel \cite{HSR03}. This follows because every entanglement-breaking channel can be written as a composition of a general pre-processing quantum channel, followed by a classical channel, which is in turn followed by a general post-processing quantum channel \cite{HSR03}. It is then clear that the pre-processing channel can be absorbed into a local operation of the receiver Bob, while the post-processing channel can be absorbed into a local operation of Alice, so that this is essentially just assistance by a classical feedback channel again, and our result thus applies.

\section{Employing the sandwiched R\'{e}nyi relative entropy}

\label{sec:employ-sandwichedR}

In this section, we explore what kinds of bounds
we can obtain on the previously defined capacities, by making use of the
sandwiched R\'{e}nyi relative entropy.

Let us define the amortized sandwiched R\'{e}nyi divergence of the completely
positive maps $\mathcal{N}_{A\rightarrow B}$ and $\mathcal{M}_{A\rightarrow
B}$ as follows \cite{WBHK20}:%
\begin{multline}
\widetilde{D}_{\alpha}^{\mathcal{A}}(\mathcal{N}_{A\rightarrow B}%
\Vert\mathcal{M}_{A\rightarrow B})\coloneqq\\
\sup_{\rho_{RA},\sigma_{RA}}\widetilde{D}_{\alpha}(\mathcal{N}_{A\rightarrow
B}(\rho_{RA})\Vert\mathcal{M}_{A\rightarrow B}(\sigma_{RA}))-\widetilde
{D}_{\alpha}(\rho_{RA}\Vert\sigma_{RA}),
\end{multline}
for $\alpha\in(0,1)\cup(1,\infty)$, where the optimization is over all density
operators $\rho_{RA}$ and $\sigma_{RA}$. By exploiting the definition of the
sandwiched R\'{e}nyi relative entropy, it follows that the quantity above does
not change if we optimize more generally over positive semi-definite operators
$\rho_{RA}$ and $\sigma_{RA}$ with strictly positive trace.

The amortized sandwiched R\'{e}nyi divergence is subadditive in the following sense:

\begin{proposition}
[Subadditivity]\label{prop:subadd-reg-sandwiched}Let $\mathcal{N}%
_{A\rightarrow B}^{1}$, $\mathcal{N}_{B\rightarrow C}^{2}$, $\mathcal{M}%
_{A\rightarrow B}^{1}$, and $\mathcal{M}_{B\rightarrow C}^{2}$ be completely
positive maps. Then%
\begin{multline}
\widetilde{D}_{\alpha}^{\mathcal{A}}(\mathcal{N}^{2}\circ\mathcal{N}^{1}%
\Vert\mathcal{M}^{2}\circ\mathcal{M}^{1})\leq\\
\widetilde{D}_{\alpha}^{\mathcal{A}}(\mathcal{N}^{1}\Vert\mathcal{M}%
^{1})+\widetilde{D}_{\alpha}^{\mathcal{A}}(\mathcal{N}^{2}\Vert\mathcal{M}%
^{2}),
\end{multline}
for all $\alpha\in(0,1)\cup(1,\infty)$.
\end{proposition}

\begin{IEEEproof}
Let $\rho_{RA}$ and $\sigma_{RA}$ be arbitrary positive semi-definite
operators. Then%
\begin{align}
&  \widetilde{D}_{\alpha}(\mathcal{N}_{B\rightarrow C}^{2}(\mathcal{N}%
_{A\rightarrow B}^{1}(\rho_{RA}))\Vert\mathcal{M}_{B\rightarrow C}%
^{2}(\mathcal{M}_{A\rightarrow B}^{1}(\sigma_{RA})))\nonumber\\
&  \qquad-\widetilde{D}_{\alpha}(\rho_{RA}\Vert\sigma_{RA})\\
&  =\widetilde{D}_{\alpha}(\mathcal{N}_{B\rightarrow C}^{2}(\mathcal{N}%
_{A\rightarrow B}^{1}(\rho_{RA}))\Vert\mathcal{M}_{B\rightarrow C}%
^{2}(\mathcal{M}_{A\rightarrow B}^{1}(\sigma_{RA})))\nonumber\\
&  \qquad-\widetilde{D}_{\alpha}(\mathcal{N}_{A\rightarrow B}^{1}(\rho
_{RA})\Vert\mathcal{M}_{A\rightarrow B}^{1}(\sigma_{RA}))\nonumber\\
&  \qquad+\widetilde{D}_{\alpha}(\mathcal{N}_{A\rightarrow B}^{1}(\rho
_{RA})\Vert\mathcal{M}_{A\rightarrow B}^{1}(\sigma_{RA}))-\widetilde
{D}_{\alpha}(\rho_{RA}\Vert\sigma_{RA})\\
&  \leq\widetilde{D}_{\alpha}^{\mathcal{A}}(\mathcal{N}^{1}\Vert
\mathcal{M}^{1})+\widetilde{D}_{\alpha}^{\mathcal{A}}(\mathcal{N}^{2}%
\Vert\mathcal{M}^{2}).
\end{align}
The desired inequality follows because $\rho_{RA}$ and $\sigma_{RA}$ are arbitrary.
\end{IEEEproof}

\bigskip

The regularized sandwiched R\'{e}nyi divergence of the completely positive
maps $\mathcal{N}_{A\rightarrow B}$ and $\mathcal{M}_{A\rightarrow B}$ is
defined for $\alpha\in(0,1)\cup(1,\infty)$ as follows:%
\begin{equation}
\widetilde{D}_{\alpha}^{\text{reg}}(\mathcal{N}_{A\rightarrow B}%
\Vert\mathcal{M}_{A\rightarrow B})\coloneqq \lim_{n\rightarrow\infty}\frac
{1}{n}\widetilde{D}_{\alpha}(\mathcal{N}_{A\rightarrow B}^{\otimes n}%
\Vert\mathcal{M}_{A\rightarrow B}^{\otimes n}),
\end{equation}
and the limit exists, as argued in \cite[Theorem~5.4]{FawziFawzi20}.

The following equality holds for all $\alpha>1$ \cite[Theorem~5.4]%
{FawziFawzi20}:%
\begin{equation}
\widetilde{D}_{\alpha}^{\mathcal{A}}(\mathcal{N}_{A\rightarrow B}%
\Vert\mathcal{M}_{A\rightarrow B})=\widetilde{D}_{\alpha}^{\text{reg}%
}(\mathcal{N}_{A\rightarrow B}\Vert\mathcal{M}_{A\rightarrow B}).
\label{eq:sandwiched-reg=amortized}%
\end{equation}
As such, by applying \eqref{eq:sandwiched-reg=amortized} and
Proposition~\ref{prop:subadd-reg-sandwiched}, it follows that%
\begin{equation}
\widetilde{D}_{\alpha}^{\text{reg}}(\mathcal{N}^{2}\circ\mathcal{N}^{1}%
\Vert\mathcal{M}^{2}\circ\mathcal{M}^{1})\leq\widetilde{D}_{\alpha
}^{\text{reg}}(\mathcal{N}^{1}\Vert\mathcal{M}^{1})+\widetilde{D}_{\alpha
}^{\text{reg}}(\mathcal{N}^{2}\Vert\mathcal{M}^{2})
\end{equation}
for all $\alpha>1$.

We can then replace the use of the geometric R\'{e}nyi relative entropy in
Theorems~\ref{thm:n-shot-upper-bnd-bipartite}\ and
\ref{thm:geo-renyi-bound-feedback-assisted-n-shot-cap}\ with the regularized
sandwiched R\'{e}nyi relative entropy and arrive at the following statements:

\begin{corollary}
\label{cor:sandwiched-bipartite-bound}The following upper bound holds for the
$n$-shot randomness transmission capacity of a bipartite channel
$\mathcal{N}_{AB\rightarrow A^{\prime}B^{\prime}}$:%
\begin{multline}
R^{n,\varepsilon}(\mathcal{N}_{AB\rightarrow A^{\prime}B^{\prime}})\leq\\
\widetilde{\Upsilon}_{\alpha}^{\operatorname{reg}}(\mathcal{N}_{AB\rightarrow
A^{\prime}B^{\prime}})+\frac{\alpha}{n\left(  \alpha-1\right)  }\log
_{2}\!\left(  \frac{1}{1-\varepsilon}\right)  ,
\end{multline}
for all $\alpha>1$ and $\varepsilon\in\lbrack0,1)$, where%
\begin{multline}
\widetilde{\Upsilon}_{\alpha}^{\operatorname{reg}}(\mathcal{N}_{AB\rightarrow
A^{\prime}B^{\prime}})\coloneqq\\
\inf_{\substack{\mathcal{M}_{AB\rightarrow A^{\prime}B^{\prime}}%
:\\\beta(\mathcal{M}_{AB\rightarrow A^{\prime}B^{\prime}})\leq1}}\widetilde
{D}_{\alpha}^{\operatorname{reg}}(\mathcal{N}_{AB\rightarrow A^{\prime
}B^{\prime}}\Vert\mathcal{M}_{AB\rightarrow A^{\prime}B^{\prime}}).
\end{multline}

\end{corollary}

\begin{corollary}
\label{cor:sandwiched-p2p-bound}The following upper bound holds for the
$n$-shot randomness transmission capacity of a point-to-point channel
$\mathcal{N}_{A\rightarrow B^{\prime}}$ assisted by a classical feedback
channel:%
\begin{equation}
C^{n,\varepsilon}(\mathcal{N}_{A\rightarrow B^{\prime}})\leq\widetilde
{\Upsilon}_{\alpha}^{\operatorname{reg}}(\mathcal{N}_{A\rightarrow B^{\prime}%
})+\frac{\alpha}{n\left(  \alpha-1\right)  }\log_{2}\!\left(  \frac
{1}{1-\varepsilon}\right)  ,
\end{equation}
for all $\alpha>1$ and $\varepsilon\in\lbrack0,1)$, where%
\begin{multline}
\widetilde{\Upsilon}_{\alpha}^{\operatorname{reg}}(\mathcal{N}_{A\rightarrow
B^{\prime}})\coloneqq\\
\inf_{\substack{\mathcal{M}_{A\rightarrow B^{\prime}}:\\\beta(\mathcal{M}%
_{A\rightarrow B^{\prime}})\leq1}}\widetilde{D}_{\alpha}^{\operatorname{reg}%
}(\mathcal{N}_{A\rightarrow B^{\prime}}\Vert\mathcal{M}_{A\rightarrow
B^{\prime}}).
\end{multline}

\end{corollary}

These bounds are not particularly useful, because the quantities
$\widetilde{\Upsilon}_{\alpha}^{\operatorname{reg}}(\mathcal{N}_{AB\rightarrow
A^{\prime}B^{\prime}})$ and $\widetilde{\Upsilon}_{\alpha}^{\operatorname{reg}%
}(\mathcal{N}_{A\rightarrow B^{\prime}})$ may be difficult to compute in
practice. However, see the discussions in \cite[Section~5.1]{FawziFawzi20} for
progress on algorithms for computing $\widetilde{D}_{\alpha}%
^{\operatorname{reg}}(\mathcal{N}\Vert\mathcal{M})$. In the next section, we
show how these bounds simplify when the channels of interest possess symmetry.

\section{Exploiting symmetries}

\label{sec:symmetries}

In this section, we discuss how to improve the upper
bounds in Corollaries~\ref{cor:sandwiched-bipartite-bound} and
\ref{cor:sandwiched-p2p-bound}\ when a bipartite channel and point-to-point
channel possess symmetries, respectively.

We begin by recalling the definition of a bicovariant bipartite channel
\cite{DBW20}. Let $G$ and $H$ be finite groups, and for $g\in G$ and $h\in H$,
let $g\rightarrow U_{A}(g)$ and $h\rightarrow V_{B}(h)$ be unitary
representations. Also, let $(g,h)\rightarrow W_{A^{\prime}}(g,h)$ and
$(g,h)\rightarrow Y_{B^{\prime}}(g,h)$ be unitary representations. A bipartite
channel $\mathcal{N}_{AB\rightarrow A^{\prime}B^{\prime}}$\ is bicovariant
with respect to these representations if the following equality holds for all
group elements $g\in G$ and $h\in H$:%
\begin{multline}
\mathcal{N}_{AB\rightarrow A^{\prime}B^{\prime}}\circ(\mathcal{U}%
_{A}(g)\otimes\mathcal{V}_{B}(h))=\label{eq:bicovariant-def}\\
(\mathcal{W}_{A^{\prime}}(g,h)\otimes\mathcal{Y}_{B^{\prime}}(g,h))\circ
\mathcal{N}_{AB\rightarrow A^{\prime}B^{\prime}},
\end{multline}
where $\mathcal{U}_{A}(g)(\cdot)\coloneqq U_{A}(g)(\cdot)U_{A}(g)^{\dag}$,
with similar definitions for $\mathcal{V}_{B}(h)$, $\mathcal{W}_{A^{\prime}%
}(g,h)$, and $\mathcal{Y}_{B^{\prime}}(g,h)$. A bipartite channel is
bicovariant if it is bicovariant with respect to groups that have
representations as unitary one-designs, i.e.,
\begin{align}
\frac{1}{\left\vert
G\right\vert }\sum_{g\in G}\mathcal{U}_{A}(g)(X) & =\operatorname{Tr}[X]\pi_{A},\\
\frac{1}{\left\vert H\right\vert }\sum_{h\in H}\mathcal{V}_{B}%
(h)(Y) & =\operatorname{Tr}[Y]\pi_{B}.
\end{align}
Two bipartite maps $\mathcal{N}%
_{AB\rightarrow A^{\prime}B^{\prime}}$ and $\mathcal{M}_{AB\rightarrow
A^{\prime}B^{\prime}}$ are jointly bicovariant if they are bicovariant with
respect to the same representations, i.e., if \eqref{eq:bicovariant-def} holds
for both $\mathcal{N}_{AB\rightarrow A^{\prime}B^{\prime}}$ and $\mathcal{M}%
_{AB\rightarrow A^{\prime}B^{\prime}}$.

\begin{proposition}
\label{prop:ups-with-symmetry}Let $\mathcal{N}_{AB\rightarrow A^{\prime
}B^{\prime}}$ be a bipartite channel that is bicovariant with respect to
unitary representations as defined above. Then%
\begin{multline}
\boldsymbol{\Upsilon}(\mathcal{N}_{AB\rightarrow A^{\prime}B^{\prime}%
})=\label{eq:ups-simplify-bicovariance}\\
\inf_{\substack{\mathcal{M}\in
\mathcal{C}:\\\beta(\mathcal{M})\leq
1}}\sup_{\psi\in\mathcal{S}}\boldsymbol{D}(\mathcal{N}_{AB\rightarrow
A^{\prime}B^{\prime}}(\psi_{RAB})\Vert
\mathcal{M}_{AB\rightarrow A^{\prime}B^{\prime}}(\psi_{RAB})),
\end{multline}
where $\mathcal{S}$ is the set consisting of every pure state $\psi_{RAB}$ such that the
reduced state $\psi_{AB}$ satisfies%
\begin{equation}
\psi_{AB}=\frac{1}{\left\vert G\right\vert \left\vert H\right\vert }\sum_{g\in
G,h\in H}(\mathcal{U}_{A}(g)\otimes\mathcal{V}_{B}(h))(\psi_{AB}%
),\label{eq:mixing-bicovariance}%
\end{equation}
and $\mathcal{C}$ is the set of all completely positive bipartite maps that
are bicovariant with respect to the unitary representations defined above. In
the case that $\mathcal{N}_{AB\rightarrow A^{\prime}B^{\prime}}$ is
bicovariant, then%
\begin{multline}
\boldsymbol{\Upsilon}(\mathcal{N}_{AB\rightarrow A^{\prime}B^{\prime}%
})=\label{eq:ups-much-simplify-bicovariance}\\
\inf_{\substack{\mathcal{M}\in
\mathcal{C}^{1}:\\\beta(\mathcal{M}
)\leq1}}\boldsymbol{D}(\mathcal{N}
(\Phi_{\hat{A}A}\otimes\Phi_{B\hat{B}})\Vert
\mathcal{M}(\Phi_{\hat{A}A}\otimes
\Phi_{B\hat{B}})),
\end{multline}
where $\Phi_{\hat{A}A}\otimes\Phi_{B\hat{B}}$ is a tensor product of maximally
entangled states and $\mathcal{C}^{1}$ is the set of all completely positive
bicovariant maps $\mathcal{M}_{AB\rightarrow A^{\prime}B^{\prime}}$ (i.e.,
covariant with respect to one-designs).
\end{proposition}

\begin{IEEEproof}
Let $\psi_{RAB}$ be an arbitrary pure state. Define%
\begin{equation}
\overline{\rho}_{AB}\coloneqq\frac{1}{\left\vert G\right\vert \left\vert
H\right\vert }\sum_{g\in G,h\in H}(\mathcal{U}_{A}(g)\otimes\mathcal{V}%
_{B}(h))(\psi_{AB}).
\end{equation}
Let $\phi_{SAB}^{\overline{\rho}}\in\mathcal{S}$ be a purification of
$\overline{\rho}_{AB}$. Another purification of $\overline{\rho}_{AB}$ is
given by%
\begin{equation}
\psi_{GHRAB}^{\overline{\rho}}\coloneqq|\psi^{\overline{\rho}}\rangle
\!\langle\psi^{\overline{\rho}}|_{GHRAB},
\end{equation}
where
\begin{multline}
|\psi^{\overline{\rho}}\rangle_{GHRAB}\coloneqq\\
\frac{1}{\sqrt{\left\vert G\right\vert \left\vert H\right\vert }}\sum_{g\in
G,h\in H}|g\rangle_{G}|h\rangle_{H}(U_{A}(g)\otimes V_{B}(h))|\psi
\rangle_{RAB}.
\end{multline}
Let $\mathcal{M}_{AB\rightarrow A^{\prime}B^{\prime}}$ be an arbitrary
completely positive\ map satisfying $\beta(\mathcal{M}_{AB\rightarrow
A^{\prime}B^{\prime}})\leq1$. Define%
\begin{multline}
\overline{\mathcal{M}}_{AB\rightarrow A^{\prime}B^{\prime}}\coloneqq\\
\frac{1}{\left\vert G\right\vert \left\vert H\right\vert }\sum_{g\in G,h\in
H}(\mathcal{W}_{A^{\prime}}(g,h)\otimes\mathcal{Y}_{B^{\prime}}(g,h))^{\dag}\\
\circ\mathcal{M}_{AB\rightarrow A^{\prime}B^{\prime}}\circ(\mathcal{U}%
_{A}(g)\otimes\mathcal{V}_{B}(h)),
\end{multline}
and observe that $\overline{\mathcal{M}}_{AB\rightarrow A^{\prime}B^{\prime}%
}\in\mathcal{C}$. Consider the development in \eqref{eq:fig-eqs-1}--\eqref{eq:fig-eqs-last}.
\begin{figure*}%
\begin{align}
&  \boldsymbol{D}(\mathcal{N}_{AB\rightarrow A^{\prime}B^{\prime}}(\phi
_{SAB}^{\overline{\rho}})\Vert\mathcal{M}_{AB\rightarrow A^{\prime}B^{\prime}%
}(\phi_{SAB}^{\overline{\rho}}))\nonumber\\
&  =\boldsymbol{D}(\mathcal{N}_{AB\rightarrow A^{\prime}B^{\prime}}%
(\psi_{GHRAB}^{\overline{\rho}})\Vert\mathcal{M}_{AB\rightarrow A^{\prime
}B^{\prime}}(\psi_{GHRAB}^{\overline{\rho}}))
\label{eq:fig-eqs-1}
\\
&  \geq\boldsymbol{D}\Bigg(\frac{1}{\left\vert G\right\vert \left\vert
H\right\vert }\sum_{g\in G,h\in H}|g,h\rangle\!\langle g,h|_{GH}%
\otimes(\mathcal{N}_{AB\rightarrow A^{\prime}B^{\prime}}\circ(\mathcal{U}%
_{A}(g)\otimes\mathcal{V}_{B}(h)))(\psi_{RAB})\Bigg\Vert\nonumber\\
&  \qquad\qquad\frac{1}{\left\vert G\right\vert \left\vert H\right\vert }%
\sum_{g\in G,h\in H}|g,h\rangle\!\langle g,h|_{GH}\otimes(\mathcal{M}%
_{AB\rightarrow A^{\prime}B^{\prime}}\circ(\mathcal{U}_{A}(g)\otimes
\mathcal{V}_{B}(h)))(\psi_{RAB})\Bigg)
\label{eq:fig-eqs-2}\\
&  =\boldsymbol{D}\Bigg(\frac{1}{\left\vert G\right\vert \left\vert
H\right\vert }\sum_{g\in G,h\in H}|g,h\rangle\!\langle g,h|_{GH}%
\otimes((\mathcal{W}_{A^{\prime}}(g,h)\otimes\mathcal{Y}_{B^{\prime}%
}(g,h))\circ\mathcal{N}_{AB\rightarrow A^{\prime}B^{\prime}})(\psi
_{RAB})\Bigg\Vert\nonumber\\
&  \qquad\qquad\frac{1}{\left\vert G\right\vert \left\vert H\right\vert }%
\sum_{g\in G,h\in H}|g,h\rangle\!\langle g,h|_{GH}\otimes(\mathcal{M}%
_{AB\rightarrow A^{\prime}B^{\prime}}\circ(\mathcal{U}_{A}(g)\otimes
\mathcal{V}_{B}(h)))(\psi_{RAB})\Bigg)
\label{eq:fig-eqs-3}\\
&  =\boldsymbol{D}\Bigg(\frac{1}{\left\vert G\right\vert \left\vert
H\right\vert }\sum_{g\in G,h\in H}|g,h\rangle\!\langle g,h|_{GH}%
\otimes\mathcal{N}_{AB\rightarrow A^{\prime}B^{\prime}}(\psi_{RAB}%
)\Bigg\Vert\nonumber\\
&  \qquad\qquad\frac{1}{\left\vert G\right\vert \left\vert H\right\vert }%
\sum_{g\in G,h\in H}|g,h\rangle\!\langle g,h|_{GH}\otimes\mathcal{M}%
_{AB\rightarrow A^{\prime}B^{\prime}}^{g,h}(\psi_{RAB})\Bigg)
\label{eq:fig-eqs-4}\\
&  \geq\boldsymbol{D}(\mathcal{N}_{AB\rightarrow A^{\prime}B^{\prime}}%
(\psi_{RAB})\Vert\overline{\mathcal{M}}_{AB\rightarrow A^{\prime}B^{\prime}%
}(\psi_{RAB})).
\label{eq:fig-eqs-last}
\end{align}
\hrule
\end{figure*}The first equality in \eqref{eq:fig-eqs-1} holds because all purifications are related by
an isometric channel acting on the purifying system, the channel
$\mathcal{N}_{AB\rightarrow A^{\prime}B^{\prime}}$ commutes with the action of
this isometric channel because they act on different systems, and the
generalized divergence is invariant under the action of isometric channels.
The first inequality in \eqref{eq:fig-eqs-2} follows by acting with a completely dephasing channel on
the systems $GH$ and then applying the data-processing inequality. The second
equality in \eqref{eq:fig-eqs-3} follows from the bicovariance of $\mathcal{N}_{AB\rightarrow
A^{\prime}B^{\prime}}$ with respect to the given representations. The third
equality in \eqref{eq:fig-eqs-4} follows by applying the unitary
\begin{equation}
\sum_{g\in G,h\in H}|g,h\rangle\!\langle g,h|_{GH}\otimes W_{A^{\prime}}%
^{\dag}(g,h)\otimes Y_{B^{\prime}}^{\dag}(g,h),
\end{equation}
and from the unitary invariance of the generalized divergence. We have also
defined%
\begin{multline}
\mathcal{M}_{AB\rightarrow A^{\prime}B^{\prime}}^{g,h}\coloneqq\\
(\mathcal{W}_{A^{\prime}}(g,h)\otimes\mathcal{Y}_{B^{\prime}}(g,h))^{\dag
}\circ\mathcal{M}_{AB\rightarrow A^{\prime}B^{\prime}}\circ(\mathcal{U}%
_{A}(g)\otimes\mathcal{V}_{B}(h)).
\end{multline}
The last inequality in \eqref{eq:fig-eqs-last} follows from tracing over the registers $GH$ and from the
data-processing inequality. Since the inequality holds for all pure states, we
conclude that%
\begin{align}
&  \sup_{\phi_{SAB}\in\mathcal{S}}\boldsymbol{D}(\mathcal{N}_{AB\rightarrow
A^{\prime}B^{\prime}}(\phi_{SAB})\Vert\mathcal{M}_{AB\rightarrow A^{\prime
}B^{\prime}}(\phi_{SAB}))\nonumber\\
&  \geq\boldsymbol{D}(\mathcal{N}_{AB\rightarrow A^{\prime}B^{\prime}}%
\Vert\overline{\mathcal{M}}_{AB\rightarrow A^{\prime}B^{\prime}}%
)\label{eq:key-ineq-symm}\\
&  \geq\boldsymbol{\Upsilon}(\mathcal{N}_{AB\rightarrow A^{\prime}B^{\prime}%
}).
\end{align}
The second inequality follows because $\overline{\mathcal{M}}_{AB\rightarrow
A^{\prime}B^{\prime}}$ satisfies $\beta(\overline{\mathcal{M}}_{AB\rightarrow
A^{\prime}B^{\prime}})\leq1$ if $\mathcal{M}_{AB\rightarrow A^{\prime
}B^{\prime}}$ does. This in turn is a consequence of the convexity of $\beta$
(Proposition~\ref{prop:convexity-beta}) and its invariance under local unitary
channels (Corollary~\ref{cor:I-LUC-beta-p-bipartite-ch}). Since the inequality
holds for all $\mathcal{M}_{AB\rightarrow A^{\prime}B^{\prime}}$ satisfying
$\beta(\mathcal{M}_{AB\rightarrow A^{\prime}B^{\prime}})\leq1$, we conclude
that%
\begin{multline}
\inf_{\substack{\mathcal{M}
:\\\beta(\mathcal{M})\leq1}}\sup
_{\phi\in\mathcal{S}}\boldsymbol{D}(\mathcal{N}_{AB\rightarrow
A^{\prime}B^{\prime}}(\phi_{SAB})\Vert
\mathcal{M}_{AB\rightarrow A^{\prime}B^{\prime}}(\phi_{SAB}))\\
\geq\boldsymbol{\Upsilon}(\mathcal{N}_{AB\rightarrow A^{\prime}B^{\prime}}).
\end{multline}
However, the definition of $\boldsymbol{\Upsilon}(\mathcal{N}_{AB\rightarrow
A^{\prime}B^{\prime}})$ implies that
\begin{multline}
\boldsymbol{\Upsilon}(\mathcal{N}_{AB\rightarrow A^{\prime}B^{\prime}})\geq\\
\inf_{\substack{\mathcal{M}
:\\\beta(\mathcal{M})\leq1}}\sup
_{\phi\in\mathcal{S}}\boldsymbol{D}(\mathcal{N}_{AB\rightarrow
A^{\prime}B^{\prime}}(\phi_{SAB})\Vert
\mathcal{M}_{AB\rightarrow A^{\prime}B^{\prime}}(\phi_{SAB})).
\end{multline}
So we conclude the equality%
\begin{multline}
\boldsymbol{\Upsilon}(\mathcal{N}_{AB\rightarrow A^{\prime}B^{\prime}})=\\
\inf_{\substack{\mathcal{M}
:\\\beta(\mathcal{M})\leq1}}\sup
_{\phi\in\mathcal{S}}\boldsymbol{D}(\mathcal{N}_{AB\rightarrow
A^{\prime}B^{\prime}}(\phi_{SAB})\Vert
\mathcal{M}_{AB\rightarrow A^{\prime}B^{\prime}}(\phi_{SAB})).
\end{multline}
Now suppose that $\mathcal{M}_{AB\rightarrow A^{\prime}B^{\prime}}$ is an
arbitrary completely positive map satisfying $\beta(\mathcal{M}_{AB\rightarrow
A^{\prime}B^{\prime}})\leq1$, and let $\phi_{SAB}\in\mathcal{S}$. Then by
\eqref{eq:key-ineq-symm}, we conclude that%
\begin{multline}
\sup_{\phi_{SAB}\in\mathcal{S}}\boldsymbol{D}(\mathcal{N}_{AB\rightarrow
A^{\prime}B^{\prime}}(\phi_{SAB})\Vert\mathcal{M}_{AB\rightarrow A^{\prime
}B^{\prime}}(\phi_{SAB}))\\
\geq\inf_{\substack{\overline{\mathcal{M}}\in\mathcal{C}:\\\beta(\overline{\mathcal{M}})\leq1}}\sup_{\phi\in\mathcal{S}}\boldsymbol{D}%
(\mathcal{N}_{AB\rightarrow A^{\prime}B^{\prime}}(\phi_{SAB})\Vert
\overline{\mathcal{M}}_{AB\rightarrow A^{\prime}B^{\prime}}(\phi_{SAB})).
\end{multline}
Since this holds for every completely positive map $\mathcal{M}_{AB\rightarrow A^{\prime}B^{\prime}}$
satisfying $\beta(\mathcal{M}_{AB\rightarrow A^{\prime}B^{\prime}})\leq1$, we
conclude that%
\begin{multline}
\boldsymbol{\Upsilon}(\mathcal{N}_{AB\rightarrow A^{\prime}B^{\prime}})\geq\\
\inf_{\substack{\overline{\mathcal{M}}
\in\mathcal{C}:\\\beta(\overline{\mathcal{M}})\leq1}}\sup_{\phi\in\mathcal{S}}\boldsymbol{D}%
(\mathcal{N}_{AB\rightarrow A^{\prime}B^{\prime}}(\phi_{SAB})\Vert
\overline{\mathcal{M}}_{AB\rightarrow A^{\prime}B^{\prime}}(\phi_{SAB})).
\end{multline}
However, from the definition of $\boldsymbol{\Upsilon}(\mathcal{N}%
_{AB\rightarrow A^{\prime}B^{\prime}})$, we have the inequality%
\begin{multline}
\boldsymbol{\Upsilon}(\mathcal{N}_{AB\rightarrow A^{\prime}B^{\prime}})\leq\\
\inf_{\substack{\overline{\mathcal{M}}
\in\mathcal{C}:\\\beta(\overline{\mathcal{M}})\leq1}}\sup_{\phi\in\mathcal{S}}\boldsymbol{D}%
(\mathcal{N}_{AB\rightarrow A^{\prime}B^{\prime}}(\phi_{SAB})\Vert
\overline{\mathcal{M}}_{AB\rightarrow A^{\prime}B^{\prime}}(\phi_{SAB})).
\end{multline}
Thus, the equality in \eqref{eq:ups-simplify-bicovariance}\ follows. The
equality in \eqref{eq:ups-much-simplify-bicovariance}\ follows because the
only state satisfying \eqref{eq:mixing-bicovariance} for one-designs is the
tensor product of maximally mixed states, and the tensor product of maximally
entangled states purifies this state.
\end{IEEEproof}

\bigskip

Recall from Section~\ref{sec:employ-sandwichedR}\ that the bounds in
Corollaries~\ref{cor:sandwiched-bipartite-bound} and
\ref{cor:sandwiched-p2p-bound}\ are not particularly useful on their own
because $\widetilde{\Upsilon}_{\alpha}^{\operatorname{reg}}(\mathcal{N}%
_{AB\rightarrow A^{\prime}B^{\prime}})$ may be difficult to compute in
practice. However, if the bipartite channel is bicovariant, then
\eqref{eq:ups-much-simplify-bicovariance}\ implies that the regularized
quantity is bounded from above by a single-letter quantity:%
\begin{equation}
\widetilde{\Upsilon}_{\alpha}^{\operatorname{reg}}(\mathcal{N}_{AB\rightarrow
A^{\prime}B^{\prime}})\leq\widetilde{\Upsilon}_{\alpha}(\mathcal{N}%
_{AB\rightarrow A^{\prime}B^{\prime}}).
\end{equation}
We then obtain the following:

\begin{corollary}
The following upper bound holds for the $n$-shot randomness transmission
capacity of a bicovariant bipartite channel $\mathcal{N}_{AB\rightarrow
A^{\prime}B^{\prime}}$:%
\begin{multline}
R^{n,\varepsilon}(\mathcal{N}_{AB\rightarrow A^{\prime}B^{\prime}})\leq\\
\widetilde{\Upsilon}_{\alpha}(\mathcal{N}_{AB\rightarrow A^{\prime}B^{\prime}%
})+\frac{\alpha}{n\left(  \alpha-1\right)  }\log_{2}\!\left(  \frac
{1}{1-\varepsilon}\right)  ,
\end{multline}
for all $\alpha>1$ and $\varepsilon\in\lbrack0,1)$.
\end{corollary}

By applying the same reasoning in the proof of
Theorem~\ref{thm:SC-RD-bipartite-ch}, we conclude the following:

\begin{corollary}
\label{cor:cap-bound-bicovariant}The following upper bound holds for the
strong converse randomness transmission capacity of a bicovariant bipartite
channel $\mathcal{N}_{AB\rightarrow A^{\prime}B^{\prime}}$:%
\begin{equation}
\widetilde{R}(\mathcal{N}_{AB\rightarrow A^{\prime}B^{\prime}})\leq
\Upsilon(\mathcal{N}_{AB\rightarrow A^{\prime}B^{\prime}}).
\end{equation}

\end{corollary}

Let $G$ be a group and let $U_{A}(g)$ and $V_{B^{\prime}}(g)$ be unitary
representations of $g$. A point-to-point channel $\mathcal{N}_{A\rightarrow
B^{\prime}}$\ is covariant with respect to these representations if the
following equality holds for all $g\in G$ \cite{Hol02}:%
\begin{equation}
\mathcal{N}_{A\rightarrow B^{\prime}}\circ\mathcal{U}_{A}(g)=\mathcal{V}%
_{B^{\prime}}(g)\circ\mathcal{N}_{A\rightarrow B^{\prime}}.
\label{eq:cov-ch-def}
\end{equation}
A point-to-point channel is covariant if it is covariant with respect to a one-design.

By applying the same reasoning as given above, we have the following results:

\begin{corollary}
The following upper bound holds for the $n$-shot classical capacity of a
covariant point-to-point channel $\mathcal{N}_{A\rightarrow B^{\prime}}$
assisted by a classical feedback channel:%
\begin{equation}
C_{\leftarrow}^{n,\varepsilon}(\mathcal{N}_{A\rightarrow B^{\prime}}%
)\leq\widetilde{\Upsilon}_{\alpha}(\mathcal{N}_{A\rightarrow B^{\prime}%
})+\frac{\alpha}{n\left(  \alpha-1\right)  }\log_{2}\!\left(  \frac
{1}{1-\varepsilon}\right)  ,
\end{equation}
for all $\alpha>1$ and $\varepsilon\in\lbrack0,1)$.
\end{corollary}

\begin{corollary}
\label{cor:strong-converse-covariant}
The following upper bound holds for the strong converse classical capacity of
a covariant point-to-point channel $\mathcal{N}_{A\rightarrow B^{\prime}}$:%
\begin{equation}
\widetilde{C}_{\leftarrow}(\mathcal{N}_{A\rightarrow B^{\prime}})\leq
\Upsilon(\mathcal{N}_{A\rightarrow B^{\prime}}).
\end{equation}

\end{corollary}

\section{Examples}

\label{sec:examples}In this section, we apply the bounds to some key examples
of bipartite and point-to-point channels. The Matlab code used to generate the
plots below is available with the arXiv posting of our paper.

\subsection{Partial swap bipartite channel}

The partial swap unitary is defined for $p\in\left[  0,1\right]  $ as
\cite{FHSSW11,audenaert2016entropy}%
\begin{align}
S_{AB}^{p}  &  \coloneqq \sqrt{1-p}I_{AB}+i\sqrt{p}S_{AB},\\
S_{AB}  &  \coloneqq \sum_{i,j=0}^{d-1}|i\rangle\!\langle j|_{A}%
\otimes|j\rangle\!\langle i|_{B}, \label{eq:swap-op}%
\end{align}
where $A\simeq B$ and $d=d_{A}=d_{B}$. The following identity holds%
\begin{equation}
S_{AB}^{p}=e^{itS_{AB}} = \cos(t) I_{AB} + i \sin(t) S_{AB},
\end{equation}
where $\sqrt{1-p}=\cos t$. Thus, we can understand the unitary operator
$S_{AB}^{p}$ as arising from time evolution according to the Hamiltonian
$S_{AB}$. We then define the bipartite partial swap channel as%
\begin{equation}
\mathcal{S}_{AB}^{p}(\cdot)\coloneqq S_{AB}^{p}(\cdot)(S_{AB}^{p})^{\dag}.
\label{eq:partial-swap-bi-channel}%
\end{equation}

Suppose that $p=1$. Then the channel $\mathcal{S}_{AB}^{p}$\ is equivalent to
a swap channel. In this case, the forward classical capacity is equal to
$2\log_{2}d$. This follows by an argument given in \cite{BHLS03}. To see that
the rate $2\log_{2}d$ is achievable, consider the following strategy. On the
first use of the channel, Alice inputs one classical dit to her input and Bob
inputs one share of a maximally entangled state. Bob can decode the classical
dit, and after the first channel use, they share a maximally entangled state
$\Phi^{d}$. Before the second channel use, Alice can employ a super-dense
coding strategy \cite{PhysRevLett.69.2881}. She applies one of the $d^{2}$
Heisenberg--Weyl unitaries to her share of $\Phi^{d}$ and transmits it through
her input to the channel. Bob again prepares $\Phi^{d}$ and sends one share
through his channel input. Bob can then decode the message Alice sent, by
performing a Bell measurement, and they again share $\Phi^{d}$. They then
repeat this procedure many times. Even though the first channel use allows for
only $\log_{2}d$ bits to be transmitted, all of the other channel uses allow
for $2\log_{2}d$ bits to be transmitted. So in the limit of many channel uses,
the rate $2\log_{2}d$ is achievable. An upper bound of $2\log_{2}d$ is argued
in \cite{BHLS03} by employing a simulation argument. Alternatively, it can be
seen from our approach by employing Theorem~\ref{thm:SC-RD-bipartite-ch} and
Proposition~\ref{prop:ups-to-cbeta-bipartite}, and picking $S_{AA^{\prime
}BB^{\prime}}=V_{AA^{\prime}BB^{\prime}}=I_{AA^{\prime}BB^{\prime}}$ in the
definition of $C_{\beta}$. These choices satisfy the constraints and $\log
_{2}\left\Vert \operatorname{Tr}_{A^{\prime}B^{\prime}}[S_{AA^{\prime
}BB^{\prime}}]\right\Vert _{\infty}=2\log_{2}d$.

At the other extreme, when $p=0$, the channel $\mathcal{S}_{AB}^{p}$ reduces
to the tensor product of identity channels. Since this channel is a product of
local channels, Theorem~\ref{thm:SC-RD-bipartite-ch} and
Proposition~\ref{prop:zero-local-chs-gen-div-ups}\ imply that $C(\mathcal{S}%
_{AB}^{p})=0$. Thus, the partial swap unitary interpolates between these two extremes.

\begin{figure}[ptb]
\begin{center}
\includegraphics[
width=\linewidth
]{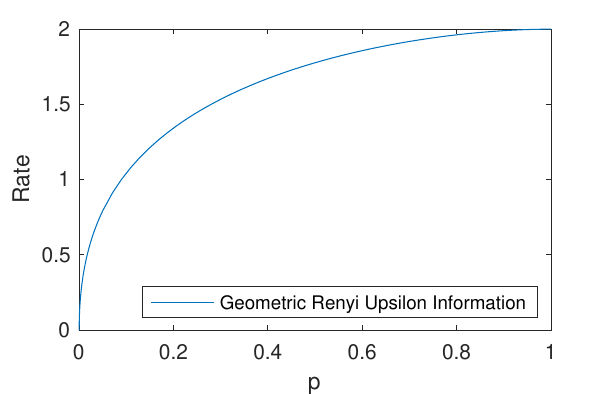}
\end{center}
\caption{Upper bound on the forward classical capacity of the partial swap
bipartite channel in \eqref{eq:partial-swap-bi-channel}, with $d=2$.}%
\label{fig:plot-partial-swap-channel}%
\end{figure}

Interestingly, the partial swap unitary is not bicovariant for $p\in(0,1)$
because the general definition involves both the identity and the swap. As
such, our bound from Theorem~\ref{thm:SC-RD-bipartite-ch} is useful in such a
case. By employing a semi-definite program to calculate $\widehat{\Upsilon
}_{\alpha}(\mathcal{S}_{AB}^{p})$ for $d=2$ and $\alpha=1+2^{-\ell}$, with
$\ell=4$, we arrive at the plot given in
Figure~\ref{fig:plot-partial-swap-channel}. The semi-definite program is included in the arXiv posting of this paper and is based on the methods mentioned in \cite[Remark~4]{Fang2019a}.

We remark that the partial swap channel is bicovariant with respect to all
unitaries of the form $U\otimes U$. As such, by applying
Proposition~\ref{prop:ups-with-symmetry}, we conclude that it suffices to
maximize $\widehat{\Upsilon}_{\alpha}(\mathcal{S}_{AB}^{p})$ over input states
$\psi_{RAB}$ possessing the following symmetry:%
\begin{equation}
\psi_{AB}=\int dU\ (U_{A}\otimes U_{B})\psi_{AB}(U_{A}\otimes U_{B})^{\dag},
\end{equation}
where $dU$ denotes the Haar measure. States possessing this symmetry are known
as Werner states \cite{Wer89}\ and can be written in terms of a single
parameter $q\in\left[  0,1\right]  $ as follows:%
\begin{equation}
W_{AB}^{(q,d)}\coloneqq \left(  1-q\right)  \frac{2}{d\left(  d+1\right)  }%
\Pi_{AB}^{+}+q\frac{2}{d\left(  d-1\right)  }\Pi_{AB}^{-},
\end{equation}
where $\Pi_{AB}^{\pm}\coloneqq \left(  I_{AB}\pm S_{AB}\right)  /2$ are the
projections onto the symmetric and antisymmetric subspaces of $A$ and $B$,
with $S_{AB}$ defined in \eqref{eq:swap-op}. Additionally, by exploiting the
same symmetry, it suffices to minimize over completely positive bipartite maps
$\mathcal{M}_{AB\rightarrow A^{\prime}B^{\prime}}$ such that%
\begin{equation}
\mathcal{M}_{AB\rightarrow A^{\prime}B^{\prime}}=\int dU\ (\mathcal{U}%
_{A^{\prime}}\otimes\mathcal{U}_{B^{\prime}})^{\dag}\circ\mathcal{M}%
_{AB\rightarrow A^{\prime}B^{\prime}}\circ(\mathcal{U}_{A}\otimes
\mathcal{U}_{B}).
\end{equation}
This is equivalent to their Choi operators satisfying%
\begin{equation}
\Gamma_{AA^{\prime}BB^{\prime}}^{\mathcal{M}}=\int dU\ (\mathcal{U}%
_{A^{\prime}}\otimes\mathcal{U}_{B^{\prime}}\otimes\overline{\mathcal{U}}%
_{A}\otimes\overline{\mathcal{U}}_{B})(\Gamma_{AA^{\prime}BB^{\prime}%
}^{\mathcal{M}}),
\label{eq:haar-haar}
\end{equation}
where $\overline{\mathcal{U}}$ denotes the complex conjugate. This further
reduces the number of parameters needed in the optimization task, which is
useful for computing $\widehat{\Upsilon}_{\alpha}(\mathcal{S}_{AB}^{p})$ for
higher-dimensional partial swap bipartite channels. We note that Haar integrals of the form in \eqref{eq:haar-haar} can be computed by generalizing the methods of \cite{EW01,JV13} (see also \cite[Section~VII]{SMKH20}).

\subsection{Noisy CNOT gate}

\begin{figure}[ptb]
\begin{center}
\includegraphics[
width=\linewidth
]{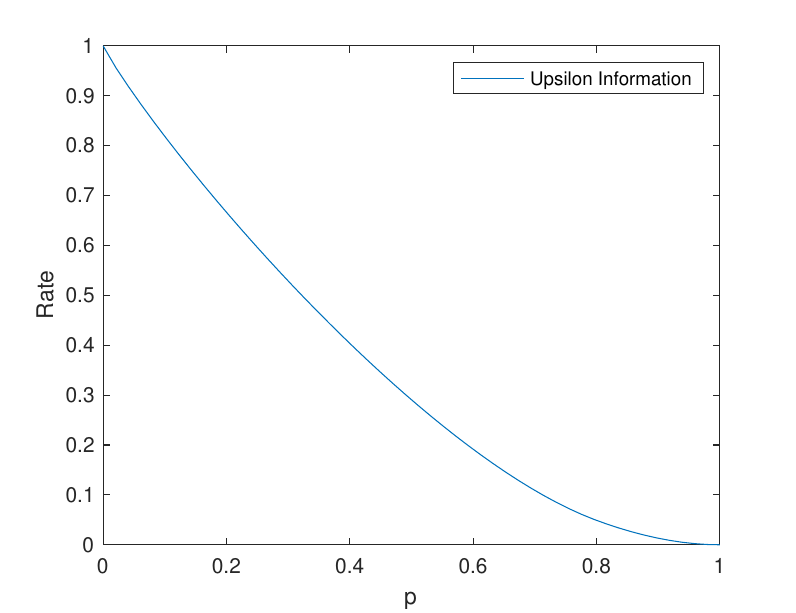}
\end{center}
\caption{Upper bound on the forward classical capacity of the noisy CNOT
bipartite channel in \eqref{eq:noisy-CNOT-channel}, with $d=2$.}%
\label{fig:noisy-CNOT}%
\end{figure}

Another example of a bipartite channel of interest is a noisy CNOT gate,
defined as follows:%
\begin{equation}
\mathcal{D}_{AB}^{p}(\cdot)\coloneqq\left(  1-p\right)  \text{CNOT}_{AB}%
(\cdot)\text{CNOT}_{AB}+p\mathcal{R}_{AB}^{\pi}(\cdot
),\label{eq:noisy-CNOT-channel}%
\end{equation}
where%
\begin{align}
\text{CNOT}_{AB} &  \coloneqq\sum_{i=0}^{d-1}|i\rangle\!\langle i|_{A}\otimes
X(i)_{B},\\
X(i)_{B} &  \coloneqq\sum_{j=0}^{d-1}|i\oplus j\rangle\!\langle j|,\\
\mathcal{R}_{AB}^{\pi}(\cdot) &  \coloneqq\operatorname{Tr}_{AB}[\cdot
]\pi_{AB},\\
\pi_{AB} &  \coloneqq\frac{I_{AB}}{d_{A}d_{B}}.
\end{align}
When $p=0$, the channel is a perfect CNOT\ gate, and when $p=1$, it is a
replacer channel. Thus, when $p=0$, the result from \cite{BHLS03} applies,
implying that $C(\mathcal{D}_{AB}^{p=0})=\log_{2}d$, and when $p=1$, the
forward classical capacity $C(\mathcal{D}_{AB}^{p=0})=0$.

This channel is bicovariant, as argued in \cite{DBW20}, and so
Corollary~\ref{cor:cap-bound-bicovariant}\ applies. Evaluating the $\Upsilon
$-information of $\mathcal{D}_{AB}^{p}$, we obtain the plot in
Figure~\ref{fig:noisy-CNOT}.

\subsection{Point-to-point depolarizing channel}

\begin{figure}[ptb]
\begin{center}
\includegraphics[
width=\linewidth
]{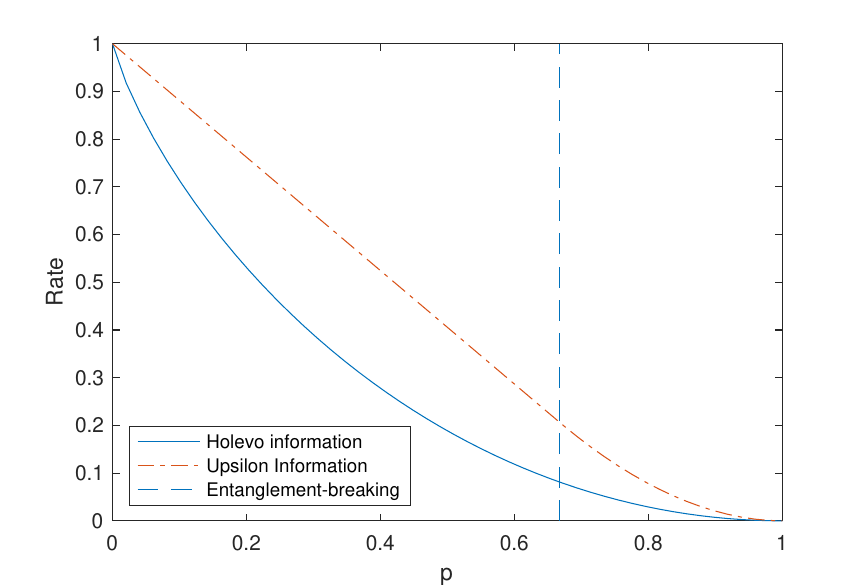}
\end{center}
\caption{Lower and upper bounds on the classical-feedback-assisted classical
capacity of the qubit depolarizing channel in \eqref{eq:dep-ch-example}, with
$d=2$. The dashed vertical line indicates that the qubit depolarizing channel
is entanglement breaking for $p\geq2/3$, so that the Holevo information is
equal to the feedback-assisted capacity for these values \cite{BN05,DingW18}.}%
\label{fig:depola-plot}%
\end{figure}

Here we consider the point-to-point depolarizing channel, defined as%
\begin{align}
\mathcal{D}^{p}(X)  &  \coloneqq \left(  1-p\right)  X+p\operatorname{Tr}%
[X]\pi,\label{eq:dep-ch-example}\\
\pi &  \coloneqq I/d.
\end{align}
It was already established in \cite{Fang2019a} that $\Upsilon(\mathcal{D}%
^{p})$ is an upper bound on its (unassisted) classical capacity, and the
Holevo information is equal to its classical capacity \cite{King03}. Our
contribution here is that $\Upsilon(\mathcal{D}^{p})$ is an upper bound on its
classical capacity assisted by a classical feedback channel.
Figure~\ref{fig:depola-plot}\ plots this upper bound and also plots the Holevo
information lower bound when $d=2$. The latter is given by $1-h_{2}(p/2)$,
where $h_{2}$ is the binary entropy function. Note that the depolarizing
channel is entanglement breaking for $p\geq\frac{d}{d+1}$. As such, the bounds
from \cite{BN05,DingW18} apply, so that, for $p\geq\frac{d}{d+1}$, the Holevo
information $1-h_{2}(p/2)$ is equal to the classical capacity assisted by
classical feedback.

\subsection{Point-to-point erasure channel}

The point-to-point quantum erasure channel is defined for $p\in[0,1]$ and integer $d\geq 2$ as \cite{GBP97}
\begin{equation}
\mathcal{E}_{p,d}(\rho) \coloneqq (1-p) \rho + p |e\rangle\!\langle e|,
\end{equation}
where $|e\rangle\!\langle e|$ is a quantum erasure symbol, orthogonal to every $d$-dimensional input state $\rho$, so that the channel output has dimension $d+1$. This channel is covariant, as defined just after \eqref{eq:cov-ch-def}. Thus, by combining \cite[Lemma~12]{WFT18} with Corollary~\ref{cor:strong-converse-covariant}, we conclude that the strong converse holds for the classical capacity of the erasure channel $\mathcal{E}_{p,d}$ assisted by classical feedback; i.e.,
\begin{equation}
C_{\leftarrow}(\mathcal{E}_{p,d}(\rho))= \widetilde{C}_{\leftarrow}(\mathcal{E}_{p,d}(\rho)) = (1-p) \log_2 d.
\end{equation}

\subsection{Other point-to-point channels}

We note here that the reader can consult  \cite[Section~6.4]{Fang2019a} to find other examples of channels for which the $\Upsilon$-information has been calculated, including dephrasure and generalized amplitude damping channels. In all cases, our results strengthen those findings, because our results imply that these quantities are upper bounds on the classical-feedback-assisted classical capacity, rather than just the unassisted classical capacity.

\section{Conclusion}

\label{sec:conclusion}

In this paper, we established several measures of
classical communication and proved that they are useful as upper bounds on the
classical capacity of bipartite quantum channels. We did so by establishing
several key properties of these measures, which played essential roles in the
upper bound proofs. One of the most critical properties is that the measures
are subadditive under serial composition of bipartite channels, which is a
property that is useful in the analysis of feedback-assisted protocols. One
important application of our results is improved upper bounds on the classical
capacity of a quantum channel assisted by classical feedback, which is a
problem that has been analyzed in the literature for some time now
\cite{BN05,BDSS06,DingW18,GMW18,DQSW19}.

Going forward from here, an open question is whether our bounds could be
improved in any way. The recent techniques of \cite{FawziFawzi20} might be
helpful in obtaining refined non-asymptotic bounds, but the main result of \cite{BSD21} implies that the sharp R\'enyi divergence of \cite{FawziFawzi20} will not be helpful in the asymptotic case. As a key example, we wonder whether classical feedback could
increase the classical capacity of the depolarizing channel. We also wonder
whether our new bounds on the classical capacity assisted by classical
feedback generally improve upon the entropy bound from \cite{DQSW19}.

\bigskip

\textit{Acknowledgements}---We thank Andreas Winter for helpful discussions
about this project and about \cite{GMW18}, and we thank Masahito Hayashi for notifying us about \cite{FN98}.
We are grateful to Bjarne Bergh, Nilanjana Datta, and Robert Salzmann for pointing out an issue with the proof of Lemma~10 of \cite{CMW14}, as well as Milan Mosonyi for discussions about this issue. We additionally thank Milan Mosonyi for extensive editorial feedback on and corrections for our paper, which helped improve it, as well as for the observation in Remark~\ref{rem:milan}.  We also thank Bjarne Bergh for a clarification about the proof of Lemma~8 of \cite{TCR09}. 
DD and MMW thank God for all of His provisions.
SK\ acknowledges support from NSERC
and the NSF\ and under Grant No.~1714215. YQ acknowledges support from a
Stanford QFARM fellowship and from an NUS Overseas Graduate Scholarship. PWS
is supported by the National Science Foundation under Grant No.~CCF-1729369
and through the NSF Science and Technology Center for Science of Information
under Grant No.~CCF-0939370. MMW\ acknowledges support from NSF\ Grant
No.~1907615, as well as Stanford QFARM and AFOSR (FA9550-19-1-0369).

\bibliographystyle{IEEEtran}
\bibliography{Ref}

\appendices

\section{Alternative formulation of measure of forward classical communication
for a bipartite channel}

\label{app:cp-formulation-beta}In this appendix, we prove the equality in
\eqref{eq:CP-exp-beta}, and we also provide the background needed to
understand it. We also provide an alternate proof of
Proposition~\ref{prop:subadditivity-bipartite-CP-maps}, in order to showcase
the utility of the expression in \eqref{eq:CP-exp-beta}.

By definition, $\mathcal{P}_{A\rightarrow B}$ is Hermiticity preserving if
$\mathcal{P}_{A\rightarrow B}(X_{A})$ is Hermitian for every Hermitian $X_{A}$.

A linear map $\mathcal{P}_{A\rightarrow B}$ is Hermiticity preserving if and
only if its Choi operator is Hermitian. Suppose that the Choi operator
$\Gamma_{RB}^{\mathcal{P}}$\ is Hermitian. Then by the standard construction,%
\begin{equation}
\mathcal{P}_{A\rightarrow B}(X_{A})=\langle\Gamma|_{AR}X_{A}\otimes\Gamma
_{RB}^{\mathcal{P}}|\Gamma\rangle_{AR}.
\end{equation}
Since $\Gamma_{RB}^{\mathcal{P}}$ is Hermitian and $X_{A}$ is also, it follows
that $\mathcal{P}_{A\rightarrow B}(X_{A})$ is Hermitian. Now suppose that
$\mathcal{P}_{A\rightarrow B}$ is Hermiticity preserving. Then%
\begin{multline}
(\Gamma_{RB}^{\mathcal{P}})^{\dag}=(\mathcal{P}_{A\rightarrow B}(\Gamma
_{RA}))^{\dag}=\mathcal{P}_{A\rightarrow B}(\Gamma_{RA}^{\dag})\\
=\mathcal{P}_{A\rightarrow B}(\Gamma_{RA})=\Gamma_{RB}^{\mathcal{P}}.
\end{multline}

To every Hermitian operator $R_{AA^{\prime}BB^{\prime}}$, there is an
associated Hermiticity-preserving map, defined as%
\begin{multline}
\mathcal{R}_{AB\rightarrow A^{\prime}B^{\prime}}(X_{AB})=\\
(\langle\Gamma|_{A\hat{A}}\otimes\langle\Gamma|_{B\hat{B}})(X_{AB}\otimes
R_{\hat{A}A^{\prime}\hat{B}B^{\prime}})(|\Gamma\rangle_{A\hat{A}}%
\otimes|\Gamma\rangle_{B\hat{B}}).
\end{multline}
Consider that%
\begin{align}
&  \mathcal{R}_{AB\rightarrow A^{\prime}B^{\prime}}(X_{AB})\nonumber\\
&  =(\langle\Gamma|_{A\hat{A}}\otimes\langle\Gamma|_{B\hat{B}})(X_{AB}\otimes
R_{\hat{A}A^{\prime}\hat{B}B^{\prime}})(|\Gamma\rangle_{A\hat{A}}%
\otimes|\Gamma\rangle_{B\hat{B}})\\
&  =(\langle\Gamma|_{A\hat{A}}\otimes\langle\Gamma|_{B\hat{B}})( T_{\hat
{A}\hat{B}}(X_{\hat{A}\hat{B}})R_{\hat{A}A^{\prime}\hat{B}B^{\prime}}%
)(|\Gamma\rangle_{A\hat{A}}\otimes|\Gamma\rangle_{B\hat{B}})\\
&  =\operatorname{Tr}_{\hat{A}\hat{B}}[T_{\hat{A}\hat{B}}(X_{\hat{A}\hat{B}%
})R_{\hat{A}A^{\prime}\hat{B}B^{\prime}}].
\end{align}
Also, if $R_{AA^{\prime}BB^{\prime}}\geq0$, then $\mathcal{R}_{AB\rightarrow
A^{\prime}B^{\prime}}$ is completely positive.\ We also make the abbreviation%
\begin{equation}
\mathcal{R}_{AB\rightarrow A^{\prime}B^{\prime}}\geq0\qquad\Leftrightarrow
\qquad\mathcal{R}_{AB\rightarrow A^{\prime}B^{\prime}}\in\text{CP}.
\end{equation}

Then consider that, for positive semi-definite $R_{AA^{\prime}BB^{\prime}}$,
\begin{align}
&  \left\Vert \operatorname{Tr}_{A^{\prime}B^{\prime}}[R_{AA^{\prime
}BB^{\prime}}]\right\Vert _{\infty}\nonumber\\
&  =\sup_{\rho_{AB}\geq0,\operatorname{Tr}[\rho_{AB}]=1}\operatorname{Tr}%
[\rho_{AB}\operatorname{Tr}_{A^{\prime}B^{\prime}}[R_{AA^{\prime}BB^{\prime}%
}]]\\
&  =\sup_{\rho_{AB}\geq0,\operatorname{Tr}[\rho_{AB}]=1}\operatorname{Tr}%
[(\rho_{AB}\otimes I_{A^{\prime}B^{\prime}})R_{AA^{\prime}BB^{\prime}}]\\
&  =\sup_{\rho_{AB}\geq0,\operatorname{Tr}[\rho_{AB}]=1}\operatorname{Tr}%
[\mathcal{R}_{AB\rightarrow A^{\prime}B^{\prime}}(\rho_{AB})]\\
&  =:\left\Vert \mathcal{R}_{AB\rightarrow A^{\prime}B^{\prime}}\right\Vert
_{1}.
\end{align}

Thus, the function $\beta(\mathcal{M}_{AB\rightarrow A^{\prime}B^{\prime}})$
for a completely positive map $\mathcal{M}_{AB\rightarrow A^{\prime}B^{\prime
}}$ can be written as%
\begin{equation}
\beta(\mathcal{M}_{AB\rightarrow A^{\prime}B^{\prime}}%
)=\label{eq:alt-beta-quantity}\\
\inf_{\substack{\mathcal{S}_{AB\rightarrow A^{\prime}B^{\prime}}%
,\\\mathcal{V}_{AB\rightarrow A^{\prime}B^{\prime}}\in\text{HermP}}}
\left\Vert \mathcal{S}_{AB\rightarrow A^{\prime}B^{\prime}}\right\Vert _{1}%
\end{equation}
subject to
\begin{equation}%
\begin{array}
[c]{c}%
\\
T_{B^{\prime}}\circ(\mathcal{V}_{AB\rightarrow A^{\prime}B^{\prime}}%
\pm\mathcal{M}_{AB\rightarrow A^{\prime}B^{\prime}})\circ T_{B}\geq0,\\
\mathcal{S}_{AB\rightarrow A^{\prime}B^{\prime}}\pm\mathcal{V}_{AB\rightarrow
A^{\prime}B^{\prime}}\geq0,\\
\operatorname{Tr}_{A^{\prime}}\circ\mathcal{S}_{AB\rightarrow A^{\prime
}B^{\prime}}=\operatorname{Tr}_{A^{\prime}}\circ\mathcal{S}_{AB\rightarrow
A^{\prime}B^{\prime}}\circ\mathcal{R}_{A}^{\pi}%
\end{array}
,
\end{equation}
where $\mathcal{R}_{A}^{\pi}(\cdot)\coloneqq \operatorname{Tr}_{A}[\cdot
]\pi_{A}$ and $\pi_{A}\coloneqq I_{A}/d_{A}$ is the maximally mixed
state.\ Note that $\mathcal{S}_{AB\rightarrow A^{\prime}B^{\prime}}\geq0$
follows because $\mathcal{S}_{AB\rightarrow A^{\prime}B^{\prime}}%
\pm\mathcal{V}_{AB\rightarrow A^{\prime}B^{\prime}}\geq0$ and adding these
allows us to conclude that $\mathcal{S}_{AB\rightarrow A^{\prime}B^{\prime}%
}\geq0$.

For a point-to-point channel $\mathcal{N}_{A\rightarrow B^{\prime}}$, this
translates to%
\begin{align}
&  \beta(\mathcal{N}_{A\rightarrow B^{\prime}})\nonumber\\
&  \coloneqq \inf_{\substack{\mathcal{S}_{A\rightarrow B^{\prime}%
},\\\mathcal{V}_{A\rightarrow B^{\prime}}\in\text{HermP}}}\left\{
\begin{array}
[c]{c}%
\left\Vert \mathcal{S}_{A\rightarrow B^{\prime}}\right\Vert _{1}:\\
T_{B^{\prime}}\circ(\mathcal{V}_{A\rightarrow B^{\prime}}\pm\mathcal{M}%
_{A\rightarrow B^{\prime}})\geq0,\\
\mathcal{S}_{A\rightarrow B^{\prime}}\pm\mathcal{V}_{A\rightarrow B^{\prime}%
}\geq0,\\
\mathcal{S}_{A\rightarrow B^{\prime}}=\mathcal{S}_{A\rightarrow B^{\prime}%
}\circ\mathcal{R}_{A}^{\pi}%
\end{array}
\right\} \\
&  =\inf_{\substack{S_{B^{\prime}}\in\text{PSD},\\\mathcal{V}_{A\rightarrow
B^{\prime}}\in\text{HermP}}}\left\{
\begin{array}
[c]{c}%
\left\Vert \mathcal{R}_{A\rightarrow B^{\prime}}^{S}\right\Vert _{1}:\\
T_{B^{\prime}}\circ(\mathcal{V}_{A\rightarrow B^{\prime}}\pm\mathcal{M}%
_{A\rightarrow B^{\prime}})\geq0,\\
\mathcal{R}_{A\rightarrow B^{\prime}}^{S}\pm\mathcal{V}_{A\rightarrow
B^{\prime}}\geq0
\end{array}
\right\} \\
&  =\inf_{\substack{S_{B^{\prime}}\in\text{PSD},\\\mathcal{V}_{A\rightarrow
B^{\prime}}\in\text{HermP}}}\left\{
\begin{array}
[c]{c}%
\operatorname{Tr}[S_{B^{\prime}}]:\\
T_{B^{\prime}}\circ(\mathcal{V}_{A\rightarrow B^{\prime}}\pm\mathcal{M}%
_{A\rightarrow B^{\prime}})\geq0,\\
\mathcal{R}_{A\rightarrow B^{\prime}}^{S}\pm\mathcal{V}_{A\rightarrow
B^{\prime}}\geq0
\end{array}
\right\}  ,
\end{align}
where $\mathcal{R}_{A\rightarrow B^{\prime}}^{S}(\cdot)=\operatorname{Tr}%
_{A}[\cdot]S_{B^{\prime}}$ is a replacer map. Thus,
\begin{multline}
\beta(\mathcal{N}_{A\rightarrow B^{\prime}})=\\
\inf_{\substack{S_{B^{\prime}}\in\text{PSD},\\\mathcal{V}_{A\rightarrow
B^{\prime}}\in\text{HermP}}}\left\{
\begin{array}
[c]{c}%
\operatorname{Tr}[S_{B^{\prime}}]:\\
T_{B^{\prime}}\circ(\mathcal{V}_{A\rightarrow B^{\prime}}\pm\mathcal{M}%
_{A\rightarrow B^{\prime}})\geq0,\\
\mathcal{R}_{A\rightarrow B^{\prime}}^{S}\pm\mathcal{V}_{A\rightarrow
B^{\prime}}\geq0
\end{array}
\right\} .
\end{multline}

\bigskip

This kind of formulation is general. For example, consider the following
SDP\ for the diamond norm:%
\begin{multline}
\frac{1}{2}\left\Vert \mathcal{N}-\mathcal{M}\right\Vert _{\diamond}=\\
\inf_{Z_{RB}\geq0}\left\{  \left\Vert \operatorname{Tr}_{B}[Z_{RB}]\right\Vert
_{\infty}:Z_{RB}\geq\Gamma_{RB}^{\mathcal{N}}-\Gamma_{RB}^{\mathcal{M}%
}\right\}  .
\end{multline}
Using the above rephrasing, we can rewrite this optimization as%
\begin{multline}
\frac{1}{2}\left\Vert \mathcal{N}-\mathcal{M}\right\Vert _{\diamond}=\\
\inf_{\mathcal{Z}_{A\rightarrow B}\in\text{CP}}\left\{  \left\Vert
\mathcal{Z}_{A\rightarrow B}\right\Vert _{1}:\mathcal{Z}_{A\rightarrow B}%
\geq\mathcal{N}_{A\rightarrow B}-\mathcal{M}_{A\rightarrow B}\right\}  .
\end{multline}

\bigskip

We can use the expression in \eqref{eq:alt-beta-quantity}\ to provide an
alternate proof of Proposition~\ref{prop:subadditivity-bipartite-CP-maps}:

\begin{proposition}
Let $\mathcal{M}_{AB\rightarrow A^{\prime}B^{\prime}}^{1}$ and $\mathcal{M}%
_{A^{\prime}B^{\prime}\rightarrow A^{\prime\prime}B^{\prime\prime}}^{2}$ be
completely positive maps. Then%
\begin{multline}
\beta(\mathcal{M}_{A^{\prime}B^{\prime}\rightarrow A^{\prime\prime}%
B^{\prime\prime}}^{2}\circ\mathcal{M}_{AB\rightarrow A^{\prime}B^{\prime}}%
^{1})\leq\\
\beta(\mathcal{M}_{A^{\prime}B^{\prime}\rightarrow A^{\prime\prime}%
B^{\prime\prime}}^{2})\cdot\beta(\mathcal{M}_{AB\rightarrow A^{\prime
}B^{\prime}}^{1}).
\end{multline}

\end{proposition}

\begin{IEEEproof}
Let $\mathcal{S}_{AB\rightarrow A^{\prime}B^{\prime}}^{1}$ and $\mathcal{V}%
_{AB\rightarrow A^{\prime}B^{\prime}}^{1}$ be Hermiticity preserving maps
satisfying the constraints in \eqref{eq:alt-beta-quantity} for $\mathcal{M}%
_{AB\rightarrow A^{\prime}B^{\prime}}^{1}$, and let $\mathcal{S}_{A^{\prime
}B^{\prime}\rightarrow A^{\prime\prime}B^{\prime\prime}}^{2}$ and
$\mathcal{V}_{A^{\prime}B^{\prime}\rightarrow A^{\prime\prime}B^{\prime\prime
}}^{2}$ be Hermiticity preserving maps satisfying the constraints in
\eqref{eq:alt-beta-quantity} for $\mathcal{M}_{A^{\prime}B^{\prime}\rightarrow
A^{\prime\prime}B^{\prime\prime}}^{2}$. Then pick%
\begin{align}
\mathcal{S}_{AB\rightarrow A^{\prime\prime}B^{\prime\prime}}^{3}  &
=\mathcal{S}_{A^{\prime}B^{\prime}\rightarrow A^{\prime\prime}B^{\prime\prime
}}^{2}\circ\mathcal{S}_{AB\rightarrow A^{\prime}B^{\prime}}^{1},\\
\mathcal{V}_{AB\rightarrow A^{\prime\prime}B^{\prime\prime}}^{3}  &
=\mathcal{V}_{A^{\prime}B^{\prime}\rightarrow A^{\prime\prime}B^{\prime\prime
}}^{2}\circ\mathcal{V}_{AB\rightarrow A^{\prime}B^{\prime}}^{1}.
\end{align}
Also, set%
\begin{equation}
\mathcal{M}_{AB\rightarrow A^{\prime\prime}B^{\prime\prime}}^{3}%
=\mathcal{M}_{A^{\prime}B^{\prime}\rightarrow A^{\prime\prime}B^{\prime\prime
}}^{2}\circ\mathcal{M}_{AB\rightarrow A^{\prime}B^{\prime}}^{1}%
\end{equation}
Then it follows that%
\begin{align}
T_{B^{\prime\prime}}\circ(\mathcal{V}_{AB\rightarrow A^{\prime\prime}%
B^{\prime\prime}}^{3}\pm\mathcal{M}_{AB\rightarrow A^{\prime\prime}%
B^{\prime\prime}}^{3})\circ T_{B}  &  \geq0,\\
\mathcal{S}_{AB\rightarrow A^{\prime\prime}B^{\prime\prime}}^{3}\pm
\mathcal{V}_{AB\rightarrow A^{\prime\prime}B^{\prime\prime}}^{3}  &  \geq0.
\end{align}
This follows from the general observation that if $\mathcal{A}\pm
\mathcal{B}\geq0$ and $\mathcal{C}\pm\mathcal{D}\geq0$, then $\mathcal{A\circ
C}\pm\mathcal{B\circ D}\geq0$. This in turn follows because%
\begin{equation}
\mathcal{A}+\mathcal{B}\geq0,\quad\mathcal{A}-\mathcal{B}\geq0,\quad
\mathcal{C}+\mathcal{D}\geq0,\quad\mathcal{C}-\mathcal{D}\geq0,
\label{eq:ABCD-1}
\end{equation}
implies that%
\begin{align}
0  &  \leq(\mathcal{A}+\mathcal{B})\circ(\mathcal{C}+\mathcal{D})\\
&  =\mathcal{A}\circ\mathcal{C}+\mathcal{A}\circ\mathcal{D}+\mathcal{B}%
\circ\mathcal{C}+\mathcal{B}\circ\mathcal{D},\\
0  &  \leq(\mathcal{A}+\mathcal{B})\circ(\mathcal{C}-\mathcal{D})\\
&  =\mathcal{A}\circ\mathcal{C}-\mathcal{A}\circ\mathcal{D}+\mathcal{B}%
\circ\mathcal{C}-\mathcal{B}\circ\mathcal{D},\\
0  &  \leq(\mathcal{A}-\mathcal{B})\circ(\mathcal{C}+\mathcal{D})\\
&  =\mathcal{A}\circ\mathcal{C}+\mathcal{A}\circ\mathcal{D}-\mathcal{B}%
\circ\mathcal{C}-\mathcal{B}\circ\mathcal{D},\\
0  &  \leq(\mathcal{A}-\mathcal{B})\circ(\mathcal{C}-\mathcal{D})\\
&  =\mathcal{A}\circ\mathcal{C}-\mathcal{A}\circ\mathcal{D}-\mathcal{B}%
\circ\mathcal{C}+\mathcal{B}\circ\mathcal{D}.
\label{eq:ABCD-last}
\end{align}
Now add the first and last to get $\mathcal{A\circ C}+\mathcal{B\circ D}\geq0$
and the second and third to get $\mathcal{A\circ C}-\mathcal{B\circ D}\geq0$.

Now consider that%
\begin{align}
&  \operatorname{Tr}_{A^{\prime\prime}}\circ\mathcal{S}_{AB\rightarrow
A^{\prime\prime}B^{\prime\prime}}^{3}\nonumber\\
&  =\operatorname{Tr}_{A^{\prime\prime}}\circ\mathcal{S}_{A^{\prime}B^{\prime
}\rightarrow A^{\prime\prime}B^{\prime\prime}}^{2}\circ\mathcal{S}%
_{AB\rightarrow A^{\prime}B^{\prime}}^{1}\\
&  =\operatorname{Tr}_{A^{\prime\prime}}\circ\mathcal{S}_{A^{\prime}B^{\prime
}\rightarrow A^{\prime\prime}B^{\prime\prime}}^{2}\circ\mathcal{R}_{A^{\prime
}}^{\pi}\circ\mathcal{S}_{AB\rightarrow A^{\prime}B^{\prime}}^{1}\\
&  =\operatorname{Tr}_{A^{\prime\prime}}\circ\mathcal{S}_{A^{\prime}B^{\prime
}\rightarrow A^{\prime\prime}B^{\prime\prime}}^{2}\circ\mathcal{P}_{A^{\prime
}}^{\pi}\circ\operatorname{Tr}_{A^{\prime}}\circ\mathcal{S}_{AB\rightarrow
A^{\prime}B^{\prime}}^{1}\\
&  =\operatorname{Tr}_{A^{\prime\prime}}\circ\mathcal{S}_{A^{\prime}B^{\prime
}\rightarrow A^{\prime\prime}B^{\prime\prime}}^{2}\circ\mathcal{P}_{A^{\prime
}}^{\pi}\circ\operatorname{Tr}_{A^{\prime}}\circ\mathcal{S}_{AB\rightarrow
A^{\prime}B^{\prime}}^{1}\circ\mathcal{R}_{A}^{\pi}\\
&  =\operatorname{Tr}_{A^{\prime\prime}}\circ\mathcal{S}_{A^{\prime}B^{\prime
}\rightarrow A^{\prime\prime}B^{\prime\prime}}^{2}\circ\mathcal{R}_{A^{\prime
}}^{\pi}\circ\mathcal{S}_{AB\rightarrow A^{\prime}B^{\prime}}^{1}%
\circ\mathcal{R}_{A}^{\pi}.
\end{align}
Since the first two lines show that $\operatorname{Tr}_{A^{\prime\prime}}%
\circ\mathcal{S}_{AB\rightarrow A^{\prime\prime}B^{\prime\prime}}%
^{3}=\operatorname{Tr}_{A^{\prime\prime}}\circ\mathcal{S}_{A^{\prime}%
B^{\prime}\rightarrow A^{\prime\prime}B^{\prime\prime}}^{2}\circ
\mathcal{R}_{A^{\prime}}^{\pi}\circ\mathcal{S}_{AB\rightarrow A^{\prime
}B^{\prime}}^{1}$, we conclude that%
\begin{equation}
\operatorname{Tr}_{A^{\prime\prime}}\circ\mathcal{S}_{AB\rightarrow
A^{\prime\prime}B^{\prime\prime}}^{3}=\operatorname{Tr}_{A^{\prime\prime}%
}\circ\mathcal{S}_{AB\rightarrow A^{\prime\prime}B^{\prime\prime}}^{3}%
\circ\mathcal{R}_{A}^{\pi}.
\end{equation}
Finally, consider that%
\begin{align}
\left\Vert \mathcal{S}_{AB\rightarrow A^{\prime\prime}B^{\prime\prime}}%
^{3}\right\Vert _{1}  &  =\left\Vert \mathcal{S}_{A^{\prime}B^{\prime
}\rightarrow A^{\prime\prime}B^{\prime\prime}}^{2}\circ\mathcal{S}%
_{AB\rightarrow A^{\prime}B^{\prime}}^{1}\right\Vert _{1}\\
&  \leq\left\Vert \mathcal{S}_{A^{\prime}B^{\prime}\rightarrow A^{\prime
\prime}B^{\prime\prime}}^{2}\right\Vert _{1}\cdot\left\Vert \mathcal{S}%
_{AB\rightarrow A^{\prime}B^{\prime}}^{1}\right\Vert _{1}.
\end{align}
The inequality follows because the trace norm on superoperators is
submultiplicative. So $\mathcal{S}_{AB\rightarrow A^{\prime\prime}%
B^{\prime\prime}}^{3}$ and $\mathcal{V}_{AB\rightarrow A^{\prime\prime
}B^{\prime\prime}}^{3}$ satisfy the constraints in
\eqref{eq:alt-beta-quantity} for $\mathcal{M}_{AB\rightarrow A^{\prime\prime
}B^{\prime\prime}}^{3}$, so we conclude that%
\begin{equation}
\beta(\mathcal{M}_{AB\rightarrow A^{\prime\prime}B^{\prime\prime}}^{3}%
)\leq\left\Vert \mathcal{S}_{A^{\prime}B^{\prime}\rightarrow A^{\prime\prime
}B^{\prime\prime}}^{2}\right\Vert _{1}\cdot\left\Vert \mathcal{S}%
_{AB\rightarrow A^{\prime}B^{\prime}}^{1}\right\Vert _{1}.
\end{equation}
Since the argument holds for all $\mathcal{S}_{AB\rightarrow A^{\prime
}B^{\prime}}^{1}$ and $\mathcal{V}_{AB\rightarrow A^{\prime}B^{\prime}}^{1}$
satisfying the constraints in \eqref{eq:alt-beta-quantity} for $\mathcal{M}%
_{AB\rightarrow A^{\prime}B^{\prime}}^{1}$, and for all $\mathcal{S}%
_{A^{\prime}B^{\prime}\rightarrow A^{\prime\prime}B^{\prime\prime}}^{2}$ and
$\mathcal{V}_{A^{\prime}B^{\prime}\rightarrow A^{\prime\prime}B^{\prime\prime
}}^{2}$ satisfying the constraints in \eqref{eq:alt-beta-quantity} for
$\mathcal{M}_{A^{\prime}B^{\prime}\rightarrow A^{\prime\prime}B^{\prime\prime
}}^{2}$, we conclude the statement of the proposition.
\end{IEEEproof}

\section{The $\alpha \to 1$ limit of R\'enyi channel divergences}

\label{app:renyi-ch-div-limits}

The following lemma was claimed in \cite{CMW14}, but the proof there is not
correct. Here, for completeness, we provide a proof, and we note that a
different proof has been derived as well \cite{Mos21}.

\begin{lemma}
\label{lem:sand-petz-renyi-to-qre-ch}Let $\mathcal{N}_{A\rightarrow B}$ be a
quantum channel, and let $\mathcal{M}_{A\rightarrow B}$ be a completely
positive map. The following limits hold%
\begin{align}
\lim_{\alpha\rightarrow1}\widetilde{D}_{\alpha}(\mathcal{N}\Vert\mathcal{M})
&  =D(\mathcal{N}\Vert\mathcal{M}),\label{eq:limit-sand-ch-a-1}\\
\lim_{\alpha\rightarrow1}D_{\alpha}(\mathcal{N}\Vert\mathcal{M})  &
=D(\mathcal{N}\Vert\mathcal{M}), \label{eq:limit-petz-ch-a-1}%
\end{align}
where $\widetilde{D}_{\alpha}(\mathcal{N}\Vert\mathcal{M})$ is the sandwiched
R\'enyi channel divergence, $D_{\alpha}(\mathcal{N}\Vert\mathcal{M})$ is the
Petz--R\'enyi channel divergence, and $D(\mathcal{N}\Vert\mathcal{M})$ is the
channel relative entropy. Specifically,%
\begin{align}
\widetilde{D}_{\alpha}(\mathcal{N}\Vert\mathcal{M})  &  \coloneqq \sup
_{\rho_{RA}}\widetilde{D}_{\alpha}(\mathcal{N}_{A\rightarrow B}(\rho
_{RA})\Vert\mathcal{M}_{A\rightarrow B}(\rho_{RA})),\notag \\
D_{\alpha}(\mathcal{N}\Vert\mathcal{M})  &  \coloneqq \sup_{\rho_{RA}%
}\widetilde{D}_{\alpha}(\mathcal{N}_{A\rightarrow B}(\rho_{RA})\Vert
\mathcal{M}_{A\rightarrow B}(\rho_{RA})),\notag \\
D(\mathcal{N}\Vert\mathcal{M})  &  \coloneqq \sup_{\rho_{RA}}D(\mathcal{N}%
_{A\rightarrow B}(\rho_{RA})\Vert\mathcal{M}_{A\rightarrow B}(\rho_{RA})),
\end{align}
where the optimizations are over every state $\rho_{RA}$, with the reference
system $R$ arbitrarily large. The sandwiched R\'enyi relative entropy $\widetilde{D}_{\alpha}(\rho\Vert\sigma)$ is defined in \eqref{eq:sandwiched-Renyi-def} and the quantum relative entropy $D(\rho\Vert\sigma)$ in \eqref{eq:q-rel-entr}. The Petz--R\'enyi relative entropy is defined for all $\alpha \in (0,1)\cup (1,\infty)$ as \cite{P85,P86}
\begin{equation}
D_{\alpha}(\rho\Vert\sigma)    \coloneqq \frac{1}{\alpha-1}\log
_{2}\operatorname{Tr}[\rho^{\alpha}\sigma^{1-\alpha}]
\end{equation}
if $\alpha \in (0,1)$ or, $\alpha \in (1,\infty)$ and $\operatorname{supp}(\rho)\subseteq \operatorname{supp}(\sigma)$. Otherwise, we define $D_{\alpha}(\rho\Vert\sigma) = +\infty$.

\end{lemma}

\begin{IEEEproof}
We first prove \eqref{eq:limit-sand-ch-a-1} and then argue that similar reasoning establishes \eqref{eq:limit-petz-ch-a-1}.

If there exists a state $\rho_{RA}$ such that $\operatorname{supp}%
(\mathcal{N}_{A\rightarrow B}(\rho_{RA}))\not \subseteq \operatorname{supp}%
(\mathcal{M}_{A\rightarrow B}(\rho_{RA}))$, then it follows that the limit on
the left-hand side of \eqref{eq:limit-sand-ch-a-1}\ and the quantity on the
right are both equal to $+\infty$. The same is true for \eqref{eq:limit-petz-ch-a-1}.

So let us instead consider the case when $\operatorname{supp}(\mathcal{N}%
_{A\rightarrow B}(\rho_{RA}))\subseteq\operatorname{supp}(\mathcal{M}%
_{A\rightarrow B}(\rho_{RA}))$ for every state $\rho_{RA}$, which is
equivalent to the single condition $\operatorname{supp}(\Gamma_{RB}%
^{\mathcal{N}})\subseteq\operatorname{supp}(\Gamma_{RB}^{\mathcal{M}})$, where
$\Gamma_{RB}^{\mathcal{N}}$ and $\Gamma_{RB}^{\mathcal{M}}$ are the Choi
operators of $\mathcal{N}_{A\rightarrow B}$ and $\mathcal{M}_{A\rightarrow B}%
$, respectively. If this is the case, then it follows that%
\begin{equation}
D_{\max}(\mathcal{N}\Vert\mathcal{M})<\infty,
\end{equation}
where%
\begin{align}
& D_{\max}(\mathcal{N}\Vert\mathcal{M})  \notag \\
&  \coloneqq \sup_{\rho_{RA}}D_{\max
}(\mathcal{N}_{A\rightarrow B}(\rho_{RA})\Vert\mathcal{M}_{A\rightarrow
B}(\rho_{RA}))\\
&  =D_{\max}(\Gamma^{\mathcal{N}}\Vert\Gamma^{\mathcal{M}}),
\end{align}
with the latter equality established in \cite{WBHK20}.

First, we show the following equality, by a straightforward argument:
\begin{equation}
\lim_{\alpha\rightarrow1^{-}}\widetilde{D}_{\alpha}(\mathcal{N}\Vert
\mathcal{M})= D(\mathcal{N}\Vert\mathcal{M}) \label{eq:renyi-ch-lower-to-1}.
\end{equation}
Indeed, consider that
\begin{align}
&  \lim_{\alpha\rightarrow1^{-}}\widetilde{D}_{\alpha}(\mathcal{N}%
\Vert\mathcal{M})\nonumber\\
&  =\sup_{\alpha\in(0,1)}\widetilde{D}_{\alpha}(\mathcal{N}\Vert\mathcal{M}%
)\label{eq:easy-chain-a-1}\\
&  =\sup_{\alpha\in(0,1)}\sup_{\rho_{RA}}\widetilde{D}_{\alpha}(\mathcal{N}%
_{A\rightarrow B}(\rho_{RA})\Vert\mathcal{M}_{A\rightarrow B}(\rho_{RA}))\\
&  =\sup_{\rho_{RA}}\sup_{\alpha\in(0,1)}\widetilde{D}_{\alpha}(\mathcal{N}%
_{A\rightarrow B}(\rho_{RA})\Vert\mathcal{M}_{A\rightarrow B}(\rho_{RA}))\\
&  =\sup_{\rho_{RA}}D(\mathcal{N}_{A\rightarrow B}(\rho_{RA})\Vert
\mathcal{M}_{A\rightarrow B}(\rho_{RA}))\\
&  =D(\mathcal{N}\Vert\mathcal{M}). \label{eq:easy-chain-a-last}%
\end{align}
The first equality is a consequence of the $\alpha$-monotonicity of~$\widetilde{D}_{\alpha}$. The fourth equality is a consequence of \eqref{eq:limit-SW-to-1} and the $\alpha$-monotonicity of $\widetilde{D}_{\alpha}$.

The following inequality%
\begin{equation}
\lim_{\alpha\rightarrow1^{+}}\widetilde{D}_{\alpha}(\mathcal{N}\Vert
\mathcal{M})\geq D(\mathcal{N}\Vert\mathcal{M}) \label{eq:renyi-ch-lower}%
\end{equation}
is straightforward, being a consequence of monotonicity in $\alpha$ of the sandwiched
R\'enyi relative entropies \cite{MDSFT13}, as well as the $\alpha\rightarrow1$
limit \cite{MDSFT13,WWY14}. To see it, let $\rho_{RA}$ be an arbitrary state.
Then it follows from $\alpha$-monotonicity that the following inequality holds
for all $\alpha>1$:%
\begin{multline}
\widetilde{D}_{\alpha}(\mathcal{N}_{A\rightarrow B}(\rho_{RA})\Vert
\mathcal{M}_{A\rightarrow B}(\rho_{RA}))\geq\\
D(\mathcal{N}_{A\rightarrow B}(\rho_{RA})\Vert\mathcal{M}_{A\rightarrow
B}(\rho_{RA})).
\end{multline}
Then%
\begin{align}
&  \lim_{\alpha\rightarrow1^{+}}\widetilde{D}_{\alpha}(\mathcal{N}%
\Vert\mathcal{M})\nonumber\\
&  =\inf_{\alpha>1}\widetilde{D}_{\alpha}(\mathcal{N}\Vert\mathcal{M}%
)\label{eq:easy-chain-1}\\
&  =\inf_{\alpha>1}\sup_{\rho_{RA}}\widetilde{D}_{\alpha}(\mathcal{N}%
_{A\rightarrow B}(\rho_{RA})\Vert\mathcal{M}_{A\rightarrow B}(\rho_{RA}))\\
&  \geq\sup_{\rho_{RA}}\inf_{\alpha>1}\widetilde{D}_{\alpha}(\mathcal{N}%
_{A\rightarrow B}(\rho_{RA})\Vert\mathcal{M}_{A\rightarrow B}(\rho_{RA}))\\
&  \geq\sup_{\rho_{RA}}D(\mathcal{N}_{A\rightarrow B}(\rho_{RA})\Vert
\mathcal{M}_{A\rightarrow B}(\rho_{RA}))\\
&  =D(\mathcal{N}\Vert\mathcal{M}). \label{eq:easy-chain-last}%
\end{align}
The first equality is a consequence of $\alpha$-monotonicity.

Let us then establish the opposite inequality. Let $\rho$ be a state and $\sigma$ a positive semi-definite operator. Recall the following bound from
Lemma~8 of \cite{TCR09} (see also Lemma~6.3 of \cite{T12}):%
\begin{equation}
D_{1+\delta}(\rho\Vert\sigma)\leq D(\rho\Vert\sigma)+4\delta\left[  \log
_{2}\nu(\rho,\sigma)\right]  ^{2}, \label{eq:AEP-ineq-petz-umeg}%
\end{equation}
which holds when $\operatorname{supp}(\rho)\subseteq \operatorname{supp}(\sigma)$  and for $\delta\in\left(  0,\frac{\ln3}{4\ln\nu(\rho,\sigma)}\right)
$, where%
\begin{align}
\nu(\rho,\sigma)  &  \coloneqq \operatorname{Tr}[\rho^{\frac{3}{2}}%
\sigma^{-\frac{1}{2}}]+\operatorname{Tr}[\rho^{\frac{1}{2}}\sigma^{\frac{1}%
{2}}]+1\\
&  =2^{\frac{1}{2}D_{\frac{3}{2}}(\rho\Vert\sigma)}+2^{-\frac{1}{2}D_{\frac{1}{2}}(\rho\Vert\sigma)}+1.
\end{align}
Note that $\nu(\rho, \sigma) \geq 3$ (as argued just after \cite[Eq.~(22)]{TCR09}), as well as
\begin{equation}
\nu(\rho, \sigma) \leq 2^{\frac{1}{2}D_{\max}(\rho\Vert\sigma)}+\sqrt{\operatorname{Tr}[\sigma]}+1,
\label{eq:nu-up-bnd}
\end{equation}
which follows because $D_{\frac{3}{2}}(\rho\Vert\sigma)\leq
D_{2}(\rho\Vert\sigma)\leq D_{\max}(\rho\Vert\sigma)$, which in turn follows
from the $\alpha$-monotonicity of the Petz--R\'enyi relative entropy and the
latter inequality was proven in \cite[Lemma~7]{BD10}.
Also, we applied the Cauchy--Schwarz inequality to conclude that $\operatorname{Tr}
[\rho^{\frac{1}{2}}\sigma^{\frac{1}{2}}]\leq \sqrt{\operatorname{Tr}[\sigma]}$.
 From the fact that
$\widetilde{D}_{1+\delta}(\rho\Vert\sigma)\leq D_{1+\delta}(\rho\Vert\sigma)$
for all $\delta>0$ \cite{WWY14}, we conclude that%
\begin{equation}
\widetilde{D}_{1+\delta}(\rho\Vert\sigma)\leq
 D(\rho\Vert\sigma)+4\delta
\left[  \log_{2}\!\left(  \nu(\rho,\sigma)\right)
\right]  ^{2}.
\end{equation}
By picking $\delta\in\left(  0,c\right)  $, where
\begin{equation}
c\coloneqq \frac{\ln3}{4  \ln  \nu(\mathcal{N}, \mathcal{M})    },
\end{equation}
and
\begin{equation}
\nu(\mathcal{N}, \mathcal{M}) \coloneqq \sup_{\rho_{RA}}\nu(\mathcal{N}_{A\to B}(\rho_{RA}), \mathcal{M}_{A\to B}(\rho_{RA})),
\end{equation}
with the optimization over every state $\rho_{RA}$,
we find that the following inequality holds for every input state~$\rho_{RA}$:%
\begin{multline}
\widetilde{D}_{1+\delta}(\mathcal{N}_{A\rightarrow B}(\rho_{RA})\Vert
\mathcal{M}_{A\rightarrow B}(\rho_{RA}))\leq\\
D(\mathcal{N}_{A\rightarrow B}(\rho_{RA})\Vert\mathcal{M}_{A\rightarrow
B}(\rho_{RA}))\\
+4\delta\left[  \log_{2}  \nu(\mathcal{N}_{A\to B}(\rho_{RA}), \mathcal{M}_{A\to B}(\rho_{RA}))  \right]  ^{2}.
\end{multline}
Note that $\nu(\mathcal{N}, \mathcal{M}) < \infty$ because $D_{\max}(\mathcal{N}\Vert\mathcal{M}) < \infty$. Indeed, from \eqref{eq:nu-up-bnd}, we conclude that
\begin{equation}
\nu(\mathcal{N}, \mathcal{M}) \leq 2^{\frac{1}{2} D_{\max}(\mathcal{N}\Vert \mathcal{M})} + \sqrt{\left\|\mathcal{M}\right\|_{\diamond}} + 1,
\end{equation}
where the diamond norm of $\mathcal{M}$ is defined as \cite{Kit97}
\begin{equation}
\left\|\mathcal{M}\right\|_{\diamond} \coloneqq \sup_{\rho_{RA}\geq 0, \operatorname{Tr}[\rho_{RA}]=1}\left\|\mathcal{M}_{A\to B}(\rho_{RA})\right\|_{1} .
\end{equation}
(We could also set
\begin{equation}
c\coloneqq \frac{\ln 3}{  4 \ln(2^{\frac{1}{2} D_{\max}(\mathcal{N}\Vert \mathcal{M})} + \sqrt{\left\|\mathcal{M}\right\|_{\diamond}} + 1)}
\end{equation}
 if desired, and we note here that an advantage of doing so is that both $D_{\max}(\mathcal{N}\Vert \mathcal{M})$ and $\left\|\mathcal{M}\right\|_{\diamond}$ are efficiently computable by semi-definite programming.)
Now taking a supremum over every input state $\rho_{RA}$, we conclude that%
\begin{equation}
\widetilde{D}_{1+\delta}(\mathcal{N}\Vert\mathcal{M})\leq 
D(\mathcal{N}%
\Vert\mathcal{M})+4\delta\left[  \log_{2}  \nu(\mathcal{N}, \mathcal{M})  \right]  ^{2}.
\label{eq:unif-bnd-sand-ch-to-umeg}
\end{equation}
Thus, by taking the limit of \eqref{eq:unif-bnd-sand-ch-to-umeg} as $\delta \to 0$, we conclude that%
\begin{equation}
\lim_{\alpha\rightarrow1^{+}}\widetilde{D}_{\alpha}(\mathcal{N}\Vert
\mathcal{M})\leq D(\mathcal{N}\Vert\mathcal{M}). \label{eq:renyi-ch-upper}%
\end{equation}
Putting together \eqref{eq:renyi-ch-lower} and \eqref{eq:renyi-ch-upper}, we
conclude \eqref{eq:limit-sand-ch-a-1}.

A proof of \eqref{eq:limit-petz-ch-a-1}
follows exactly the same approach, but we finally use \eqref{eq:AEP-ineq-petz-umeg}
directly instead and similar reasoning as above to establish that $\lim_{\alpha\rightarrow1^{+}} D_{\alpha}(\mathcal{N}\Vert
\mathcal{M})\leq D(\mathcal{N}\Vert\mathcal{M})$.
\end{IEEEproof}

\bigskip 

Now we discuss how to generalize this development to the geometric R\'enyi and
Belavkin--Staszewski relative entropies. We first begin with the following
simple extension of Lemma~8 of \cite{TCR09}:

\begin{lemma}
\label{lem:geo-renyi-to-bs-uniform}Let $\rho$ be a quantum state and $\sigma$
a positive semi-definite operator. Then%
\begin{equation}
\widehat{D}_{1+\delta}(\rho\Vert\sigma)\leq\widehat{D}(\rho\Vert
\sigma)+4\delta\left[  \log_{2}\widehat{\upsilon}(\rho,\sigma)\right]  ^{2},
\label{eq:geo-BS-a-plus-bnd}%
\end{equation}
holds for $\delta\in\left(  0,\frac{\ln3}{4\ln\widehat{\upsilon}(\rho,\sigma
)}  \right)$, where%
\begin{equation}
\widehat{\upsilon}(\rho,\sigma)\coloneqq 2^{\frac{1}{2}\widehat{D}_{3/2}%
(\rho\Vert\sigma)}+2^{-\frac{1}{2}\widehat{D}_{1/2}(\rho\Vert\sigma)}+1 \geq 3.
\end{equation}

\end{lemma}

\begin{IEEEproof}
The proof below follows the proof of Lemma~8 of \cite{TCR09} quite closely,
with some slight differences to account for the different entropies involved.
We provide a detailed proof for completeness. First, suppose that
$\operatorname{supp}(\rho)\not \subseteq \operatorname{supp}(\sigma)$. Then
both the left-hand side and right-hand side of \eqref{eq:geo-BS-a-plus-bnd}
are equal to $+\infty$, so that there is nothing to prove in this case.

Now suppose that $\operatorname{supp}(\rho)\subseteq\operatorname{supp}%
(\sigma)$, which implies that we can restrict the development to the support
of $\sigma$, and on this space, $\sigma$ is invertible. Let us suppose furthermore for now that
$\operatorname{supp}(\rho)=\operatorname{supp}(\sigma)$, and then we
apply a limit at the end of the proof. As observed in \cite[Proposition~72]%
{KW20}, we can write%
\begin{equation}
\widehat{D}_{1+\delta}(\rho\Vert\sigma)=\frac{1}{\delta}\log_{2}\langle
\varphi^{\rho}|X^{\delta}|\varphi^{\rho}\rangle,
\end{equation}
where%
\begin{align}
X &  \coloneqq\rho^{\frac{1}{2}}\sigma^{-1}\rho^{\frac{1}{2}}\otimes I,\\
|\varphi^{\rho}\rangle &  \coloneqq(  \rho^{\frac{1}{2}}\otimes I)
|\Gamma\rangle,
\end{align}
and $|\Gamma\rangle$ is defined in \eqref{eq:gamma-def-unnorm}.
Then consider that%
\begin{equation}
\frac{1}{\delta}\log_{2}\langle\varphi^{\rho}|X^{\delta}|\varphi^{\rho}%
\rangle\leq\frac{1}{\delta\ln2}\left(  \langle\varphi^{\rho}|X^{\delta
}|\varphi^{\rho}\rangle-1\right)  ,
\end{equation}
where we have applied the inequality $\ln x\leq x-1$, which holds for all
$x>0$. Now expand $X^{\delta}$ as%
\begin{equation}
X^{\delta}=I+\delta\ln X+r_{\delta}(X),
\end{equation}
where $r_{\delta}(X)\coloneqq X^{\delta}-\delta\ln X-I$. Then it follows that%
\begin{align}
& \widehat{D}_{1+\delta}(\rho\Vert\sigma) \notag \\
&  \leq\frac{1}{\delta\ln2}\left(
\delta\langle\varphi^{\rho}|\ln X|\varphi^{\rho}\rangle+\langle\varphi^{\rho
}|r_{\delta}(X)|\varphi^{\rho}\rangle\right)  \\
&  =\widehat{D}(\rho\Vert\sigma)+\frac{1}{\delta\ln2}\langle\varphi^{\rho
}|r_{\delta}(X)|\varphi^{\rho}\rangle.
\end{align}
Now, again applying the inequality $\ln x\leq x-1$ for $x>0$, consider that%
\begin{align}
r_{\delta}(x) &  =e^{\delta\ln x}-\delta\ln x-1\\
&  =e^{\delta\ln x}+\ln\!\left(  \frac{1}{x^{\delta}}\right)  -1\\
&  \leq e^{\delta\ln x}+\frac{1}{x^{\delta}}-2\\
&  =e^{\delta\ln x}+e^{-\delta\ln x}-2\\
&  =2\left(  \cosh(\delta\ln x)-1\right)  \\
&  =:s_{\delta}(x).
\end{align}
Since $\frac{\partial}{\partial x}s_{\delta}(x)=2\delta\sinh(\delta\ln x)/x$
and thus $\frac{\partial}{\partial x}s_{\delta}(x)\geq0$ for $x\geq1$, it
follows that $s_{\delta}(x)$ is monotonically increasing in $x$ for $x\geq1$.
Also, since $\frac{\partial^{2}}{\partial x^{2}}s_{\delta}(x)=2\delta
(\delta\cosh(\delta\ln x)-\sinh(\delta\ln x))/x^{2}$ and thus $\frac
{\partial^{2}}{\partial x^{2}}s_{\delta}(x)\leq0$ for all $\delta\leq1/2$ and
$x\geq3$, it follows that $s_{\delta}(x)$ is concave in $x$ for all
$\delta\leq1/2$ and $x\geq3$. Furthermore, we have that 
\begin{align}
s_{\delta}(x) & =s_{\delta}(1/x),
\label{eq:s-delta-prop-inv}\\
s_{\delta}(x^{2}) & =s_{2\delta}(x).
\label{eq:s-delta-prop-squaring}
\end{align}
 Then we find,
for all $x>0$, that%
\begin{align}
s_{\delta}(x) &  \leq s_{\delta}\!\left(  x+\frac{1}{x}+2\right)  \\
&  =s_{\delta}\!\left(  \left(  \sqrt{x}+\frac{1}{\sqrt{x}}\right)
^{2}\right)  \\
&  =s_{2\delta}\!\left(  \sqrt{x}+\frac{1}{\sqrt{x}}\right)  \\
&  \leq s_{2\delta}\!\left(  \sqrt{x}+\frac{1}{\sqrt{x}}+1\right)  .
\end{align}
The first inequality follows from monotonicity of $s_{\delta}(x)$ in $x$ for
$x\geq1$, as well as $s_{\delta}(x)=s_{\delta}(1/x)$. Indeed, for $x\geq1$, we
apply monotonicity to conclude that $s_{\delta}(x)\leq s_{\delta}\!\left(
x+\frac{1}{x}+2\right)  $. For $x\in(0,1)$, it follows that $1/x>1$, and so \eqref{eq:s-delta-prop-inv} and 
monotonicity imply that $s_{\delta}(x)=s_{\delta}(1/x)\leq s_{\delta
}\!\left(  x+\frac{1}{x}+2\right)  $. The second equality follows from
applying \eqref{eq:s-delta-prop-squaring}. The last inequality again follows
from the facts that $\sqrt{x}+\frac{1}{\sqrt{x}}\geq1$ for $x>0$ and from
applying monotonicity of $s_{\delta}(x)$ in $x$ for $x\geq1$. Now consider
that%
\begin{align}
&  \langle\varphi^{\rho}|r_{\delta}(X)|\varphi^{\rho}\rangle\nonumber\\
&  \leq\langle\varphi^{\rho}|s_{\delta}(X)|\varphi^{\rho}\rangle\\
&  \leq\langle\varphi^{\rho}|s_{2\delta}\!\left(  \sqrt{X}+\frac{1}{\sqrt{X}%
}+I\right)  |\varphi^{\rho}\rangle\\
&  \leq s_{2\delta}(\widehat{\upsilon}(\rho,\sigma)).
\end{align}
The first inequality follows because the scalar inequality $r_{\delta}(x)\leq
s_{\delta}(x)$ extends to the operator inequality $r_{\delta}(X)\leq
s_{\delta}(X)$, holding for all positive definite $X$. The second inequality
follows for a similar reason, but using the scalar inequality $s_{\delta
}(x)\leq s_{2\delta}\!\left(  \sqrt{x}+\frac{1}{\sqrt{x}}+1\right)  $. The
final inequality follows from Jensen's inequality (see \cite[Lemma~11]{TCR09}) and the fact that \cite[Eq.~(H.172)]{KW20}
\begin{equation}
\widehat{\upsilon}(\rho,\sigma)=\langle\varphi^{\rho}|(\sqrt{X}+1/\sqrt
{X}+I)|\varphi^{\rho}\rangle,
\end{equation}
and also because $\sqrt{X}+\frac{1}{\sqrt{X}%
}+I$ has its eigenvalues in $[3,\infty)$. Note that this latter statement justifies the inequality $\widehat{\upsilon}(\rho,\sigma)\geq 3$, which implies that $2\delta< \frac{\ln 3}{2 \ln \widehat{\upsilon}(\rho,\sigma)} \leq \frac{1}{2}$.  Letting $f(y)\coloneqq2\left(
\cosh(y)-1\right)  $, Taylor's theorem implies that there exists a constant
$c\in\left[  0,y\right]  $ such that%
\begin{align}
f(y) &  =f(0)+f^{\prime}(0)y+\frac{f^{\prime\prime}(c)}{2}y^{2}\\
&  =\frac{f^{\prime\prime}(c)}{2}y^{2}\\
&  =\cosh(c)y^{2}\\
&  \leq\cosh(y)y^{2}.
\end{align}
Using this and the fact that $s_{2\delta}(x)=f(2\delta\ln x)$, we find that%
\begin{align}
&  \frac{1}{\delta\ln2}s_{2\delta}(\widehat{\upsilon}(\rho,\sigma))\nonumber\\
&  \leq\frac{1}{\delta\ln2}\cosh(2\delta\ln\widehat{\upsilon}(\rho
,\sigma))\left(  2\delta\ln\widehat{\upsilon}(\rho,\sigma)\right)  ^{2}\\
&  =4\delta\left(  \log_{2}\widehat{\upsilon}(\rho,\sigma)\right)  ^{2}%
\ln2\cosh(2\delta\ln\widehat{\upsilon}(\rho,\sigma))\\
&  \leq4\delta\left(  \log_{2}\widehat{\upsilon}(\rho,\sigma)\right)  ^{2}.
\end{align}
The last inequality follows from the assumption that $\delta\leq\frac{\ln
3}{4\ln\widehat{\upsilon}(\rho,\sigma)}$, so that 
\begin{equation}
\ln2\cosh(2\delta
\ln\widehat{\upsilon}(\rho,\sigma))
\leq \ln2\cosh\!\left(\frac{\ln 3}{2}\right)
 \leq1.
\end{equation}

In the case that $\operatorname{supp}(\rho)\subseteq\operatorname{supp}%
(\sigma)$, we define $\rho_{\lambda}=\left(  1-\lambda\right)  \rho+\lambda
\pi_{\sigma}$, where $\lambda\in\left[  0,1\right]  $ and $\pi_{\sigma
}\coloneqq \Pi_{\sigma}/\operatorname{Tr}[\Pi_{\sigma}]$. Then applying the
above development we find that%
\begin{equation}
\widehat{D}_{1+\delta}(\rho_{\lambda}\Vert\sigma)\leq\widehat{D}(\rho
_{\lambda}\Vert\sigma)+4\delta\left[  \log_{2}\widehat{\upsilon}(\rho
_{\lambda},\sigma)\right]  ^{2}.
\end{equation}
The inequality in \eqref{eq:geo-BS-a-plus-bnd}\ then follows by taking the
limit $\lambda\rightarrow0$.
\end{IEEEproof}

\bigskip 
Now by applying Lemma~\ref{lem:geo-renyi-to-bs-uniform}, and an argument similar to that given for
Lemma~\ref{lem:sand-petz-renyi-to-qre-ch}, so that%
\begin{equation}
\widehat{D}_{1+\delta}(\mathcal{N}\Vert\mathcal{M})\leq\widehat{D}%
(\mathcal{N}\Vert\mathcal{M})+4\delta\left[  \log_{2}\widehat{\nu}(\mathcal{N}, \mathcal{M})  \right]  ^{2},
\label{eq:geo-renyi-ch-BS-up-bnd-uniform}%
\end{equation}
where
\begin{equation}
\widehat{\nu}(\mathcal{N}, \mathcal{M}) \coloneqq \sup_{\rho_{RA}}\widehat{\nu}(\mathcal{N}_{A\to B}(\rho_{RA}), \mathcal{M}_{A\to B}(\rho_{RA})),
\end{equation}
 we conclude the following:

\begin{lemma}
\label{lem:ch-rel-ent-limit}
Let $\mathcal{N}_{A\rightarrow B}$ be a quantum channel, and let
$\mathcal{M}_{A\rightarrow B}$ be a completely positive map. The following
limit holds
\begin{equation}
\lim_{\alpha\rightarrow1}\widehat{D}_{\alpha}(\mathcal{N}\Vert\mathcal{M}%
)=\widehat{D}(\mathcal{N}\Vert\mathcal{M}),
\end{equation}
where $\widehat{D}_{\alpha}(\mathcal{N}\Vert\mathcal{M})$ is the geometric
R\'enyi channel divergence and $\widehat{D}(\mathcal{N}\Vert\mathcal{M})$ is
the Belavkin--Staszewski channel relative entropy, both defined from  \eqref{eq:gen-ch-div}.
\end{lemma}

Finally, we have the following:

\begin{proposition}
\label{prop:lim-optimized-ch-div}
Let $\mathcal{N}_{A\rightarrow B}$ be a quantum channel, and let $\mathcal{C}$
be a compact set of completely positive maps. Then%
\begin{align}
\lim_{\alpha\rightarrow1}\inf_{\mathcal{M}\in\mathcal{C}}D_{\alpha
}(\mathcal{N}\Vert\mathcal{M})  &  =\inf_{\mathcal{M}\in\mathcal{C}%
}D(\mathcal{N}\Vert\mathcal{M}),\\
\lim_{\alpha\rightarrow1}\inf_{\mathcal{M}\in\mathcal{C}}\widetilde{D}%
_{\alpha}(\mathcal{N}\Vert\mathcal{M})  &  =\inf_{\mathcal{M}\in\mathcal{C}%
}\widetilde{D}(\mathcal{N}\Vert\mathcal{M}),\\
\lim_{\alpha\rightarrow1}\inf_{\mathcal{M}\in\mathcal{C}}\widehat{D}_{\alpha
}(\mathcal{N}\Vert\mathcal{M})  &  =\inf_{\mathcal{M}\in\mathcal{C}}%
\widehat{D}(\mathcal{N}\Vert\mathcal{M}).
\end{align}

\end{proposition}

\begin{IEEEproof}
First, if there does not exist $\mathcal{M}\in\mathcal{C}$ such that $D_{\max
}(\mathcal{N}\Vert\mathcal{M})<\infty$, then all quantities are equal to
$+\infty$. This is because the condition $D_{\max
}(\mathcal{N}\Vert\mathcal{M})<\infty$ holds if and only if $\operatorname{supp}(\Gamma_{RB}%
^{\mathcal{N}})\subseteq\operatorname{supp}(\Gamma_{RB}^{\mathcal{M}})$, where
$\Gamma_{RB}^{\mathcal{N}}$ and $\Gamma_{RB}^{\mathcal{M}}$ are the Choi
operators of $\mathcal{N}_{A\rightarrow B}$ and $\mathcal{M}_{A\rightarrow B}%
$, respectively, and all of the underlying quantities are equal to $+\infty$ if this condition does not hold (this is the case for $D_{\alpha}$, $\widetilde{D}_{\alpha}$, and $\widehat{D}_{\alpha
}$ for $\alpha > 1$ and it is also the case for these quantities in the limit $\alpha \to 1^-$).

So let us suppose that there is such an $\mathcal{M}\in\mathcal{C}%
$. We conclude that
\begin{align}
\lim_{\alpha\rightarrow1^-}\inf_{\mathcal{M}\in\mathcal{C}}D_{\alpha
}(\mathcal{N}\Vert\mathcal{M})  &  =\inf_{\mathcal{M}\in\mathcal{C}%
}D(\mathcal{N}\Vert\mathcal{M}),\\
\lim_{\alpha\rightarrow1^-}\inf_{\mathcal{M}\in\mathcal{C}}\widetilde{D}%
_{\alpha}(\mathcal{N}\Vert\mathcal{M})  &  =\inf_{\mathcal{M}\in
\mathcal{C}}\widetilde{D}(\mathcal{N}\Vert\mathcal{M}),\\
\lim_{\alpha\rightarrow1^-}\inf_{\mathcal{M}\in\mathcal{C}}\widehat{D}_{\alpha
}(\mathcal{N}\Vert\mathcal{M})  &  =\inf_{\mathcal{M}\in\mathcal{C}%
}\widehat{D}(\mathcal{N}\Vert\mathcal{M}),
\end{align}
by applying Lemmas~\ref{lem:sand-petz-renyi-to-qre-ch} and \ref{lem:ch-rel-ent-limit}, the $\alpha$-monotonicity of the underlying R\'enyi divergences, as well as \cite[Corollary~A2]{MH11}, along with the facts that $D_{\alpha
}(\mathcal{N}\Vert\mathcal{M})$, $\widetilde{D}%
_{\alpha}(\mathcal{N}\Vert\mathcal{M})$, and $\widehat{D}_{\alpha
}(\mathcal{N}\Vert\mathcal{M})$ are lower semi-continuous in $\mathcal{M}$ (see Lemma~\ref{lem:ch-rel-ent-lower-semicont} below).

By employing the fact that the channel relative entropies are ordered with respect to $\alpha$, so that the limit as $\alpha \to 1^+$ is the same as the infimum over $\alpha>1$, and applying Lemmas~\ref{lem:sand-petz-renyi-to-qre-ch} and \ref{lem:ch-rel-ent-limit}, we conclude that%
\begin{align}
\lim_{\alpha\rightarrow1^+}\inf_{\mathcal{M}\in\mathcal{C}}D_{\alpha
}(\mathcal{N}\Vert\mathcal{M})  &  =\inf_{\mathcal{M}\in\mathcal{C}%
}D(\mathcal{N}\Vert\mathcal{M}),\\
\lim_{\alpha\rightarrow1^+}\inf_{\mathcal{M}\in\mathcal{C}}\widetilde{D}%
_{\alpha}(\mathcal{N}\Vert\mathcal{M})  &  =\inf_{\mathcal{M}\in
\mathcal{C}}\widetilde{D}(\mathcal{N}\Vert\mathcal{M}),\\
\lim_{\alpha\rightarrow1^+}\inf_{\mathcal{M}\in\mathcal{C}}\widehat{D}_{\alpha
}(\mathcal{N}\Vert\mathcal{M})  &  =\inf_{\mathcal{M}\in\mathcal{C}%
}\widehat{D}(\mathcal{N}\Vert\mathcal{M}).
\end{align}
This concludes the proof.
\end{IEEEproof}

\begin{lemma}
\label{lem:ch-rel-ent-lower-semicont}
Let $\mathcal{N}$ be a quantum channel and $\mathcal{M}$ a completely positive
map. The channel divergences $D_{\alpha}(\mathcal{N}\Vert\mathcal{M})$,
$\widetilde{D}_{\alpha}(\mathcal{N}\Vert\mathcal{M})$, and $\widehat
{D}_{\alpha}(\mathcal{N}\Vert\mathcal{M})$\ are lower semi-continuous in
$\mathcal{N}$ and $\mathcal{M}$ for the values of $\alpha$ for which the
data-processing inequality holds.
\end{lemma}

\begin{IEEEproof}
For a state $\rho$ and a positive semi-definite operator~$\sigma$, it is known
that the underlying divergences $D_{\alpha}(\rho\Vert\sigma)$, $\widetilde
{D}_{\alpha}(\rho\Vert\sigma)$, and $\widehat{D}_{\alpha}(\rho\Vert\sigma
)$\ are lower semi-continuous in $\rho$ and $\sigma$ for the values of $\alpha$ for which
the data-processing inequality holds. This follows from the reasoning in
 \cite[Lemma~A.3]{FawziFawzi20}. We can then use this prove the desired statement for the channel
divergences, and we show the proof explicitly for $D_{\alpha}(\mathcal{N}%
\Vert\mathcal{M})$, with the proofs for the other quantities following the same line of reasoning. Let $\mathcal{N}_{n}$ be a sequence of channels that
converge to $\mathcal{N}$, and let $\mathcal{M}_{n}$ be a sequence of
completely positive maps that converge to $\mathcal{M}$ (we can take the
convergence to be in the diamond norm, but it is not so relevant since we are
in the finite-dimensional case). Then the desired statement is equivalent to
proving that%
\begin{equation}
\liminf_{n\rightarrow\infty}D_{\alpha}(\mathcal{N}_{n}\Vert\mathcal{M}%
_{n})\geq D_{\alpha}(\mathcal{N}\Vert\mathcal{M}).
\end{equation}
To this end, let $\rho_{RA}$ be an arbitrary state. It then follows that
$(\operatorname{id}_{R}\otimes\mathcal{N}_{n})(\rho_{RA})\rightarrow
(\operatorname{id}_{R}\otimes\mathcal{N})(\rho_{RA})$ and $(\operatorname{id}%
_{R}\otimes\mathcal{M}_{n})(\rho_{RA})\rightarrow(\operatorname{id}_{R}%
\otimes\mathcal{M})(\rho_{RA})$. From the lower semi-continuity of $D_{\alpha
}$, we conclude that%
\begin{multline}
\liminf_{n\rightarrow\infty}D_{\alpha}((\operatorname{id}_{R}\otimes
\mathcal{N}_{n})(\rho_{RA})\Vert(\operatorname{id}_{R}\otimes\mathcal{M}%
_{n})(\rho_{RA}))\\
\geq D_{\alpha}((\operatorname{id}_{R}\otimes\mathcal{N})(\rho_{RA}%
)\Vert(\operatorname{id}_{R}\otimes\mathcal{M})(\rho_{RA})).
\end{multline}
Since this holds for every state $\rho_{RA}$, we conclude that%
\begin{align}
& D_{\alpha}(\mathcal{N}\Vert\mathcal{M})\notag \\
& =\sup_{\rho_{RA}}D_{\alpha}((\operatorname{id}_{R}\otimes\mathcal{N}%
)(\rho_{RA})\Vert(\operatorname{id}_{R}\otimes\mathcal{M})(\rho_{RA}))\\
& \leq\sup_{\rho_{RA}}\liminf_{n\rightarrow\infty}D_{\alpha}%
((\operatorname{id}_{R}\otimes\mathcal{N}_{n})(\rho_{RA})\Vert
(\operatorname{id}_{R}\otimes\mathcal{M}_{n})(\rho_{RA}))\\
& \leq\liminf_{n\rightarrow\infty}\sup_{\rho_{RA}}D_{\alpha}%
((\operatorname{id}_{R}\otimes\mathcal{N}_{n})(\rho_{RA})\Vert
(\operatorname{id}_{R}\otimes\mathcal{M}_{n})(\rho_{RA}))\\
& =\liminf_{n\rightarrow\infty}D_{\alpha}(\mathcal{N}_{n}\Vert\mathcal{M}%
_{n}).
\end{align}
The second inequality follows because the quantity can only increase with the
supremum on the inside. 
\end{IEEEproof}

\begin{remark}
\label{rem:milan}
One can extend the statement of Lemma~\ref{lem:ch-rel-ent-lower-semicont} to values of $\alpha$ beyond those for which data processing holds, by the following argument. For all $\alpha\in(0,\infty)$ and $\varepsilon >0$, the relative entropies $D_\alpha(\rho\Vert \sigma + \varepsilon I)$ and $\widetilde{D}_\alpha(\rho\Vert \sigma + \varepsilon I)$ are continuous in $(\rho,\sigma)$ and monotone decreasing in $\varepsilon$. Furthermore, 
\begin{align}
D_\alpha(\rho\Vert \sigma ) & = \sup_{\varepsilon >0 }D_\alpha(\rho\Vert \sigma + \varepsilon I),
\\ \widetilde{D}_\alpha(\rho\Vert \sigma ) & = \sup_{\varepsilon >0 }\widetilde{D}_\alpha(\rho\Vert \sigma + \varepsilon I).
\end{align}
Since the supremum of a set of lower semi-continuous functions is lower semi-continuous, it follows that $D_\alpha(\rho\Vert \sigma )$ and $\widetilde{D}_\alpha(\rho\Vert \sigma )$ are lower semi-continuous in $(\rho,\sigma)$. Since this is all that Lemma~\ref{lem:ch-rel-ent-lower-semicont} relies upon, the desired statement follows for $D_\alpha$ and $\widetilde{D}_\alpha$. A similar conclusion can be made for $\widehat{D}_\alpha$ by invoking Theorem~5.5 of \cite{Hiai19}, where it was shown that the maximal $f$-divergence is lower semi-continuous for an arbitrary operator convex function $f$, and also noting that $\widehat{D}_\alpha$ is an example of a maximal $f$-divergence.
\end{remark}

\begin{IEEEbiographynophoto}{Dawei Ding} received the Ph.D. degree from Stanford University in 2019. He is currently a Research Scientist with the Alibaba Quantum Laboratory in Sunnyvale, California. His research interests span quantum Shannon theory, superconducting quantum computing, quantum computer systems, and quantum gravity.
\end{IEEEbiographynophoto}

\begin{IEEEbiographynophoto}{Sumeet Khatri} received the PhD degree from Louisiana State University in 2021. He is currently a postdoctoral researcher at Freie Universitat Berlin. His research interests are quantum communication, quantum computing, and machine learning.
\end{IEEEbiographynophoto}

\begin{IEEEbiographynophoto}{Yihui Quek} received her PhD from Stanford University in 2021. She is currently a Humboldt fellow at the Free University of Berlin, and will begin a Harvard Quantum Initiative Prize Postdoctoral Fellowship at Harvard University in late 2022. She is interested in quantum Shannon theory, quantum algorithms and applying learning theory to quantum settings, and is excited to contribute to building a new theory of physical computer science.
\end{IEEEbiographynophoto}

\begin{IEEEbiographynophoto}{Peter W.~Shor} received a B.S.~in Mathematics from Caltech in 1981. He then
got a Ph.D.~in Applied Mathematics from MIT in 1985. After a one year
post-doctoral fellowship at the Mathematical Sciences Research Institute in
Berkeley, he took a job at AT\&T Bell Laboratories. He worked there until
2003, when he went to MIT as the Morss Professor of Applied Mathematics.
In 1994, he discovered an algorithm for factoring large integers into primes
on a (still hypothetical) quantum computer. Since then, he has spent most of
his research time investigating quantum computing and quantum information
theory.
\end{IEEEbiographynophoto}

\begin{IEEEbiographynophoto}{Xin Wang} received the Ph.D. degree (Chancellor's List for Best Thesis) in quantum information from the University of Technology Sydney in 2018. He was a Hartree Post-Doctoral Fellow with the Joint Center for Quantum Information and Computer Science, University of Maryland, College Park. He is currently a Staff Researcher with the Institute for Quantum Computing, Baidu Research. His research investigates a broad range of perspectives of quantum computing and quantum information, including quantum communication, entanglement theory, fault-tolerant quantum computing, near-term quantum algorithms, and quantum machine learning.
\end{IEEEbiographynophoto}

\begin{IEEEbiographynophoto}{Mark M.~Wilde} (M'99--SM'13--F'22) was born in Metairie, Louisiana, USA. He received the Ph.D.~degree in electrical engineering from the University of Southern California, Los Angeles, California, in 2008. He is an Associate Professor in the Department of Physics and Astronomy and the Center for Computation and Technology at Louisiana State University, and, in July 2022, he will begin as Associate Professor of Electrical and Computer Engineering at Cornell University. His current research interests are in quantum Shannon theory, quantum computation, quantum optical communication, quantum computational complexity theory, and quantum error correction.
\end{IEEEbiographynophoto}

\end{document}